\documentclass[11pt]{article}
\usepackage{cite}
\usepackage{amsmath,amscd,amsfonts,mathtools,amssymb}
\usepackage[small,bf,hang]{caption}
\usepackage{slashed}
\usepackage{mathabx}
\usepackage{latexsym,epsfig}
\usepackage{stmaryrd}

\usepackage[vcentermath]{youngtab}

\usepackage[usenames,dvipsnames]{color}
\usepackage{amssymb}
\usepackage{breqn}

\def\hybrid{
        \topmargin -20pt
        \oddsidemargin 0pt
        \headheight 0pt \headsep 0pt
        \textwidth 6.25in 
        \textheight 9.5in 
        \marginparwidth .875in
        \parskip 5pt plus 1pt \jot = 1.5ex}

\hybrid

 \csname
@addtoreset\endcsname{equation}{section}

\def\moth{\mathsurround=0pt}
\newdimen\zo \zo=0pt

\def\tick{\leaders\hrule height 0.5ex depth 0pt \hskip 0.5pt}
\def\upboxfill{$\moth \setbox\zo\hbox{\tick}%
  \hskip 3pt\hbox to 0pt{$\tick$\hss}\hrulefill \hbox to 7.5pt{$\tick$\hss}$}

\def\dtick{\leaders\hrule height .34pt depth 0.5ex \hskip 0.5pt}
\def\downboxfill{$\moth \setbox\zo\hbox{\dtick}%
  \hskip 2pt\hbox to 0pt{$\dtick$\hss}\hrulefill \hbox to 2pt{$\dtick$\hss}$}

\def\bec{\begin{center}}
\def\ec{\end{center}}
\def\a{\alpha}  

\def\b{\beta}

\def\d{\delta} 
\def\D{\Delta}

\def\L{\Lambda}
\def\m{\mu}
\def\n{\nu}

\def\z{\zeta}
\def\O{\Omega}

\def\cL{{\cal L}}
\def\cD{{\cal D}}
\def\cF{{\cal F}}

\def\cM{{\cal M}}

\def\del{\partial}

\def\nn{\nonumber}

 \def\det{{\rm det\,}}
\def\be{\begin{equation}}
\def\ee{\end{equation}}
\newcommand{\beq}{\begin{equation}\begin{aligned}}
\newcommand{\eeq}{\end{aligned}\end{equation}}
\def\bea{\begin{eqnarray}}
\def\eea{\end{eqnarray}}
\def\ba{\begin{array}}
\def\ea{\end{array}}

\allowdisplaybreaks[2]

\begin{document}

\begin{titlepage}

\rightline\today
\begin{center}
\vskip 1.6cm
{\Large \bf {$L_\infty$ algebras and Tensor Hierarchies \\ ~ \\ in Exceptional Field Theory and Gauged Supergravity }}\\
 \vskip 2.0cm
{\large {Yago Cagnacci, Tomas Codina and Diego Marques}}
\vskip 0.5cm

{\it Instituto de Astronom\'ia y F\'isica del Espacio (CONICET-UBA)}\\ {\it Buenos Aires, Argentina} \\[1ex]

\vskip 0.5cm

{\small \verb"{yago, tcodina, diegomarques}@iafe.uba.ar"}

\vskip 1cm
{\bf Abstract}	
\end{center}

\vskip 0.2cm

\noindent
\begin{narrower}

{\small

We show how the gauge and field structure of the tensor hierarchies in Double and $E_{7(7)}$ Exceptional Field Theory fits into $L_\infty$ algebras. Special attention is paid to redefinitions, the role of covariantly constrained fields and intertwiners. The results are connected to Gauged Supergravities through generalized Scherk-Schwarz reductions. We find that certain gauging-dependent parameters generate trivial gauge transformations, giving rise to novel symmetries for symmetries that are absent in their ungauged counterparts. }

\end{narrower}

\vskip 1.5cm

\end{titlepage}

\tableofcontents




\vspace{0.5cm}

\section{Introduction}

While the homotopy associative $A_\infty$ algebra \cite{Stasheff:1963} is the mathematical structure  underlying the classical open string field theory sector, closed string field theory is organized by a homotopy Lie algebra, an $L_\infty$ algebra whose axioms and identities were given in \cite{Zwiebach:1992ie}-\cite{Lada:1992wc}. These algebras feature in a vast range of field theories \cite{Barnich:1997ij}, including consistent truncations of closed string field theory \cite{Sen:2016qap}, higher-spin theories \cite{Berends:1984rq} and gauge theories \cite{gauge}. Given this context it was recently suggested in \cite{Hohm:2017pnh} that $L_\infty$ algebras provide a classification of perturbative gauge invariant classical field theories. This expectation was supported in that paper by a number additional examples of field theories fitting into $L_\infty$ algebras: Chern-Simons theories, Einstein gravity, etc. Later on the scope was even expanded \cite{Blumenhagen:2017ogh}. More aligned with our plan is the fact that  Double Field Theory (DFT) \cite{Siegel:1993th}-\cite{Hull:2009mi} also enjoys an underlying $L_\infty$ structure: Courant algebroids can be cast in this language \cite{Roytenberg:1998vn} as well as their duality covariant counterparts \cite{Hull:2009zb}-\cite{Deser:2016qkw}, but more generally the full interacting theory exhibits this structure \cite{Hohm:2017pnh}.

The explicit relation between the elements appearing in an $L_\infty$ algebra and those in field theories has been systematized in \cite{Hohm:2017pnh}. The fields, gauge transformations and equations of motion belong to distinct graded subspaces in the algebra. Certain products can be read from the gauge transformations of the fields, the equations of motion, the closure identities, etc. Other products are then chosen to satisfy the $L_\infty$ identities, which eventually force the inclusion of additional graded subspaces.

This paper is devoted to discuss how tensor hierarchy \cite{deWit:2008ta} algebras fit into $L_\infty$ algebras\footnote{Interesting papers with similar goals are  \cite{Baraglia:2011dg},\cite{Cederwall:2018aab},\cite{Arvanitakis:2018cyo},\cite{Hohm:2018ybo}. Tensor hierarchies are also discussed in \cite{Riccioni:2007au},\cite{Greitz:2013pua} and in EFT in \cite{Aldazabal:2013via},\cite{Wang}}. We consider a number of representative examples in four space-time dimensions, such as the tensor hierarchy entering the Kaluza-Klein formulation of DFT (KK-DFT) \cite{Hohm:2013nja}, the one in $E_{7(7)}$ Exceptional Field Theory (EFT) \cite{Hohm:2013uia}, and those of gauged supergravities (in particular half-maximal \cite{Schon:2006kz}, maximal \cite{deWit:2002vt} and we also give a general discussion based on \cite{Weidner:2006rp}). For general reviews see \cite{reviews}.

\noindent A number of caveats are in order:
\begin{itemize}
\item We will only discuss the tensor hierarchy sector, namely $p$-form fields, and ignore other fields such as the graviton and scalars. The reason for this is that the metric must be invertible, and in duality covariant theories the scalars are grouped into a group-valued tensor, so they  necessarily involve a background field expansion making the analysis cumbersome. Considering the tensor hierarchy fields only involves finite expansions and so can be dealt with exactly and non-iteratively. It is important to emphasize that the products and identities involving field perturbations of the metric and scalars will be ignored, so those considered here are only a subset of the full story.

\item Tensor hierarchies are saturated by space-time dimensionality, namely, in $n$ space-time dimensions they can only include up to $n$-forms. However, the hierarchy can be projected to end at any given level $\leq n$, forming a subalgebra that closes exactly. In some cases the tower of $p$-forms ends before space-time saturation, this is the case of KK-DFT, where the tower has a single unprojected two-form and the tensor hierarchy algebra closes exactly. Other cases, like in EFT and some gauged supergravities the hierarchy is space-time saturated, but only a projection of it (via intertwiners) is dynamical and enters the action. Democratic formulations including all $p$-forms in the action are possible \cite{Bergshoeff:2009ph}. All these situations will be discussed here.

\item We will mostly focus on $L_\infty^{\rm gauge}$ or $L_\infty^{\rm gauge+fields}$, subalgebras of the full $L_\infty^{\rm full}$ algebra that includes interactions and dynamics. There will be exceptions such as when analyzing the KK-DFT tensor hierarchy, where we discuss the full $L_\infty^{\rm full}$ algebra including the equations of motion. Crucial to our analysis will be a theorem stating sufficient criteria for an algebra to fit into $L_{\infty}^{\rm gauge}$ \cite{Hohm:2017cey}, and a small extension introduced here to the $L_\infty^{\rm gauge + fields}$ case.
\end{itemize}

Let us briefly anticipate some points to be discussed in the paper:
\begin{itemize}
\item It is well known that Double and Exceptional Field Theories are restricted by a section condition, and as a consequence the gauge transformations admit non-vanishing trivial parameters. We show that something similar happens in gauged supergravities: the quadratic constraints imply that the gauge transformations admit non-vanishing trivial parameters that depend on the gaugings and are then absent in the ungauged theories. We will show how these two facts are connected through generalized Scherk-Schwarz compactifications.

\item It is a common saying that in Double and Exceptional Field Theories the gauge algebra closes with respect to a bracket whose Jacobiator is a trivial parameter. This is only true for a specific choice of variables. While field redefinitions leave the gauge algebra intact, parameter redefinitions change the brackets and their field dependence in such a way that the Jacobiator is no longer a trivial parameter. Any two sets of variables are of course equivalent, and must correspond to an isomorphism of the $L_\infty$ algebra, so the fact that the gauge algebra is an $L_\infty$ algebra remains intact, but the way in which the products are defined changes. In Double and Exceptional Field Theories, and in gauged supergravities there are two interesting set of parameters. One in which the gauge transformations of the fields can be cast in a manifestly covariant form (with respect to generalized diffeomorphisms and gauge transformations respectively): this is the usual set chosen in ExFT and gauged supergravities. For this set the brackets and trivial parameters are field dependent, and the Jacobiator in not a trivial parameter. There is a different set of parameters, related to the previous one through field dependent redefinitions, for which the brackets and trivial parameters do not depend on the fields, and the Jacobiator is a trivial parameter.

\item When dynamics is taken into account -so as to obtain the $L_\infty^{\rm full}$- the tensor hierarchy couples to the other fields through the equations of motion. It is then impossible to disentangle the tensor hierarchy from the rest of the theory and, even if the perturbations of the metric and scalars are ignored, the $L_\infty$ products and identities must depend on their background values. We will show this explicitly with some examples.

\item Some ExFT feature the presence of ``covariantly constrained fields'' satisfying section-type conditions. These are required by closure of the algebra and covariance of the field strengths, and so play a crucial role in the $L_{\infty}$ analysis. They are also crucial to establish how to extend the tensor hierarchy to higher forms, which also require new covariantly constrained fields.
\end{itemize}

The paper is organized as follows. Section \ref{sec-lft} is a self-contained review of $L_\infty$ algebras, closely following \cite{Hohm:2017pnh}. We review sufficient criteria  \cite{Hohm:2017cey} for a theory to fit into $L_\infty^{\rm gauge}$, and slightly generalized the theorem to the case $L_\infty^{\rm gauge + fields}$. In Section \ref{sec::DFT} we review the KK-formulation of DFT and show how the tensor hierarchy sector embeds into $L_\infty^{\rm full}$. The gauge algebra of the tensor hierarchy in $E_{7(7)}$ EFT  is discussed in Section \ref{sec::EFT}, together with its embedding in $L_\infty^{\rm gauge + fields}$ algebras. Finally in Section \ref{sec::gs} we show how tensor hierarchies in gauged supergravities fit into $L_\infty$ algebras. The Appendices contain complementary computations.

\section{$L_\infty$ algebra and its field theory} \label{sec-lft}

In this section we give a brief self-contained review of $L_\infty$ algebras in the $\ell$-picture and their relation to field theories, their gauge sector and dynamics, as presented in \cite{Hohm:2017pnh}. We refer to that paper and references therein for more details.

To have an $L_\infty$ algebra we first define a vector graded space $X$ which is the direct sum of vector spaces $X_n$, each of which has degree $n$
\beq
X = \bigoplus_{n}  X_n \,, \quad  n \in \mathbb{Z} \ .
\eeq
We will denote by $x$ an element of $X$ with definite degree, $i.e$, $x\in X_p$ for some fixed $p$.
Next we introduce multilinear products $\ell_k$
\beq
\ell_k:  X^{\otimes k} \rightarrow X \ ,
\eeq
with intrinsic degree $k-2$, meaning that when acting on elements $x_i$  we obtain
\beq
\hbox{deg}(\ell_k(x_1,x_2,...,x_k))= k-2 + \sum_{i=1}^k  \hbox{deg} (x_i) \ .
\eeq
It is useful to note that $\ell_1$ acts as a mapping as follows $\ell_1 : X_p \to X_{p-1}$.  The products are graded commutative and obey the $L_\infty$ relations, which constitute a deformation of Jacobi identities and will be introduced soon. The  graded commutative property means that a sign might appear when exchanging the arguments. For a permutation $\sigma$ of $k$ labels we have
\beq
\ell_k ( x_{\sigma(1)} , \ldots , x_{\sigma(k)} ) \ = \ (-1)^\sigma \epsilon(\sigma;x ) \,
\ell_k (x_1 , \ldots \,, x_k) \ .
\eeq
The $(-1)^\sigma$ factor gives a plus or minus sign if the permutation is even or odd, respectively. The $\epsilon(\sigma;x )$ factor is the Koszul sign. We first take a graded commutative algebra $\Lambda (x_1, x_2, \cdots )$ with
\be
x_i \wedge x_j \ = \ (-1)^{{\rm deg}(x_i) {\rm deg}(x_j)}   \,  x_j \wedge x_i \,,   \quad \forall i, j\  ,
\ee
and then define the Koszul sign for a general permutation as
\be
 x_1\wedge \ldots  \wedge x_k   =  \epsilon (\sigma; x)  \   x_{\sigma(1)} \wedge \ldots   \wedge \, x_{\sigma(k)} \ .
\ee
For example, for $\ell_2$ we get
 \be
  \ell_2(x_1,x_2) \ = \  (-1)^{1+ x_1 x_2}\,  \ell_2(x_2,x_1)\ ,
   \ee
where we introduced the notation
\beq
\ (-1)^{{\rm deg}(x_i) {\rm deg}(x_j)}\equiv(-1)^{x_i x_j}\ ,
\eeq
that is, within exponents the $x_i$ always refer to its degree.

The $L_\infty$ relations are labeled by a positive integer $n$ given by the number of inputs. Explicitly they take the general form
\be
\label{main-Linty-identity}
\sum_{i+j= n+1}  (-1)^{i(j-1)} \sum_\sigma  (-1)^\sigma \epsilon (\sigma; x) \, \ell_j \, \bigl( \, \ell_i ( x_{\sigma(1)}  \,, \, \ldots\,, x_{\sigma(i)} ) \,, \, x_{\sigma(i+1)}, \, \ldots \, x_{\sigma (n)} \bigr) \ = \ 0\ .
\ee
The sum over $\sigma$ is a sum over ``unshuffles'', it includes only the terms which satisfy
\be
\sigma(1) <  \, \cdots \, <  \, \sigma(i) \,,  \qquad
\sigma(i+1) <  \, \cdots \, <  \, \sigma(n) \ .
\ee
As a shorthand notation, it is common to write these relations as
\be
\label{main-Linty-identity-schem}
\sum_{i+j= n+1}  (-1)^{i(j-1)}  \ell_j \, \ell_i \ = \ 0\ ,
\ee
such that
\bea
n = 1 \ \ \ \ \ \ \ \ 0 &=& \ell_1 \ell_1 \\
n = 2 \ \ \ \ \ \ \ \ 0 &=& \ell_1 \ell_2 - \ell_2 \ell_1 \\
n= 3 \ \ \ \ \ \ \ \  0 &=& \ell_1 \ell_3 + \ell_2 \ell_2 + \ell_3 \ell_1 \\
n = 4 \ \ \ \ \ \ \ \ 0 &=& \ell_1 \ell_4 - \ell_2 \ell_3 + \ell_3 \ell_2 - \ell_4 \ell_1 \ , \ \dots
\eea

For $n=1$ we have
  \beq
	\label{l1-identity}
  \ell_1 ( \ell_1 (x)) \ = \ 0\ .
  \eeq
	This means that $\ell_1$ is a nilpotent operator, sometimes called $Q$ as the BRST operator.

The $n = 2$ identity is
  \be
  \label{L2L1}
 \ell_1(\ell_2(x_1,x_2)) \ = \ \ell_2(\ell_1(x_1),x_2) + (-1)^{x_1}\ell_2(x_1,\ell_1(x_2))\ .
 \ee
It implies that $\ell_1$ acts as a (graded) distributive operator on the arguments of $\ell_2$.

For $n=3$ we have
 \bea
  0  & = & \ell_2(\ell_2(x_1,x_2),x_3) + (-1)^{(x_1+ x_2) x_3}\ell_2(\ell_2(x_3,x_1),x_2)
   +(-1)^{(x_2+ x_3) x_1 }\ell_2(\ell_2(x_2,x_3),x_1) \label{L3L1}\\
   &&\!\!\!\!\!\!\!\!\!\!\!\!\!\!\!\!\!\!\!\! +  \ell_1(\ell_3 (x_1,x_2, x_3))    + \ell_3(\ell_1 (x_1) ,x_2, x_3)
  +  (-1)^{x_1}  \ell_3( x_1 ,\ell_1(x_2), x_3)
    +  (-1)^{x_1+ x_2}  \ell_3( x_1 ,x_2, \ell_1(x_3))  \ . \nn
 \eea
In the following we will see that when the arguments of $\ell_2$ are the gauge parameters of some field theory, it will be related to the bracket of the gauge algebra. As such,  the first line above will become the Jacobiator. For this reason, the last line characterizes a {\it deformation} of a strict  Lie algebra. Since the Jacobi identity
holds modulo a BRST exact term $(\ell_1\ell_3+\ell_3\ell_1)$, in the language of homological algebra $\ell_3$ is a
chain homotopy, so the saying is that $\ell_2$ satisfies the Jacobi identity ``up to homotopy'', or that this is an homotopy Lie algebra \cite{Lada:1992wc}.

We also display the $n=4$ identity, as will be needed later
\be\label{ell4ell1}
\begin{split}
0 \ = \ & \  \ \ell_1 ( \, \ell_4 ( x_1, x_2, x_3, x_4))  \\[1.0ex]
& - \ell_2 ( \, \ell_3 (x_1, x_2, x_3) , x_4)
+ \, (-1)^{x_3 x_4}\, \ell_2  ( \, \ell_3 (x_1, x_2, x_4) , x_3)
 \\
 & + (-1)^{(1+x_1)x_2} \ell_2 (x_2, \ell_3 (x_1, x_3, x_4))
 \, - (-1)^{x_1} \ell_2 (x_1, \ell_3 (x_2, x_3, x_4) ) \\[1.5ex]
 & +  \ell_3 ( \ell_2 (x_1, x_2 ) , x_3, x_4)
 \ + (-1)^{1 + x_2 x_3} \   \ell_3 ( \ell_2 (x_1, x_3 ) , x_2, x_4) \\
 &  +  (-1)^{x_4 (x_2 + x_3)} \ell_3 ( \ell_2 (x_1, x_4 ) , x_2, x_3)
 \ - \ell_3 ( x_1, \ell_2 (x_2, x_3 ) ,  x_4)  \\
 &  + (-1)^{x_3x_4}\ell_3 (x_1,  \ell_2 (x_2, x_4 ) , x_3)
\  + \ell_3 (x_1,  x_2, \ell_2 (x_3, x_4 ) )  \\[1.5ex]
&  - \ell_4  (\ell_1 (x_1), x_2, x_3, x_4)
 \ - (-1)^{x_1}  \ell_4 ( x_1, \ell_1 (x_2), x_3, x_4)   \\
 & - (-1)^{x_1+x_2}  \ell_4 ( x_1, x_2, \ell_1 (x_3), x_4)
 \ - (-1)^{x_1+ x_2 + x_4}  \ell_4 ( x_1, x_2, x_3, \ell_1 (x_4)) \ .
 \end{split}
\ee
Notice that, in the same sense that for $n=3$ the ``Jacobiator'' $\ell_2 \ell_2$ vanishes ``up to homotopy'', the same is true in $n = 4$ for $\ell_2 \ell_3 - \ell_3 \ell_2$.

After this brief self-contained introduction to $L_\infty$ algebras, we now show how to relate these results with field theories \cite{Hohm:2017pnh}. In the first place, we must assign a given degree $p$ to gauge parameters, fields, EOM, etc. and so specify to what vector subspace $X_p$ they belong. The general rule is to take the gauge parameters $\z$ as vectors of degree $p=0$, the dynamical fields $\Psi$ as vectors of degree $p=-1$ and the EOM $\cF$ as vectors of degree $p = -2$. The EOM's will form a subset of $X_{-2}$ and, when needed, we will refer to a general element of $X_{-2}$ as ``$E$".
\begin{table}[ht]
\begin{center}
    \begin{tabular}{|l|l|l|}
    \hline  $X_0$  & $X_{-1}$   &  $X_{-2}$  \\ \hline
Gauge Parameters
$\z$ & Fields $\Psi$ & $E \ (\ni  EOM  {\cal F})$ \\ \hline    \end{tabular}
    \end{center}
    \caption{Graded subspaces and the elements they contain.}
\end{table}
If the field theory exhibits symmetries for symmetries, this picture is incomplete and requires an extra graded subspace $X_{1}$ with elements parameterizing such ambiguity, and possibly further graded subspaces $X_{2},\,X_{3},\,...$.
Here, $\z$, $\Psi$ and $\cF$ stand for direct sums in case there are more than one of each. After these assignments, one can readily read some brackets from certain equations in the field theory. Lets see some examples.

\begin{itemize}
\item The gauge transformations {\it define} the brackets $\ell_{n+1}(\z,  \Psi^n)$ as follows
\be
        \delta_{\z}\Psi =\sum_{n\ge 0}   \dfrac{1}{n!}
      (-1)^{{n(n-1)}\over{2}}\,			
			 \ell_{n+1}(\z,  \Psi^n)
			 =\ell_1(\z)+\ell_2(\z,\Psi)-\dfrac{1}{2}\ell_3(\z,\Psi^2)-... \ , \label{gaugel}
\ee
where we use the exponential notation for short
\beq
\Psi^k = \underbrace{\Psi,...,\Psi}_{k\;\text{times}} \ .
\eeq
It can be checked that $\delta_{\z}\Psi$ so defined consistently  belongs to the same vector subspace than the fields, namely $X_{-1}$.

\item The EOM instead {\it define} the $\ell_n (\Psi^n)$ brackets as follows\footnote{As a side remark we point out that it might also be possible to go a step further and define an action from which the EOM are obtained, study the gauge algebra, the covariance of the EOM under gauge transformations, etc. For this one should be able to define an inner product
$ \langle x_1, x_2 \rangle $ with the properties
	\beq
\label{vmglrsbtt}
\begin{split}
\langle x_1, x_2 \rangle \ = \ &    (-1)^{x_1 x_2 } \langle x_2 , x_1 \rangle \,,\\
\langle x , \, \ell_n (x_1, \ldots x_{n}) \rangle
 \  =  \  &  (-1)^{x x_1  +1}  \langle x_1 , \, \ell_n (x, x_2  \ldots x_{n}) \rangle \ .
\end{split}
\eeq
The first one accounts for its graded symmetry,  and the second one implies that it is a multilinear graded-commutative function of all the arguments.
With this inner product the action is given by
\be
S \ =\ \sum_{n=1}^\infty  {(-1)^{n(n-1)\over 2}\over (n+1)!} \, \langle
\Psi,   \ell_n ( \Psi^n) \, \rangle \,,
\ee
and it can be checked that the EOM are obtained from it.}
\be
  {\cal F}(\Psi)  =  \sum_{n=1}^\infty
   {(-1)^{n(n-1)\over 2}\over n!}  \ell_n(\Psi^n)\ = \
   \ell_1(\Psi)  - \tfrac{1}{2}\ell_2(\Psi^2) - \tfrac{1}{3!} \ell_3(\Psi^3)
  \, +\tfrac{1}{4!}\ell_4(\Psi^4)+\cdots\ .\ \ \ \ \ \label{eoml}
  \ee

\item The study of the gauge algebra leads to interesting features. Taking the commutator of two gauge transformations and using the $L_\infty$ identities one finds
\beq
\label{commurel}
[\delta_{\z_1},\delta_{\z_2}] \Psi=\delta_{-\mathbf C(\z_1,\z_2, \mathbf \Psi)} \Psi+  \delta^{{}^T}_{\z_1, \z_2 } \ ,\Psi.
\eeq
with
\beq
\label{expan2}
            \mathbf C(\z_1,\z_2, \mathbf \Psi) \equiv\sum_{n\ge 0} {1\over n!}
            (-1)^{{n(n-1)\over 2}}
            \;\ell_{n+2}(\z_1,\z_2,\Psi^n)\ ,
\eeq
and
\beq
 \delta^{{}^T}_{\z_1, \z_2 } \Psi\,\equiv  \sum_{n\ge 0} {1\over n!}
            (-1)^{{(n-2)(n-3)\over 2}}
            \;\ell_{n+3}(\z_1,\z_2, \cF, \Psi^n)\ . \ \label{onshellclosure}
\eeq
The supralabel in $\delta^{{}^T}_{\z_1, \z_2 }$ stands for ``trivial" as it is a term that vanishes on-shell.
Suppose we want to rewrite in the $L_\infty$ framework some gauge algebra. If it closes off-shell, then $\delta^{{}^T}_{\z_1, \z_2 }\Psi=0$ and we obtain a gauge algebra that closes under the bracket given by $\mathbf C(\z_1,\z_2, \mathbf \Psi)$, which might be field dependent. If furthermore it turns out to be field independent, then we get
\beq
\label{closure}
                      [\delta_{\z_1},\delta_{\z_2}] \Psi
                      =\delta_{-\ell_2(\z_1,\z_2)} \Psi \ , \ \ \ \ \ell_{n+2}(\z_1,\,\z_2,\, \Psi^n) = 0 \ ,
\eeq
and the closure bracket is simply given by $\ell_2$, as anticipated under \eqref{L3L1}.

\item With respect to the gauge algebra, one can also consider the {\it gauge Jacobiator} ${\cal J}$, given by
\be
\label{gaugejacobiator}
{\cal J} (\z_1, \z_2, \z_3) \ \equiv \
\sum_{\hbox{\tiny\rm cyc}}  \bigl[ \delta_{\z_3}\,,  \, [  \delta_{\z_2}\,,
\, \delta_{\z_1} ]  \bigr] = 0 \, \ .
\ee
This vanishes by definition, as can be seen by expanding the terms and acting on a probe field from right to left, {\it i.e.} $
\d_{\z_1}\d_{\z_2}\d_{\z_3}\Psi=\d_{\z_1}\left(\d_{\z_2}\left(\d_{\z_3}\Psi\right)\right) $. Using the $L_\infty$ relations over \eqref{gaugejacobiator} one finds
\beq
\label{gjac}
\sum_{\hbox{\tiny\rm cyc}}  \bigl[ \delta_{\z_3}\,,  \, [  \delta_{\z_2}\,,
\, \delta_{\z_1} ]  \bigr] \,  \Psi  =  \,  -\delta_{\,
Q' \chi    } \, \Psi -
\, \delta^{{}^T}_ {
\chi } \Psi
\ ,
\eeq
with $\chi \in X_1$ given by
\beq
\chi=\sum_{n\ge 0} {1\over n!}
            (-1)^{{(n-2)(n-3)\over 2}}
            \;\ell_{n+3}(\z_1,\z_2, \z_3, \Psi^n)\ , \label{Prodxi3Psi}
\eeq
and
\beq
Q'\chi=\sum_{n\ge 0} {1\over n!}
            (-1)^{{n(n+1)\over 2}}
            \;\ell_{n+1}(\chi,\Psi^n) \label{trivparamQ}
\eeq
\beq
\delta^{{}^T}_ {\chi } \Psi= \sum_{n\ge 0} {1\over n!}
            (-1)^{{n(n-1)\over 2}}
            \;\ell_{n+2}(\cF,\chi,\Psi^n)\ .
\eeq
It can be shown that the RHS of \eqref{gjac} vanishes identically when using the $L_\infty$ relations, as it must be.

\item It will be useful in what follows to consider the gauge transformation of the equations of motion, which gives
	\beq
	\label{covf}
	\delta_\z\cF=\sum_{n\ge 0} {1\over n!}
            (-1)^{{n(n-1)\over 2}}
            \;\ell_{n+2}(\z_1,\cF, \mathbf \Psi)\,=\ell_2(\z,\cF)+\ell_3(\z,\cF,\Psi)-\frac{1}{2}\ell_4(\z,\cF,\Psi^2)+...
	\eeq
\end{itemize}

\begin{table}[ht]
\begin{center}
    \begin{tabular}{|l|l|l|}
    \hline  {\bf Product } \ \ \ \ \ \ \  \ \ \ \   & {\bf From} \ \ \ \  \ \ \ \ \ \ \  \ \ \ \   &  {\bf Equation} \ \ \ \ \ \   \\ \hline
$\ell_n(\Psi^n)$ & Equations of motion & (\ref{eoml}) \\ \hline
$\ell_{n+1} (\z ,\, \Psi)$ & Field gauge transformations & (\ref{gaugel}) \\ \hline
$\ell_{n+2} (\z^2 ,\, \Psi)$ & Closure & (\ref{expan2}) \\ \hline
$\ell_{n+3} (\z^3 ,\, \Psi)$ & Jacobiator & (\ref{Prodxi3Psi}) \\ \hline
$\ell_{n + 2}(\z ,\, {\cal F},\, \Psi^n)$ & EOM gauge transformations & (\ref{covf})\\ \hline
$\ell_{n + 3}(\z^2 ,\, {\cal F},\, \Psi^n)$ & On-shell closure & (\ref{onshellclosure})\\ \hline
    \end{tabular}
    \caption{Products that can be read from kinematic and dynamical equations in a field theory.}
    \end{center}
\end{table}

We collect in a Table 2 some of the products that can be directly read from standard expressions in a field theory. To see if a particular theory can be written in this framework one has to determine all $\ell_n$ products acting over all vectors (fields, gauge parameters and equations of motion) and then check all $L_\infty$ relations. It could also be possible that the vector subspaces defined so far are not enough (namely that Table 1 is incomplete) and one has to consider additional ones, as it happens for example in DFT where one has to add a vector subspace $X_1$. The fundamental formulas are \eqref{gaugel} and \eqref{eoml}. Knowing the specific form of the gauge transformations of the particular theory we are interested in, we can immediately read off the products $\ell_{n+1}(\z, \Psi^n)$ and $\ell_{n}( \Psi^n)$ respectively. In principle with these products and the $L_\infty$ relations we can determine all products. However, this can be a tedious work. At this point formulas like \eqref{commurel}, \eqref{gjac} and \eqref{covf} come to our rescue. They are a consequence of the previous ones plus the $L_\infty$ relations but they are also important as they  allow to read off certain products immediately. We insist however that whatever the route to identify products is, in the end all $L_\infty$ identities must be checked explicitly.

Take for example the case of DFT \cite{Hull:2009zb}. The gauge algebra closes under the C-bracket, which does not depend on fields
\beq
\left[\delta_{\z_1},\delta_{\z_2}\right] \Psi  =\delta_{-\left[\z_1,\z_2\right]_{(C)}} \Psi \ .
\eeq
Comparing this expression with \eqref{closure} we can readily make the identification
\beq
\ell_2(\z_1,\z_2)=\left[\z_1,\z_2\right]_{(C)} \ .
\eeq
Now we could evaluate the identity for three gauge parameters \eqref{L3L1}. We have pointed out that the last line of \eqref{L3L1} in this particular case gives the Jacobiator, which in the case of DFT is famously given by
\beq
J (\z_1, \z_2, \z_3 ) \ \equiv  \ 3\bigl[ \, [\z_{[1} , \z_2 ] \,, \z_{3]} \bigr] \ =\del \,  N(\z_1, \z_2, \z_3) \ ,
\eeq
with the so-called Nijenhuis scalar defined by
\beq
N(\z_1, \z_2, \z_3) \ \equiv \ \tfrac{1}{2}
 \, [\z_{[1}, \z_2 ]_{(C)} \cdot \z_{3]}\ .
\eeq
After setting $\ell_3(\ell_1(\z_i),\z_j,\z_k)\equiv0$ we can identify
\beq
\ell_3(\z_1,\z_2,\z_3)&=-N(\z_1, \z_2, \z_3)\ ,\\
\ell_1(N(\z_1, \z_2, \z_3))&=\del N(\z_1, \z_2, \z_3)\ .
\eeq
These two last identifications done after evaluating a particular $L_\infty$ identity could also be obtained by directly comparing the DFT gauge Jacobiator with \eqref{gjac}
\beq
\sum_{\hbox{\tiny\rm cyc}}  \bigl[ \delta_{\Lambda_3}\,,  \, [  \delta_{\Lambda_2}\,,
\, \delta_{\Lambda_1} ]  \bigr] \,  \Psi  =  \,  \delta_{\del N(\z_1, \z_2, \z_3)}\Psi \ ,
\eeq
which, of course, is zero when evaluating the gauge variation of the field after using the strong constraint.

A clarification is in order. We have said that we need to determine all products acting over all vectors and that some equations make this possible, for example \eqref{eoml} leads to the knowledge of $\ell_{n}(\Psi^n)$. However, we would like to know not only the diagonal part of $\ell_{n}$ acting on fields, but $\ell_{n}(\Psi_1, \Psi_2, ...,\Psi_n)$ with non diagonal entries. This can be achieved with the help of polarization identities. For the simplest case, $\ell_2$, we get
\beq
2\ell_2 (\Psi_1, \Psi_2) \ = \ & \   \ell_2 (\Psi_1 + \Psi_2,\Psi_1 + \Psi_2 ) - \ell_2 (\Psi_1,\Psi_1) - \ell_2 (\Psi_2,\Psi_2)\ ,
\eeq
and this expression can be generalized to any $\ell_n$.

After we have identified the products, we must check the $L_\infty$ identities. However, there is a large subset of identities which are automatically fulfilled due to the general construction of an $L_\infty$ field theory and it is not necessary to do an explicit check. As shown in \cite{Hohm:2017pnh},	the identities acting over a list of fields $(\Psi_1, \, \ldots\, , \Psi_n)$ hold true provided we define
	\beq
	\ell_{n+1} ( E, \Psi_1, \, \ldots\, , \Psi_n) \ = \ 0 \,,   \quad n \geq 0, \quad E \in X_2 \ .
	\eeq
The identities acting over a list $(\z_1,\z_2, \Psi_1, \, \ldots\, , \Psi_n)$,   and $(\z, \Psi_1, \, \ldots\, , \Psi_n)$ with $n\geq 1$ also hold true. These identities come from closure of the gauge algebra over fields and the gauge transformations of the equations of motion, respectively.

\subsection{Sufficient criteria for the gauge sector to fit into $L_\infty$} \label{sec::theorem}

In \cite{Hohm:2017cey} a theorem was presented, stating sufficient conditions for a gauge algebra to lie within an $L_\infty^{\rm gauge}$ structure. Since we will refer to it often, we dedicate a separate subsection to review it briefly. The theorem only refers to the gauge algebra, not the complete field theory. Given an algebra $(V,[\, \cdot,\cdot\,])$ with bilinear antisymmetric 2-bracket and a vector space $U$ with a linear map ${\cal D}:\, U\rightarrow V$
satisfying
  \be\label{bracketCOMP}
  [{\rm Im}({\cal D}), V] \ \subset \ {\rm Im}({\cal D})\ ,
  \ee
and
 \be\label{JACisMAP}
  \forall v_1, v_2, v_3 \in V : \; \; {\rm Jac}(v_1,v_2,v_3) \ \in \ {\rm Im}({\cal D}) \ ,
 \ee
where ${\rm Im}({\cal D})$ denotes the image of ${\cal D}$, there exists a 3-term L$_{\infty}$ structure with $\ell_2(v,w)=[v,w]$ on the graded vector space with
 \be\label{longerchain}
  X_2 \;\xrightarrow{\ell_1=\iota } \; X_1 \; \xrightarrow{\ell_1={\cal D}}
  \; X_{0}\ ,
 \ee
where $X_0=V$, $X_1=U$, $X_2={\rm Ker}({\cal D})$, ${\rm Ker}({\cal D})$ denotes the kernel of ${\cal D}$ and $\iota$ denotes the
inclusion of ${\rm Ker}({\cal D})$ into $U$.

In a more physical language, the hypothesis of this theorem are roughly
\begin{itemize}
\item The field transformations admit trivial parameters. The gauge algebra closes with respect to a given bracket. The bracket between a trivial parameter and a generic one is itself a trivial parameter.
\item The Jacobiator computed from the bracket is a trivial parameter.
\end{itemize}
When these hypothesis are met, the theorem states that the gauge algebra fits into an $L_\infty$ structure, with a few non-vanishing products written in equation (4.26) in \cite{Hohm:2017cey}.

We will show that all the cases considered in this paper satisfy these hypothesis. However, this will prove not to be always a simple task: the form of the brackets and the trivial parameters is highly sensitive to redefinitions, and the criteria outlined above is only useful when a specific set of parameters is considered. Meeting the criteria involves finding an appropriate set of variables, and we will comment on what happens when other sets  are considered.

In the next subsection we point out that the consequences of these hypothesis are in fact a little broader: when the hypothesis are met the algebra fits into $L_\infty^{gauge+fields}$.

\subsection{Sufficient criteria for the gauge + field sectors to fit into $L_\infty$} \label{sec::theorem2}

The  theorem of the previous Subsection \ref{sec::theorem} can be extended so as to consider in addition the graded space of fields, in what was called in \cite{Hohm:2017pnh} $L_\infty^{\text{gauge+fields}}$. To be more precise, we will take the fields to belong to the graded subspace $X_{-1}$ and consider their gauge transformations given by \eqref{gaugel}. We will not consider their EOM, so all the products in \eqref{eoml}, namely $\ell_n(\Psi^n)$, are taken to be null and thus there is no need for a graded subspace $X_{-2}$. Under the same conditions, we show that the algebra is in fact that of $L_\infty^{\rm gauge + fields}$ and the only additional non-vanish products (apart from those in equation (4.26) in \cite{Hohm:2017cey}) are those obtained from the gauge transformations of the fields \eqref{gaugel}, namely
\be
        \delta_{\z}\Psi =\sum_{n\ge 0}   \dfrac{1}{n!}
      (-1)^{{n(n-1)}\over{2}}\,			
			 \ell_{n+1}(\z,  \Psi^n)
			 =\ell_1(\z)+\ell_2(\z,\Psi)-\dfrac{1}{2}\ell_3(\z,\Psi^2)-... \ .
\ee

We have to deal now with products involving fields $\Psi\in X_{-1}$ and other elements from other subspaces. We can consider the $L_\infty$ relations acting over the following lists:
\begin{itemize}	
\item Two  $\z's \in X_{0}$ and any number $n\geq0$ of $\Psi's \in X_{-1}$: $(\z_1,\z_2,\Psi_1,\Psi_2,...)$.
	\item One  $\z \in X_{0}$ and any number $n>0$ of $\Psi's \in X_{-1}$: $(\z,\Psi_1,\Psi_2,...)$.
		\item Only one $\z \in X_{0}$. There is no $\Psi\in X_{-1}$ in this list, but we have to consider it because the product $\ell_1$ maps to the subspace of fields: $\ell_1(\z)\in X_{-1} $.
		\item At least one $c \in X_2$ (and any other vector on the list, including at least one $\Psi \in X_{-1}$): $(c,...)$.
		\item At least one $\chi \in X_1$ (and any other vector on the list, including at least one $\Psi \in X_{-1}$): $(\chi,...)$.
	\item Three or more  $\z's \in X_{0}$ and any number $n>0$ of $\Psi's \in X_{-1}$: $(\z_1,\z_2,\z_3...,\z_n, \Psi_1,\Psi_2,...)$.	
\end{itemize}

As shown in \cite{Hohm:2017pnh}, the products given by the first case satisfy the $L_\infty$ relations as a consequence of closure of the gauge algebra over fields. The second item is trivially fulfilled once we make the choice $\ell_1(x)=0$, with $x\in X_{-1}$.
The third one gives rise to the identity $\ell_1(\ell_1(\z))=0$ which is verified trivially for the same reason.

For the other cases, we will construct all possible terms of the identities, which are of the form $\ell_i(...\ell_j(...)) $ and show that they are null or compensate each other. We will make use of the nontrivial products defined in equation (4.26) in \cite{Hohm:2017cey}.

\paragraph{At least one $c \in X_2$ (and any other vector on the list, including at least one $\Psi \in X_{-1}$).} There are only two nontrivial products involving a $c$ in $X_2$: $\ell_1(c)$, $\ell_2(c,\z)$. First consider that we have at least two $c's$. We would get the products
\bea
	&&\ell_i(c_1,\Psi,...\underbrace{\ell_2(c_2,\z)}_{\equiv \tilde c \in X_{2}})\Rightarrow \ell_i(c_1,\Psi,...\tilde c)=0, \\
		&&\ell_i(c_1,\Psi,...\underbrace{\ell_1(c_2)}_{\equiv  \chi \in X_{1}})\Rightarrow \ell_i(c_1,\Psi,...\chi)=0,\\
	&&\ell_i(c_1,c_2...\ell_j(...))=0\ .
	\eea
		For only one $c$ we get
	\bea
	&&\ell_i(\Psi,...\underbrace{\ell_1(c)}_{\equiv \chi \in X_{1}})\Rightarrow \ell_i(\Psi,...\chi)=0, \\
		&&\ell_i(\Psi,...\underbrace{\ell_2(c,\z)}_{\equiv \tilde c \in X_{2}})\Rightarrow \ell_i(\Psi,...\tilde c)=0, \\
	&&\ell_i(c,\underbrace{\ell_j(\Psi...)}_{\text{must be in } X_0.})\ ,
	\eea
but it is not possible to form a nontrivial $\ell_j(\Psi...) \in X_0$, thus the last product is also zero.

\paragraph{At least one $\chi \in X_1$ (and any other vector on the list, including at least one $\Psi \in X_{-1}$).} There are only four nontrivial products involving $\chi$: $\ell_1(\chi)$, $\ell_2(\chi,\z)$, $\ell_2(\chi_1,\chi_2)$, $\ell_3(\chi,\z_1,\z_2)$ . This means that we can have at most two $\chi$'s.

 If we have more than two $\chi$'s we would necessary have to group at least one with a $\Psi$ inside a product (as we cannot have more than two $\chi's$ together in the same product), which is zero. Take for example the case of three $\chi$'s:
\beq
\ell_i(\chi_1,\chi_2,...\underbrace{\ell_j(\chi_3, \Psi...)}_{=0})\qquad\text{or}\qquad \ell_i(\chi_1,...\underbrace{\ell_j(\chi_2,\chi_3, \Psi...)}_{=0})\ .
\eeq
Now take the list $(\chi_1,\chi_2,...)$. The possibilities are:
	\bea
		&&\ell_i(\chi_1...\underbrace{\ell_j(\chi_2, \Psi...)}_{=0})=0\\
		&&\underbrace{\ell_i(\chi_1, \Psi...\ell_j(\chi_2, ...))}_{=0}=0\\
		&&\ell_i(\Psi...\underbrace{\ell_2(\chi_1,\chi_2)}_{\equiv c\in X_2})= \ell_i(\Psi,...c)=0\ ,
	\eea
	where we have used in the last line that there is no product mixing a $c\in X_2$ and a $\Psi\in X_{-1}$.
	
	Now take the list with only one $\chi$ and consider it as an argument of the second product. As we cannot take together a $\Psi\in X_{-1}$ and a $\chi\in X_{1}$ inside the same product, the possibilities are
	\bea
	&&\ell_i(\Psi...\underbrace{\ell_3(\chi,\z_1,\z_2)}_{\equiv c\in X_2})=\ell_i(\Psi...c)=0\\
	&&\ell_i(\Psi...\underbrace{\ell_2(\chi,\z)}_{\equiv \tilde \chi \in X_1})=\ell_i(\Psi...\tilde\chi)=0\\
		&&\ell_i(\Psi...\underbrace{\ell_1(\chi)}_{\equiv \z_{triv.} \in X_0})\Rightarrow \ell_{n+1}(\Psi^n,\z_{triv.})=0\ .
	\eea
	
The last line gives zero, as these products come from the gauge transformations of the fields with a trivial parameter.	
	
Considering that $\chi$ is in the outermost product, we get
	\beq
	\ell_i(\chi,\ell_j(\Psi...))\Rightarrow
	\begin{dcases}
   \ell_j(\Psi...)\in X_0,\quad\text{not possible},\\
   \ell_j(\Psi...)\in X_1,\quad\text{not possible}\ .
\end{dcases}
	\eeq
	
		\beq
	\ell_i(\chi,\z,\ell_j(\Psi...))\Rightarrow
   \ell_j(\Psi...)\in X_0,\quad\text{not possible} \ .
	\eeq
	
	\paragraph{Three or more  $\z's \in X_{0}$ and any number $n>0$ of $\Psi's \in X_{-1}$.} We start by considering three gauge parameters. We first notice that we cannot use the products $\ell_{n+1} (\z,\Psi^n)$, which give an element $\tilde \Psi\in X_ {-1}$. If they were in the second product $\ell_j$ we would have
\beq
&\ell_i(\z_1,\z_2,\ell_{n+1} (\z,\Psi^n))=\ell_3(\z_1,\z_2,\tilde\Psi)=0\ .
\eeq
Trying to use them in the first product does not work either. For $\ell_1 (\z)$ we would need the second product to give an element of $ X_0$:
\beq
\ell_1 (\underbrace{\ell_{n+3}(\z_1,\z_2,\z_3, \Psi^n)}_{\notin X_0})\ .
\eeq
For $\ell_2 (\z ,\Psi)$ we would need the second product to give an element of $ X_0$ or $ X_{-1}$ :
\beq
\ell_2 (\underbrace{\ell_{n+2}(\z_1,\z_2,\z_3,\Psi^{n-1}}_{\notin X_0}) ,\Psi),
\qquad
\ell_2 (\z_1,\underbrace{ \ell_{n+2}(\z_2,\z_3,\Psi^{n}}_{\notin X_{-1}}) )\ .
\eeq
For $\ell_3 (\z,\Psi,\Psi)$ we would need the second product to give an element of $ X_0$ or $ X_{-1}$ :
\beq
\ell_3 (\underbrace{\ell_{n+1}(\z_1,\z_2,\z_3,\Psi^{n-2}}_{\notin X_0}) ,\Psi,\Psi),
\qquad
\ell_3 (\z_1,\underbrace{ \ell_{n+1}(\z_2,\z_3,\Psi^{n-1}}_{\notin X_{-1}}),\Psi )\ .
\eeq
Thus, one of the products must be $\ell_3(\z_1,\z_2,\z_3)\equiv\chi\in X_1$. If it was in the second product $\ell_j$ we would have
\beq
&\ell_i(...\ell_j(...))=\ell_{n+1}(\Psi^n,\ell_3(\z_1,\z_2,\z_3))=\ell_{n+1}(\Psi^n,\chi)=0\ .
\eeq
 If it was in the first product, we would need the second one to give a $X_0$ element to get a nontrivial term, but this is not the case
\beq
\ell_3(\z_1,\z_2,\underbrace{\ell_{n+1}(\z_3,\Psi^n)}_{\notin X_0})\ .
\eeq
Considering more than three gauge parameters and repeating the arguments as before one also finds that the $L_\infty$ identities are fulfilled trivially.

\section{$L_\infty$ and Double Field Theory} \label{sec::DFT}

The embedding of DFT into $L_\infty^{\rm full}$ was performed in \cite{Hohm:2017pnh}, and so there will be no surprises in this section. Here we consider the KK-formulation of DFT \cite{Hohm:2013nja} with the goal of learning how its tensor hierarchy fits into $L_\infty$. The advantage of this formulation is that it is similar in structure with EFT and gauge supergravities, and so will serve as a platform to extract lessons for future discussions.

\subsection{Review of the Kaluza-Klein formulation of Double Field Theory}
In this section we review the KK-formulation of DFT \cite{Hohm:2013nja}. We present in the Appendix the relation of this formulation with the generalized metric approach \cite{Hohm:2010pp}. The fields and gauge parameters depend on external and internal coordinates $(x^\mu\, , \, X^{M})$, with $\mu$ an external $GL(n)$ index, and $M$ an internal fundamental $O(d,d)$ index. The duality invariant metric $\eta_{M N}$ and its inverse raise and lower internal indices, and the internal coordinate dependence is restricted by the strong constraint that states that fields and parameters and their products are annihilated by the duality invariant Laplacian
\be
 \partial_M \otimes \partial^M \dots = 0 \ .
\ee
The fields are
\be
g_{\mu \nu} \ , \ \ \ B_{\mu \nu} \ , \ \ \ \phi \ , \ \ \ A_\mu{}^M \ , \ \ \ {\cal M}_{M N} \ ,
\ee
the scalar matrix ${\cal M}_{M N}$ being a duality group-valued constrained field (i.e. ${\cal M}_M{}^P {\cal M}_P{}^N = \delta^N_M$). The gauge fields $A_\mu{}^M$ and $B_{\mu \nu}$ will be referred to as the fields of the {\it tensor hierarchy}.

\noindent The symmetries of the theory are:
\begin{itemize}
\item A global $O(d,d|\mathbb{R})$ symmetry.
\item A local $O(1,n-1|\mathbb{R}) \times O(d|\mathbb{R}) \times O(d|\mathbb{R})$ which is trivial in this formulation.
\item External diffeomorphisms, parameterized by a vector $\xi^{\mu}$.
\item Gauge transformations of the two-form, parameterized by a one-form $\Xi_{\mu}$.
\item Internal generalized diffeomorphisms, parameterized by an $O(d,d)$ vector $\Lambda^{M}$.
\end{itemize}

We will deal with two different sets of parameters, related by field-redefinitions. Those noted with a hat $(\widehat \Lambda \, , \, \widehat \Xi)$ are such that the gauge transformations can be written in a covariant form with respect to internal generalized diffeomorphisms. An alternative set of parameters $(\Lambda \, , \, \Xi)$ are more convenient in order to make contact with the usual formulation of DFT (see Appendix) and to explore how the tensor hierarchy gauge algebra is that of $L_\infty$. Both sets of parameters are related as follows
\bea
\widehat \Lambda^M &=& \Lambda^M + \xi^\mu  A_\mu{}^M \ , \label{ParamRedefDFT}\\
\widehat \Xi_\mu &=& \Xi_\mu - \left(B_{\mu \nu} - \frac 1 2 A_\mu{}^M A_{\nu M}\right)\, \xi^\nu + A_\mu{}^M \Lambda_M\ . \nn
\eea

\noindent The covariant form of the local symmetries is the following
\bea
\delta g_{\mu \nu} &=& \left({\cal L}_\xi^{\cal D} + \widehat {\cal L}_{\widehat \Lambda} \right) g_{\mu \nu} = \xi^{\rho} {\cal D}_{\rho} g_{\mu \nu} + 2\, {\cal D}_{(\mu}{\xi^{\rho}} g_{\nu) \rho} + \widehat \Lambda^M \partial_{M} g_{\mu \nu} \ , \label{from}\\
\delta B_{\mu \nu} &=& A_{[\mu}{}^M \, \delta A_{\nu]M} + \xi^\rho \, {\cal H}_{\rho \mu \nu} + 2 {\cal D}_{[\mu} \widehat \Xi_{\nu]} - {\cal F}_{\mu \nu}{}^M \, \widehat\Lambda_M \ , \label{TransfB}\\
\delta e^{-2\phi} &=& \xi^{\rho}{\cal D}_{\rho} e^{-2\phi} + \partial_{M}\left(\widehat \Lambda^M e^{- 2\phi} \right) \ , \ \ \ \ \delta \phi \ =\  \xi^\rho {\cal D}_{\rho} \phi + \widehat \Lambda^M \partial_M \phi - \frac 1 2 \partial_M \widehat \Lambda^M \ , \\
\delta A_{\mu}{}^M &=& \xi^\rho\, {\cal F}_{\rho \mu}{}^M + g_{\mu \rho} \, {\cal M}^{M N} \partial_N \xi^\rho + {\cal D}_\mu \widehat \Lambda^M + \partial^M \widehat \Xi_\mu \ , \label{TransfA}\\
\delta {\cal M}_{M N} &=& \xi^\rho {\cal D}_{\rho} {\cal M}_{M N} + \widehat {\cal L}_{\widehat \Lambda} {\cal M}_{M N} \ . \label{to}
\eea

Before defining the quantities that appear in these transformations, let us briefly go through a review of generalized diffeomorphisms. First we define the D-bracket
\be \left[ \Lambda \, , \, V\right]_{(D)}^M = \Lambda^{P} \partial_{P} V^M + \left(\partial^M \Lambda_P - \partial_P \Lambda^M \right) V^P \ ,\label{Dbracket}
\ee
which is not antisymmetric but does satisfy the Jacobi identity. In terms of it we can define the generalized Lie derivative
\be
\widehat {\cal L}_{\Lambda} V^M = \left[ \Lambda \, , \, V\right]_{(D)}^M  + \lambda (V) \, \partial_P \Lambda^P V^M\ .
\ee
Here $\lambda(V)$ is called a weight: all tensors in this section have vanishing weight except for the generalized dilaton. Extending the action of the generalized Lie derivative to tensors of higher rank is straightforward. Crucially, it admits trivial gauge parameters of the form
\be
\Lambda_{trivial}^M = \partial^M \chi \ , \ \ \ \ \ \ \widehat {\cal L}_{\Lambda_{trivial}} \dots = 0 \ , \label{TrivialDFT}
\ee
parameterized by functions $\chi(x,X)$. Then, we have a situation of symmetries for symmetries in DFT: two apparently different gauge parameters generate the same transformation if they are related by a trivial parameter. There is even a symmetry for symmetries for symmetries situation, given by constant shifts in the space of functions that leave the trivial parameters invariant.

While the symmetric part of the D-bracket is in fact a trivial parameter
\be \left[\Lambda_{(1} \, , \, \Lambda_{2)}\right]_{(D)}^M = \partial^M \left( \frac 1 2 \Lambda_{1}^P \Lambda_{2 P} \right) \ ,
\ee
the antisymmetric part is called the C-bracket
\be
\left[\Lambda_1 \, , \, \Lambda_2 \right]_{(C)}^M = \left[\Lambda_{[1} \, , \, \Lambda_{2]}\right]_{(D)}^M = 2 \Lambda_{[1}^P\partial_P \Lambda_{2]}^M - \Lambda_{[1}^P\partial^M \Lambda_{2]P} \ ,
\ee
and doesn't satisfy the Jacobi identity.

External derivatives $\partial_\mu$ of objects that transform tensorially with respect to these transformations are not covariant, so making these symmetries manifest requires the introduction of new derivatives
\be
{\cal D}_\mu = \partial_\mu - \widehat {\cal L}_{A_\mu}  \ \ \ \ \rightarrowtail \ \ \ \ \delta_\Lambda \, D_{\mu} \dots = \widehat {\cal L}_{\Lambda} \, D_\mu \dots \ , \label{CovDer}
\ee
which are covariant because the connection transforms as follows with respect to internal generalized diffeomorphisms (\ref{TransfA})\footnote{Since the gauge vector $A_\mu$ enters the covariant derivative as a generator of the generalized Lie derivative, its gauge transformation is only determined up to trivial parameters of the form (\ref{TrivialDFT}). However, all possible ``orbits'' are related through redefinitions of the parameters $\Xi$, as can be seen from (\ref{TransfA}).}
\be
\delta_\Lambda A_{\mu}{}^M = \partial_\mu \Lambda^M + \left[\Lambda \, , \, A_\mu\right]{}_{(D)}^M \ .
\ee

The dilaton $e^{-2\phi}$ transforms as a density (it is in fact the only tensor in KK-DFT with non-vanishing weight $\lambda\left(e^{-2\phi}\right) = 1$)
\be
\delta_\Lambda e^{-2 \phi} = \partial_{M}\left(\Lambda^M e^{-2\phi}\right) \ ,
\ee
and so its covariant derivative reads explicitly\footnote{We note in passing that this covariant derivative can be integrated by parts in the presence of the measure $e^{- 2\phi}$
\be
\int d^n x d^{2d} X\, e^{-2\phi}\, V^{\mu M_1 \dots M_k}\, {\cal D}_\mu U_{M_1 \dots M_k} = - \int d^n x d^{2d} X \, {\cal D}_{\mu}\left(e^{-2\phi} V^{\mu M_1 \dots M_k}\right)\,  U_{M_1 \dots M_k} + {\rm T.D.} \nn
 \ee}
\bea
{\cal D}_\mu e^{-2\phi} &=& \partial_\mu e^{-2\phi} - \partial_M\left(A_\mu{}^M \, e^{-2\phi} \right) \ , \nn\\
{\cal D}_\mu \phi &=& \partial_\mu \phi - A_{\mu}{}^M \partial_M \phi + \frac 1 2 \partial_M A_{\mu}^M \ , \label{CovariantDerivatived}
\eea
such that ${\cal D}_\mu \phi$ transforms as a scalar with vanishing weight
\bea
\delta_\Lambda {\cal D}_{\mu} e^{- 2 \phi} &=& \partial_M\left(\Lambda^M \, {\cal D}_\mu e^{-2\phi}\right) \ , \\
\delta_\Lambda {\cal D}_{\mu} \phi &=& \Lambda^M \, \partial_M {\cal D}_\mu \phi \ .
\eea

The covariant curvatures of the gauge fields are given by
\bea
{\cal F}_{\mu \nu}{}^M &=& 2 \partial_{[\mu} A_{\nu]}{}^M - \left[ A_\mu \, , \, A_\nu \right]_{(C)}^M - \partial^M B_{\mu \nu} \ ,  \label{FieldstrengthsDFT} \\
{\cal H}_{\mu \nu \rho} &=& 3 \left({\cal D}_{[\mu} B_{\nu \rho]} + A_{[\mu}{}^M \partial_\nu A_{\rho]M} - \frac 1 3 A_{[\mu}{}^M \left[A_\nu \, , \, A_{\rho]}\right]_{(C)M}\right) \ , \nn
\eea
transforming under local symmetries as follows\footnote{A useful intermediate step is
\bea
\delta {\cal F}_{\mu \nu}{}^M &=& 2\, {\cal D}_{\mu} \delta A_{\nu]}{}^M - \partial^M \Delta B_{\mu \nu} \ , \\
\delta {\cal H}_{\mu \nu \rho} &=& 3 \, {\cal D}_{[\mu} \Delta B_{\nu \rho]} + 3 \, \delta A_{[\mu}{}^M \, {\cal F}_{\nu \rho] M} \ ,
\eea
with $\Delta B_{\mu \nu} = \delta B_{\mu \nu} - A_{[\mu}^M\, \delta A_{\nu] M}$.}
\bea
\delta {\cal F}_{\mu \nu}{}^M &=& \left({\cal L}^{\cal D}_\xi + \widehat {\cal L}_{\widehat \Lambda}\right) {\cal F}_{\mu \nu}{}^M  - {\cal H}_{\mu \nu \rho} \, \partial^M \xi^\rho + 2 {\cal D}_{[\mu} \left( g_{\nu]\rho} {\cal M}^{M N} \partial_N \xi^\rho\right)\ , \\
\delta {\cal H}_{\mu \nu \rho} &=& \left({\cal L}^{\cal D}_\xi + \widehat {\cal L}_{\widehat \Lambda}\right) {\cal H}_{\mu \nu \rho} + 3 {\cal F}_{[\mu \nu}{}^M g_{\rho]\sigma} {\cal M}_{M N} \partial^N \xi^\sigma \ ,
\eea
and they are related through Bianchi identities
\bea
3\, {\cal D}_{[\mu} {\cal F}_{\nu \rho]}{}^M + \partial^M {\cal H}_{\mu \nu \rho} &=& 0 \ , \label{BIsDFT}\\
D_{[\mu} {\cal H}_{\nu \rho \sigma]} -\frac 3 4 {\cal F}_{[\mu \nu}{}^M {\cal F}_{\rho \sigma] M} &=& 0 \ . \nn
\eea

Now that we have defined all the ingredients appearing in the gauge transformations, we can verify closure. The gauge transformations obviously close
\be
\left[ \delta_1 \, , \, \delta_2 \right] = - \delta_{\xi_{12}} - \delta_{\Lambda_{12}} - \delta_{\Xi_{12}} \ ,
\ee
but only for the particular set of un-hatted parameters the brackets are field-independent
\bea
\xi_{12}^\mu &=& \left[\xi_1 \, , \, \xi_2 \right]^\mu + \Lambda_{[1}^P \partial_P \xi_{2]}^\mu \ , \\
\Lambda_{12}^M &=& \left[\Lambda_1 \, , \, \Lambda_2 \right]^M_{(C)} + 2 \xi_{[1}^\rho \partial_\rho \Lambda_{2]}^M + \xi_{[1}^\rho \partial^M \Xi_{2]\rho} + \Xi_{[1\rho} \partial^M \xi_{2]}^\rho \ , \label{BracketLambda}\\
\Xi_{12\, \mu} &=& 2 \Lambda_{[1}^P \partial_P \Xi_{2] \mu} + \Lambda_{[1}^P \partial_\mu \Lambda_{2]P} + \xi_{[1}^\rho \partial_\rho \Xi_{2]\mu}  - \xi_{[1}^\rho \partial_\mu \Xi_{2]\rho} - \Xi_{[1\rho} \partial_\mu \xi_{2]}^\rho \ . \label{BracketXi} \eea
This fact makes this set of parameter particularly convenient to explore how the DFT tensor hierarchy fits into an $L_\infty$ algebra.

We can now turn to dynamics. The action of KK-DFT is given by
\be
S = \int d^n x\, d^{2d} X\, e^{- 2 d} {\cal L} \ , \label{KKDFT}
\ee
where $d$ depends as usual on $g= \det[g_{\mu \nu}]$ and $\phi$
\be
e^{-2d} = \sqrt{- g} \, e^{-2 \phi} \ ,
\ee
and the Lagrangian is
\bea
{\cal L} &=& \widehat {\cal R} - 4 g^{\mu \nu} {\cal D}_{\mu}{\phi} {\cal D}_{\nu} \phi + 4 \nabla_{\mu} \left(g^{\mu \nu} {\cal D}_{\nu} \phi\right) - \frac 1 {12} {\cal H}_{\mu \nu \rho} {\cal H}^{\mu \nu \rho} \nn \\
&& + \ \frac 1 8 g^{\mu \nu} {\cal D}_{\mu}{\cal M}_{M N}\, {\cal D}_{\nu} {\cal M}^{M N} - \, \frac 1 4 {\cal M}_{M N} {\cal F}_{\mu \nu}{}^M {\cal F}^{\mu \nu N} - V \ .
\eea
Up to a single term, the ``scalar potential'' $V$ is defined as minus the generalized Ricci scalar in DFT, but with the generalized metric replaced by the scalar matrix $\cal M$
\bea
- V &=& 4\, {\cal M}^{M N} \partial_{M N} d - \partial_{M N} {\cal M}^{M N} - 4\, {\cal M}^{M N} \partial_M d \partial_N d + 4\, \partial_M {\cal M}^{M N} \partial_N d  \\ &&+ \, \frac 1 8 {\cal M}^{M N} \partial_M {\cal M}^{P Q} \partial_{N}{\cal M}_{P Q} + \frac 1 2 {\cal M}^{M N} \partial_M {\cal M}^{P Q} \partial_P {\cal M}_{Q N} + \, \frac 1 4 {\cal M}^{M N} \partial_M g^{\mu \nu} \partial_{N} g_{\mu \nu}\ . \nn
\eea
Given that the metric transforms as a scalar wrt internal diffeomorphisms, within the Ricci scalar $\widehat {\cal R}$ all derivatives $\partial_\mu$ must be replaced by $A$-covariantized derivatives ${\cal D}_{\mu}$ in the following way
\bea
\widehat {\cal R} &=& g^{\mu \nu}\, \widehat {\cal R}^\rho{}_{\mu \rho \nu} \ , \\
\widehat {\cal R}^\rho{}_{\sigma \mu \nu} &=& {\cal D}_\mu \widehat \Gamma_{\nu \sigma}^\rho - {\cal D}_\nu \widehat \Gamma_{\mu \sigma}^\rho + \widehat \Gamma_{\mu \delta}^\rho \widehat \Gamma_{\nu \sigma}^\delta -\widehat \Gamma_{\nu \delta}^\rho \widehat \Gamma_{\mu \sigma}^\delta \ , \\
\widehat \Gamma^\rho_{\mu \nu} &=& \frac 1 2 g^{\rho \sigma} \left({\cal D}_{\mu} g_{\nu \sigma} + {\cal D}_{\nu} g_{\mu \sigma} - {\cal D}_\sigma g_{\mu \nu}\right) \ ,
\eea
and we also defined a new covariant derivative $\nabla_{\mu} = {\cal D}_{\mu} + \widehat \Gamma_\mu$ that covariantizes the part of the external diffeomorphisms that involves external derivatives only.

Written like this, each term in the Lagrangian is independently and manifestly a $\Lambda$-scalar, $\Xi$-invariant and duality invariant (the scalar potential is the only exception, as the invariance with respect to internal diffeomorphisms is far from being manifest). The relative coefficients between the terms are fixed by external diffeomorphism invariance, and as a result
\be
\delta {\cal L} = \xi^\mu {\cal D}_\mu {\cal L} + \widehat \Lambda^M \partial_M {\cal L} = \xi^\mu \partial_\mu {\cal L} + \Lambda^M \partial_M {\cal L}\ .
\ee
 On the other hand, while $\sqrt{- g}$ transforms as a density under external diffeos and as a scalar under internal ones, $e^{-2\phi}$ plays the opposite role, and hence $e^{- 2 d}$ acts as a density wrt both
 \be
  \delta e^{- 2 d} = {\cal D}_{\mu} \left(\xi^\mu e^{-2 d} \right) + \partial_M \left(\widehat \Lambda^M e^{-2d}\right) = \partial_\mu\left(\xi^\mu e^{-2d}\right) + \partial_M \left(\Lambda^M e^{-2d}\right)  \ ,
 \ee
 where ${\cal D}_{\mu}$ acts on $d$ the same way as on $\phi$ in (\ref{CovariantDerivatived}). As a consequence, the action is invariant wrt to all the local symmetries.

 Regarding equations of motion, we find that under variations of the action wrt the fields one has
 \be
 \delta S = \int d^n x\, d^{2d} X\, \left(\delta g^{\mu \nu} \Delta g_{\mu \nu} + \delta B_{\mu \nu} \Delta B^{\mu \nu} + \dots\right) \ ,
 \ee
 with
 \bea
       e^{2d} \Delta g_{\mu \nu} &=& \frac 1 4 g_{\mu \nu}\, \Delta \phi + \widehat {\cal R}_{\mu \nu} + 2 \nabla_{(\mu} {\cal D}_{\nu)}\phi - \frac 1 4 {\cal H}_{\mu\rho \sigma} {\cal H}_\nu{}^{\rho \sigma} - \frac 1 2 {\cal F}_{\mu \rho}{}^M {\cal F}_\nu{}^{\rho N} {\cal M}_{M N} \nn\\ &&  + \frac 1 8 {\cal D}_\mu {\cal M}_{M N} {\cal D}_\nu {\cal M}^{M N} - \frac 1 2 \partial_M \left({\cal M}^{M N} \partial_N g_{\mu \nu}\right)\ , \\
       \Delta B^{\mu \nu} &=& \frac 1 2 {\cal D}_{\rho} \left(e^{-2d} {\cal H}^{\rho \mu \nu}\right) - \frac 1 2 \partial^M \left(e^{- 2 d} {\cal M}_{M N} {\cal F}^{\mu \nu N}\right) \ ,  \label{eomB}\\
       e^{2d} \Delta \phi &=& -2 {\cal L} \ , \\
       \Delta A^\mu{}_M &=& \Delta B^{\mu \nu} A_{\nu M} + {\cal D}_{\rho} \left(e^{-2d} {\cal M}_{M N} {\cal F}^{\rho \mu N}\right) - \frac 1 2 e^{-2d} {\cal H}^{\mu \nu \rho} {\cal F}_{\nu\rho M} \label{eomA} \\
       && -\, \partial_P \left(e^{-2d} {\cal M}^{P N} g^{\mu \nu} {\cal D}_{\nu} {\cal M}_{N M}\right) + {\cal D}_{\nu}\left( e^{-2d} \partial_M g^{\mu \nu}\right) \nn \\
       && \!\!\!\!\!\!\!\!\!\!\!\!\!\!\!\! \!\!\!\!\!\!\!\!\!\!+\, e^{-2d} \left(- 4 g^{\mu \nu} \partial_M {\cal D}_\nu d - \frac 1 4 g^{\mu \nu} {\cal D}_{\nu} {\cal M}^{P Q} \partial_M {\cal M}_{P Q} - \frac 1 2 g^{\mu \nu} {\cal D}_{\nu} g^{\rho \sigma} \partial_M g_{\rho \sigma} + {\cal D}_{\nu} g^{\mu \rho} \partial_M g_{\rho \sigma} g^{\sigma \nu}\right) \nn \\
       e^{2 d} \Delta {\cal M}_{M N} &=& - \frac 1 2 {\cal D}_{\mu}\left(e^{-2d} g^{\mu \nu} {\cal D}_\nu {\cal M}\right){}_{(\overline M \underline N)} - \frac 1 2 {\cal F}_{\mu \nu (\overline M} {\cal F}^{\mu \nu}{}_{\underline N)} \nn\\
       && + {\cal R}_{(\overline M \underline N)}[{\cal M}, d] + \frac 1 4 \partial_{(\overline M} g^{\mu \nu} \partial_{\underline N)} g_{\mu \nu}\ .
  \eea
  In the last line we used the standard notation for left-right projections\footnote{Defining $P_{M N} = \frac 1 2 \left(\eta_{M N} - {\cal M}_{M N}\right)$ and $\bar P_{M N} = \frac 1 2 \left(\eta_{M N} + {\cal M}_{M N}\right)$, the notation is such that $V_{\underline M} = P_M{}^N V_N$  and $V_{\overline M} = \bar P_M{}^N V_N$.}, and wrote the variation of the scalar potential in the same way as the generalized Ricci tensor because the former is equal to minus the generalized Ricci scalar.

\subsection{The tensor hierarchy}

Here we restrict attention to the degrees of freedom involved in the tensor hierarchy, namely $A_\mu{}^M$ and $B_{\mu \nu}$, and so set the gravitational and scalar degrees of freedom to background values (which we write with an overline). The symmetries relevant in the discussion of the tensor hierarchy are internal generalized diffeos $\Lambda^M$, and gauge transformations of the two-form $\Xi_\mu$, so we will also ignore external diffeomorphisms from now on $\xi^\mu = 0$. This in particular implies that $\widehat \Lambda = \Lambda$ in (\ref{ParamRedefDFT}), and
\be
\widehat \Xi_\mu = \Xi_\mu + A_{\mu}{}^M \, \Lambda_M \ . \label{redefDFT}
\ee

When the transformations  (\ref{TransfB}) and (\ref{TransfA})
\bea
\delta A_{\mu}^M &=& {\cal D}_{\mu} \Lambda^M + \partial^M \widehat \Xi_\mu \ ,  \label{TransfHierarchyDFT}\\
\delta B_{\mu \nu} &=& A_{[\mu}{}^M \delta A_{\nu]M} + 2 {\cal D}_{[\mu} \widehat \Xi_{\nu]} - \Lambda_M \, {\cal F}_{\mu \nu}{}^M \ , \nn
\eea
are written in terms of the un-hatted parameters the gauge covariance is no longer manifest
\bea
\delta A_{\mu}{}^M &=& \partial_{\mu} \Lambda^M + \widehat {\cal L}_\Lambda A_{\mu}{}^M + \partial^M \Xi_\mu \ , \label{gaugetransfDFT} \\
\delta B_{\mu \nu} &=& \Lambda^M  \partial_{M} B_{\mu \nu} + 2 \partial_{[\mu} \Xi_{\nu]} + \partial_M  \Xi_{[\mu}\, A_{\nu]}{}^M - A_{[\mu}{}^M \partial_{\nu]} \Lambda_M \ . \nn
\eea

These gauge transformations are annihilated by the following trivial parameters
\be
\Lambda^M = \partial^M \chi \ , \ \ \ \ \Xi_\mu = - \partial_\mu \chi \ , \ \ \ \ \widehat \Xi_\mu = - {\cal D}_\mu \chi \ ,\label{trivialparam}
\ee
where we define the arbitrary function $\chi$ to carry vanishing weight.

Let us now discuss the gauge algebra. We saw in (\ref{BracketLambda})-(\ref{BracketXi}) that closure with respect to the parameters $\Lambda^M$ and $\Xi_\mu$ was achieved through the brackets
\bea
\Lambda_{12}^M &=& \left[\Lambda_1 \, , \, \Lambda_2 \right]^M_{(C)}  \ , \label{Brackets}\\
\Xi_{12\, \mu} &=& 2 \Lambda_{[1}^P \partial_P \Xi_{2] \mu} + \Lambda_{[1}^P \partial_\mu \Lambda_{2]P}  \ . \nn \eea
We can now collectively define the gauge parameters
\be
\zeta = \left( \Lambda^M \, , \Xi_\mu \right) \ ,
\ee
and note the brackets as
\bea
\Lambda_{12}^M &=& \zeta_{12}^M = \left[\zeta_1 \, , \, \zeta_2 \right]^M \ , \\
\Xi_{12\, \mu} &=& \zeta_{12\, \mu} = \left[\zeta_1 \, , \, \zeta_2 \right]_\mu \ .
\eea
This notation is useful to compute the Jacobiator
\bea
J(\zeta_1,\, \zeta_2,\, \zeta_3) = \left[\left[\zeta_1,\, \zeta_2\right],\, \zeta_3 \right] + {\rm cyc.} = 3\, \left[\left[\zeta_{[1},\, \zeta_2\right],\, \zeta_{3]} \right] \ , \label{DefJacobiator}
\eea
the components of which are known to be trivial parameters
\be
J^M = \partial^M N \ ,  \ \ \ \ J_\mu = - \partial_\mu N \ , \label{Jacobiators}
 \ee
 where we introduced the Nijenhuis scalar
 \be
N(\zeta_1,\, \zeta_2,\, \zeta_3) = \frac 1 6 \left[\Lambda_1\, , \Lambda_2\right]^M_{(C)} \Lambda_{3M} + {\rm cyc.} = \frac 1 2 \left[\Lambda_{[1}\, , \Lambda_2\right]^M_{(C)} \Lambda_{3]M}\ . \label{Nijenhuis}
 \ee

If instead we choose to work with the redefined parameters
\be
\widehat \zeta = \left( \Lambda^M \, , \, \widehat \Xi_\mu \right)\ ,
\ee
we then find
\bea
\widehat \zeta_{12}^M &=& \left[\widehat \zeta_1\, , \, \widehat \zeta_2\right]^M = \left[\Lambda_1 \, , \, \Lambda_2\right]^M_{(C)} \ , \\
\widehat \zeta_{12\, \mu} &=& \left[\widehat \zeta_1\, , \, \widehat \zeta_2\right]_\mu = {\cal D}_\mu \Lambda_{[1}^M \, \Lambda_{2]M} \ ,
\eea
and now
\be
\widehat J^M = \partial^M N \ , \ \ \ \ \widehat J_\mu = - {\cal D}_\mu N + 2 \, {\cal D}_{\mu} \left[\Lambda_{[1} \, , \, \Lambda_2\right]^M_{(C)} \Lambda_{3]M} -  \left[\Lambda_{[1} \, , \, \Lambda_2\right]^M_{(C)} {\cal D}_{\mu}\Lambda_{3]M} \ ,
\ee
and so we see that the Jacobiator is no longer a trivial parameter. We will discuss this further and its relevance for EFT later.

\subsection{The $L_\infty$ structure of the tensor hierarchy} \label{sct-Linf}

Here we discuss how the algebraic structure of the tensor hierarchy in DFT fits into $L_\infty$. We begin by discussing the kinematics, and later move to consider the dynamics. The first step is to define the graded vector spaces, which divide here into the following subspaces
\beq
X_1:&\quad\text{functions:}\quad\chi(x^\mu,\, X^M)\\
X_0:&\quad\text{gauge parameters (generically noted $\zeta$):}\quad \L^M,\,\Xi_\m \\
X_{-1}:&\quad\text{fields (generically noted $\Psi$):}\quad A_\m{}^M,\,B_{\m\n}\\
X_{-2}:&\quad\text{field equations (generically noted ${\cal F}$): }\quad \Delta A^\mu{}_M,\, \Delta B^{\mu \nu}\ .
\eeq
The space of functions might come as a surprise at the moment, but it turns out to be a general feature in theories that exhibit symmetries for symmetries. We will sometimes add indices to indicate which particular subspace we refer to within a given level. For example: for $\Psi \in X_{-1}$, $\Psi_\m{}^M$
refers to a field $A_\m{}^M$ and $\Psi_{\mu \nu}$ to a field $B_{\mu \nu}$. The same notation will be used over products,  for instance $\chi\in X_1$, $\ell_1(\chi)_{\m}$
indicates a parameter $\Xi_{\m}$ as $\ell_1$ maps from $X_1$ to $X_{0}$, etc.

We will first describe thoroughly how to construct the pure gauge part and then consider the dynamics of the fields.  The pure gauge structure of DFT is given by the $O(D,D)$ covariantization of the Courant algebroid \cite{Hull:2009zb}. The relation between this structure and $L_\infty$ algebras was done in \cite{Roytenberg:1998vn} and later extended to the $O(D,D)$ case in \cite{Deser:2016qkw} and \cite{Hohm:2017pnh}. As discussed in Subsection \ref{sec::theorem}, a recent paper \cite{Hohm:2017cey} established sufficient conditions for a gauge algebra to have an $L_\infty$ structure. In this section we review these results, closely following the line of reasoning in \cite{Hohm:2017cey} for a general bracket and then specializing to our case of interest.

For the sake of simplicity we begin by focussing on the subspace $X_0$ with elements $\z=\L^M+\Xi_\m$, ignoring for the moment to the maximum extent the spaces $X_{-1}$ and lower. We will proceed in an order that might seem capricious, but is only intended to reach the relevant results quickly.  The first identity we consider is  $n = 3$
 \beq
\label{jacidentity}
-\ell_1(\ell_3 (\zeta_1,\zeta_2, \zeta_3))=& \ \ell_2(\ell_2(\zeta_1,\zeta_2),\zeta_3) + \ell_2(\ell_2(\zeta_3,\zeta_1),\zeta_2)
   +\ell_2(\ell_2(\zeta_2,\zeta_3),\zeta_1) \;\\
  & + \ell_3(\ell_1 (\zeta_1) ,\zeta_2, \zeta_3)
  +    \ell_3( \zeta_1 ,\ell_1(\zeta_2), \zeta_3)
    + \ell_3( \zeta_1 ,\zeta_2, \ell_1(\zeta_3)) \ .
\eeq
We have already all the required ingredients to evaluate the RHS. As explained before, $\ell_2(\zeta_1,\, \zeta_2)$ must be read from the crossing of \eqref{closure} and (\ref{Brackets}), yielding
\begin{eqnarray}
\ell_2(\zeta_1,\, \zeta_2) = \left[\Lambda_1 \, , \, \Lambda_2 \right]^M_{(C)}  + 2 \Lambda_{[1}^P \partial_P \Xi_{2] \mu} + \Lambda_{[1}^P \partial_\mu \Lambda_{2]P} \ .
\label{eq:}
\end{eqnarray}
The first line in the RHS of (\ref{jacidentity}) then becomes the Jacobiator (\ref{DefJacobiator}-\ref{Jacobiators})
\begin{eqnarray}
J(\z_1,\z_2,\z_3)=\partial^M N(\z_1,\z_2,\z_3)- \partial_\m N(\z_1,\z_2,\z_3)\equiv \cD N(\z_1,\z_2,\z_3) \ ,
\label{jac}
\end{eqnarray}
where $N$ is the Nijenhuis scalar (\ref{Nijenhuis}) and we have defined the operator
\beq
\cD \equiv \partial^M - \partial_\m \ .
\eeq
The second line in (\ref{jacidentity}) is schematically of the form $\ell_3(\zeta_i,\, \zeta_j,\, \ell_1(\zeta_k))$. Given that $\ell_1$ maps the last argument into $X_{-1}$ these terms are of the form $\ell_3(\zeta_1,\zeta_2,\Psi)$ and vanish in light of the discussion in (\ref{closure}). Then, the $n=3$ equation (\ref{jacidentity}) can be written as
\be
- \ell_1 (\ell_3(\zeta_1,\zeta_2, \zeta_3)) =  {\cal D} \, N(\zeta_1,\zeta_2, \zeta_3)  \in X_0 \ .\label{n3schematics}
\ee
Now note that the RHS of (\ref{n3schematics}) is not zero, and so $\ell_3(\zeta_1,\,\zeta_2,\,\zeta_3) \in X_1$ must also be non vanishing, implying as anticipated that a new graded subspace $X_1$ of functions $\chi(x,X)$ is required (see Table 3).
\begin{table}[ht]
\begin{center}
    \begin{tabular}{|l|l|l|l|}
    \hline  $X_1$ & $X_0$  & $X_{-1}$   &  $X_{-2}$  \\ \hline Functions $\chi$&
Gauge Parameters
$\z$ & Fields $\Psi$ & $E \ (\ni  EOM  {\cal F})$ \\ \hline    \end{tabular}
    \end{center}
    \caption{New subspace $X_1$ required by the failure of Jacobi.}
\end{table}

From here it is trivial to see that  (\ref{jacidentity}) is satisfied for the following choice of products
\beq
\ell_3(\zeta_1,\zeta_2,\zeta_3)&=-N(\L_1,\L_2,\L_3)\;\in X_1\\
\ell_1(\chi)&=\cD\chi,\;\chi\in X_1 \ .
\label{elesgauge}
\eeq

Having introduced a new space of functions we now have to define new products and verify identities, that include such elements as arguments. We begin the verification of the first identity for $\chi \in X_1$
\beq
  n=1:\quad \ell_1 ( \ell_1 (\chi)) \ = \ell_1 ( \cD\chi)\stackrel{?}{=}0 \ .\\
 \eeq
Since $\cD \chi$ belongs to the space of parameters, and is in fact a trivial one (\ref{trivialparam}) one then finds from (\ref{gaugel}) that
\beq
  \delta_{\cD \chi}\Psi =\ell_1(\cD \chi)+\ell_2(\cD \chi,\Psi)-\dfrac{1}{2}\ell_3(\cD \chi,\Psi^2)-... = 0 \ .
	\label{ojito}
\eeq
Since each term contains different powers in field perturbations, each one must vanish separately, in particular the first one which implies that $\ell_1(\cD \chi) = 0$, which is what we wanted to verify.

Now we consider the next identity for $n=2$ \eqref{L2L1}, with $x_1=\zeta$ and $x_2=\chi$
\beq
\ell_1(\ell_2(\zeta,\chi))=& \ \ell_2(\ell_1(\zeta),\chi) + \ell_2(\zeta,\ell_1(\chi)) \ .
\eeq
 An extra hypothesis was considered in \cite{Hohm:2017cey}: for any $\zeta\in X_0$ we will need $
  \left[\text{Im}(\cD), \zeta\right] \subset \ \text{Im}(\cD)$
$i.e.$, $\text{Im}(\cD)$ is an ideal of $X_0$ or, stated in other way, the commutator of a trivial gauge parameter and any gauge parameter gives back a trivial gauge parameter. A short computation shows that this is exactly what happens in the tensor hierarchy of DFT
\beq
\left[\cD\chi,\, \zeta \right]  = \cD \left(-\frac 1 2 \Lambda^P \partial_P \chi\right)\ .
\eeq
Comparing this with the following rewriting of the $n=2$ identity
\be
{\cal D} \ell_2(\zeta,\, \chi) = \ell_2(\ell_1(\zeta),\chi) - \left[{\cal D}\chi,\,\zeta \right] \ ,
\ee
we conclude that $\ell_2(\Psi,\chi) = 0$ and
\be
\ell_2(\zeta,\, \chi) = \frac 1 2 \Lambda^P \partial_P \chi \ .
\ee

So far we obtained the following non-vanishing products with entries belonging to the graded subspaces $X_1$ and $X_0$
\beq
\label{ALLclousureproducts}
\ell_1 (\chi) &= \ \partial^M \chi - \partial_\mu \chi \ , \\
\ell_1(\z) & =\ \partial_\mu \Lambda^M + \partial^M \Xi_\mu + 2 \partial_{[\mu} \Xi_{\nu]}\ , \\
\ell_2(\z_1,\z_2) & = \ [\z_1,\z_2] = \left[\Lambda_1 \, , \, \Lambda_2 \right]^M_{(C)}  + 2 \Lambda_{[1}^P \partial_P \Xi_{2] \mu} + \Lambda_{[1}^P \partial_\mu \Lambda_{2]P} \ , \\
\ell_2(\z,\chi) & =\ \L(\chi)\equiv\frac 1 2 \Lambda^P \partial_P \chi \ , \\
    \ell_3(\z_1,\z_2,\z_3) & =\ - N(\Lambda_1,\Lambda_2,\Lambda_3) = - \frac 1 2 \left[\Lambda_{[1}\, , \Lambda_2\right]^P_{(C)} \Lambda_{3]P} \ .
 \eeq

These are the only non trivial products coming from the pure gauge sector and are enough to show that this sector fits into an $L_\infty$ structure (with the help of the observation made in Subsection \ref{sec::theorem}). The remaining identities that have not been checked (involving two $\chi's \in X_1$ and for $\z's \in X_0$) are presented in the Appendix. The outcome of the forthcoming analysis is that the products obtained so far will not be modified when considering entries with lower grade. In particular, the absence of a $4$-product $\ell_4$ indicates that $\ell_2 \ell_3 - \ell_3 \ell_2$ vanishes exactly as opposed to ``up to homotopy'', meaning that the gauge sector of the hierarchy tensor in DFT is governed by an $L_3$ algebra.

We now take into account the fields, their gauge transformations and their dynamics. Some products that are trivial to identify are the following:

\begin{itemize}
\item From the gauge transformations of DFT (\ref{gaugetransfDFT}) and the identifications (\ref{gaugel}) we rapidly find
\beq
\ell_2(\z,\, \Psi)& = \ \left[\L,A_\mu\right]_{(D)}^M + \del_P\Xi_{[\m}A_{\nu]}{}^P -A_{[\m}{}^P\del_{\n]}\L_P+ \L^P\del_P B_{\m\n} \ ,\\
\ell_{n+1}(\z,\Psi^n)& = \ 0, \quad n>1 \ .
\eeq

\item From the equations of motion of the gauge fields (\ref{eomA}) and two-form (\ref{eomB}) and crossing with (\ref{eoml}) we read off the products involving only fields $\ell_n(\Psi^n)$. As mentioned previously, we will only consider the fields of the tensor hierarchy and ignore the fluctuations around the background for the metric $g_{\mu \nu} = \overline g_{\mu \nu}$, the dilaton $\phi = \overline \phi$ and ${\cal M}_{M N} = \overline {\cal M}_{M N}$. Note that these backgrounds are by no means assumed to be constant nor covariantly constant: they are only forced to satisfy the equations of motion. In this case, the non-trivial products are of the form  $\ell_n(\Psi^n)$ with $n \leq 3$ and $\Psi = A_\mu{}^M + B_{\mu \nu}$.  Here we only write explicitly the single product as will be needed later. Expanding the EOM \eqref{eomA} and \eqref{eomB} in powers of $A_\m{}^M$ and $B_{\m\n}$ and keeping only the leading order gives
\beq
\label{productseom}
\ell_1(\Psi)=&-\del^M\left(e^{-2\overline d}\overline \cM_{M N}\overline g^{\mu \rho} \overline g^{\nu \sigma}\del_{[\rho}A_{\sigma]}{}^N\right)+\frac 3 2\del_\rho\left(e^{-2\overline d}\overline g^{\rho \sigma} \overline g^{\mu \delta} \overline g^{\nu \epsilon}\del_{[\sigma}B_{\delta\epsilon]}\right)  \\ &+\frac 1 2 \del^M\left(e^{-2\overline d}\overline \cM_{M N} \overline g^{\mu \rho} \overline g^{\nu \sigma} \del^N B_{\rho\sigma}\right)
+2\del_\rho\left(e^{-2\overline d}\overline \cM_{M N} \overline g^{\rho \sigma} \overline g^{\mu \delta}\del_{[\sigma}A_{\delta]}{}^N \right) \\ & + \partial_P \left(e^{-2\widebar d} \widebar {\cal M}^{P N} \widebar g^{\mu \nu} \widehat {\cal L}_{A_\nu} \widebar {\cal M}_{N M}\right) - \widehat {\cal L}_{A_\nu}\left( e^{-2 \widebar d} \partial_M \widebar g^{\mu \nu}\right)  \\
&-\del_\rho\left(e^{-2\overline d}\overline \cM_{M N} \overline g^{\rho \sigma} \overline g^{\mu \delta}\del^N B_{\sigma\delta}\right)\ -\, e^{-2\widebar d} \left(- 4 \widebar g^{\mu \nu} \partial_M \widehat {\cal L}_{A_\nu} \widebar d \right. \\
&\left. - \frac 1 4 \widebar g^{\mu \nu} \widehat {\cal L}_{A_\nu} \widebar {\cal M}^{P Q} \partial_M \widebar {\cal M}_{P Q} - \frac 1 2 \widebar g^{\mu \nu} \widehat {\cal L}_{A_\nu} \widebar g^{\rho \sigma} \partial_M \widebar g_{\rho \sigma} + \widehat{\cal L}_{A_\nu} \widebar g^{\mu \rho} \partial_M \widebar g_{\rho \sigma} \widebar g^{\sigma \nu}\right) .\\
\eeq
Higher products of fields are obtained in the same way, we don't show the expressions here in order to lighten the presentation, and also because the identities in which they appear are fulfilled by construction as explained at the end of Section \ref{sec-lft}).

\item We can use the gauge transformation of the EOM \eqref{covf} to determine products of the form $\ell_{n + 2} (\zeta ,\, {\cal F},\, \Psi^n)$, $n \geq 0$. First let us point out that the EOM for $B_{\m\n}$ \eqref{eomB} is covariant, namely it transforms as a tensorial density
\be
\delta \Delta B_{\mu \nu} = \widehat {\cal L}_\Lambda \Delta B^{\mu \nu} = \partial_M \left(\Lambda^M \Delta B^{\mu \nu}\right) \ , \ \ \  \lambda (\Delta B^{\mu \nu}) = 1 \ ,
 \label{deltaeomB}
\ee
and the EOM for $A^\mu{}_M$ includes a covariant part also of weight one (which we denote $\D_c$, $c.f$ (\eqref{eomA}))
\beq
\Delta A^\mu{}_M = \Delta B^{\mu \nu} A_{\nu M} + \Delta_c A^\mu{}_M\ , \ \  \ \delta \Delta_c A^\mu{}_M = \widehat {\cal L}_\Lambda \Delta_c A^\mu{}_M  \ ,  \ \ \ \ \lambda (\Delta_c A^\mu{}_M) =  1 \ .
\label{deltaeomA}
\eeq
Then, from \eqref{covf} we read
 \beq
\label{lf}
\ell_2(\z,{\cal F})& =\  \widehat  \cL_\L \Delta B^{\m\n} + \Delta B^{\m\n} \, (\del_\n\L_M+\del_M\Xi_\n) + \widehat \cL_\L \Delta_{c} A^\mu{}_M\ ,
\\
\ell_3(\z, {\cal F} ,\Psi) & =\ \widehat \cL_\L  \Delta B^{\mu \nu}\,  A_{\nu M} + \Delta B^{\mu \nu} \, \widehat \cL_{\L} A_\nu{}_M \ , \\
\ell_{n+2} (\z, {\cal F},\Psi^n) &=\ 0 \ ,  \ \ \ \ \ {\rm for }\ n \geq 2 \ .
\eeq
\end{itemize}

The products presented so far are enough to show that the theory lies within an $L_\infty$ structure (modulo a subtlety to be discussed in what follows). The direct way to verify this  is to write down all possible non-trivial identities with their corresponding arguments and to check that there is no need for new non-zero products. This is done in detail in the Appendix. The strategy we follow consists in writing all possible lists of vectors with definite degree and then apply the identities over them. Such possibilities are given by the following lists:
\begin{itemize}	
	\item Any number of $\Psi's \in X_{-1}$: $(\Psi_1,\Psi_2,...)$.
	\item One  $\z \in X_{0}$ and any number $n>0$ of $\Psi's \in X_{-1}$: $(\z,\Psi_1,\Psi_2,...)$.
	\item Two  $\z's \in X_{0}$ and any number $n\geq0$ of $\Psi's \in X_{-1}$: $(\z_1,\z_2,\Psi_1,\Psi_2,...)$.
	\item Three or more  $\z's \in X_{0}$ and any number $n\geq0$ of $\Psi's \in X_{-1}$: $(\z_1,\z_2,\z_3...,\z_n, \Psi_1,\Psi_2,...)$.
	\item At least one $\chi \in X_1$ (and any other vector on the list): $(\chi,...)$.
	\item At least one $E \in X_{-2}$ (and any other vector on the list): $(E,...)$.
	\item Only one $\z \in X_{0}$.
\end{itemize}
The first three lists fulfill the $L_\infty$ identities by construction as pointed out at the end of Section \ref{sec-lft}. The first one is associated to the non-existence of a graded subspace $X_{-3}$, the second one to gauge transformations of the equations of motion, and the third one to closure of the gauge algebra over fields. The fourth, fifth and sixth ones fulfill the identities straightforwardly (see Appendix).

The last one has an interesting implication: the dynamics of the tensor hierarchy can not fit alone within an $L_\infty$ algebraic structure, it requires the extra field perturbations and their background values. So far we considered mostly the gauge sector and decoupled the degrees of freedom of the fields not belonging to the tensor hierarchy. In particular, we found that if the graded subspace $X_{-2}$ is assumed to be absent (namely, the dynamics is ignored) then from (\ref{ALLclousureproducts}) one finds that the following choice for $\ell_1$ over gauge parameters is consistent with the $L_\infty$ structure
\be
\ell_1(\z)  =\ \partial_\mu \Lambda^M + \partial^M \Xi_\mu + 2 \partial_{[\mu} \Xi_{\nu]}\ .
\ee
Since this product maps into the space of fields, in order to verify the $\ell_1(\ell_1(\z)) = 0$ identity we must insert the above into (\ref{productseom}). We rapidly get to the conclusion that it does not work. What we are missing are the contributions to $X_{-1}$ coming from the gauge transformations of the other fields. This is in fact to be expected because on the one hand the EOM mix all the fields, and on the other because there are non-vanishing $\ell_1(\z)$ components along the fields that do not belong to the tensor hierarchy. Consider as an example the gauge transformation of the scalar fields
\beq
\d_\L\cM_{PQ}&=\widehat \cL_\L \cM_{PQ}= \underbrace{\widehat \cL_\L \overline \cM_{PQ}}_{\ell_1(\L)_{PQ}}
+ \underbrace{\widehat \cL_\L m_{PQ}}_{\ell_2(\L, m)_{PQ}}+...
\eeq
where $m_{PQ} = m_{\overline P \underline Q}$ is the scalar first order perturbation. Only when these components of $\ell_1(\L)$ are considered, namely
\beq
\ell_1(\z)=\del_\m\L^M   + \partial^M \Xi_\mu + 2 \partial_{[\mu} \Xi_{\nu]} +                      \widehat \cL_\L \overline\cM_{PQ} +\widehat \cL_\L \overline d +\widehat \cL_\L \overline g_{\m\n}\ , \label{background}
\eeq
the following identity is obtained
\beq
\ell_1(\ell_1(\L))^\m{}_{M}= \widehat \cL_\L \left(\D A^\m{}_{M}\right)_{bckg} = 0 ,
\eeq
where $\left(\D A^\m{}_{M}\right)_{bckg}$ stands for the equations of motion evaluated on background fields, and we used that background fields are by definition constrained to satisfy the EOM.

This concludes the proof on how the tensor hierarchy in KK-DFT fits into an $L_\infty$ algebra. We remark that it is not enough to restrict the analysis to the tensor hierarchy components of the different graded subspaces, but one should also consider other components that might contribute through dependence on background fields, even if their perturbations are ignored. The same situation arises when one considers the list $(\z,\Psi_1,\Psi_2,...)$: the corresponding $L_\infty$ identity can be seen as coming from the gauge transformation of the equations of motion, and we have made use of the gauge transformations over all fields (including $\cM_{PQ}$, $d$ and $g_{\m\n}$) in order to get the products in \eqref{lf}.

\subsection{Redefinitions}

When discussing KK-DFT we considered two sets of parameters. Ones with a hat $(\widehat \Lambda^M,\ \widehat \Xi_\mu)$ that are convenient to write transformations covariantly with respect to internal generalized diffeomorphisms. This set is interesting as it is the one usually considered in Exceptional Field Theories and gauged supergravities. A property of this set of parameters is that, as shown before, the brackets with respect to which the gauge transformations close are field dependent, and the Jacobiator is not a trivial gauge parameter, and so identifying symmetries for symmetries in this set is a rather non-trivial task. We circumvented these issues by performing the redefinitions (\ref{ParamRedefDFT}) that connect the hatted parameter with the un-hatted ones, which have the disadvantage of spoiling the manifest generalized diffeomorphism covariance of the gauge transformations, but are the specific set in which the brackets are field independent and the Jacobiator is a trivial parameter, making them easier to analyze in the context of $L_\infty$ algebras.

Given that the redefinition (\ref{ParamRedefDFT}) is field dependent, the graded subspace to which the hatted parameters belong must involve mixing between the graded subspaces of the other set of parameters. Both sets define equivalent algebras and so there must exist an isomorphism that connects them. Here we discuss how things fit into $L_\infty$ in the hatted ``frame''. We start by considering the gauge transformations in the form of interest, given by
\beq
\d_\z A_\m{}^M&=\cD_\m\L^M+\del^M\widehat\Xi_\mu\\
\d_\z B_{\m\n}&=2\cD_{[\m}\widehat\Xi_{\n]}-\L_M\cF_{\m\n}{}^M+A_{[\m}{}^M\d A_{\n]M}\\
\eeq
from which we can identify the non-trivial products $\ell_1(\z)$, $\ell_2(\z,\Psi)$, $\ell_3(\z,\Psi_1,\Psi_2)$, $\ell_4(\z,\Psi_1,\Psi_2,\Psi_3)$.

We saw that closure
\beq
\left[\d_1,\d_2\right]=-\d_{\L_{12}}-\d_{\widehat
\Xi_{12}} \ ,
\eeq
holds with respect to field dependent brackets
\beq
\L_{12}^M&=\left[\L_1,\L_2\right]_{(C)}^M \nn\\
\widehat\Xi_{12\m}&=\cD_\m\L_{[1}^M\L_{2]M} \ .
\eeq
Comparing with \eqref{commurel} we now get the identifications
\beq
\ell_2(\zeta_1,\zeta_2)&=\left[\L_1,\L_2\right]_{(C)}^M+\del_\m\L_{[1}^M\L_{2]M}\\
\ell_3(\z_1,\z_2,A)&=-\widehat \cL_{A_\m}\L_{[1}^M\L_{2]M} \ ,
\eeq
which differ from the previous ones, the novelty being that we have a non-vanishing $\ell_3(\z_1, \z_2, \Psi)$, as expected for a field dependent bracket. There are however certain products that still take the same value. For example, consider the identity for $n=3$
\beq
-\ell_1(\ell_3 (\zeta_1,\zeta_2, \zeta_3))=&  \ell_2(\ell_2(\zeta_1,\zeta_2),\zeta_3) + \ell_2(\ell_2(\zeta_3,\zeta_1),\zeta_2)
   +\ell_2(\ell_2(\zeta_2,\zeta_3),\zeta_1) \;\\
  & + \ell_3(\ell_1 (\zeta_1) ,\zeta_2, \zeta_3)
  +    \ell_3( \zeta_1 ,\ell_1(\zeta_2), \zeta_3)
    + \ell_3( \zeta_1 ,\zeta_2, \ell_1(\zeta_3))\ .
\eeq
This time, the $\ell_2\ell_2$ terms on the RHS of the first line are not anymore the Jacobiator of the bracket, because the bracket now depends on fields, and now the second line does not vanish. It would be convenient to separate this expression according to its index structure. For the $M$ component, we get
\beq
\left[\ell_1(\ell_3(\z_1,\z_2,\z_3))\right]^M&= - \text{Jac}(\L_1,\L_2,\L_3) = -\partial^M N(\L_1,\L_2,\L_3)\ ,
\eeq
suggesting
\beq
\ell_3(\z_1,\z_2,\z_3)&= - N(\L_1,\L_2,\L_3),\quad \in X_1 \\
\left[\ell_1(\chi)\right]^M&=\del^M\chi, \quad \chi \in X_1\ .
\eeq
For the $\m$ part we get
\beq
	\left[\ell_1(\ell_3(\z_1,\z_2,\z_3))\right]_\m&= - \frac 3 2 \del_\m\left[\L_{[1},\L_2\right]_{(C)}^P \L_{3]P} + \frac 3 2 \del_\m\L_{[3\,P} \left[\L_{1},\L_{2]}\right]_{(C)}^P \\
&+3\del_\m\L_{[1}^P\del_P\L_{2}^N\L_{3]N}+6\L_{[3}^P\del_P\del_\m\L_{1N}\L_{2]}^N\ ,
\eeq
which can be rewritten as\footnote{The following identity is useful for this computation\beq
	\a \del_\m\left[\L_{[1},\L_2\right]_{(C)}^P \L_{3]P}+\b \del_\m\L_{[3\,P} \left[\L_{1},\L_{2]}\right]_{(C)}^P&=(2\a+\b)\del_\m\L_{[1}^Q\del_Q\L_2^P\L_{3]P} \\
&+ 3\a \L_{[1}^Q\del_\m\del_Q\L_2^P\L_{3]P}-(\a+2\b)\del_\m\L_{[1}^Q\del^P\L_{2Q}\L_{3]P}
\eeq
}
\beq\left[\ell_1(\ell_3(\z_1,\z_2,\z_3))\right]_\m&=\partial_\m N (\L_1,\L_2,\L_3)\ .
\eeq
Thus we see that this identity gives  the same identifications we had before
\beq
\ell_3(\z_1,\z_2,\z_3)&= -N (\L_1,\L_2,\L_3),\quad \in X_1 \\
\ell_1(\chi)&=\cD\chi, \quad \chi \in X_1 \ .
\eeq
Even though now we needed a non-trivial $\ell_3 \ell_1$ to complement $\ell_2\ell_2$ to obtain the above identifications, we emphasize that now $\ell_1(\chi)$ is {\it not} a trivial gauge parameter. Instead, the product form of the trivial parameter must be read from (\ref{trivparamQ}).

\section{$L_\infty$ and $E_{7(7)}$ Exceptional Field Theory} \label{sec::EFT}

In this section we present a self-contained review of $E_{7(7)}$ EFT \cite{Hohm:2013uia} and discuss how it fits into $L_\infty$.  The fields and gauge parameters depend on external and internal coordinates $(x^\mu\, , \, X^{M})$, with $\mu$ an external $GL(n)$ index, and $M$ an internal fundamental $E_{7(7)}$ index. The invariant tensors are the generators $(t_\alpha)_M{}^N$ and the symplectic metric $\Omega_{M N}$, which is antisymmetric and raises and lowers internal indices through the north-west south-east convention\footnote{The only two $E_{7(7)}$ identities that are needed to reproduce the computations in this section are
\begin{eqnarray}
 (t_\alpha)_M{}^K (t^\alpha)_N{}^L &=& \frac 1 {24} \delta^K_M \delta^L_N + \frac 1 {12} \delta^L_M \delta^K_N + (t_\alpha)_{M N} (t^\alpha)^{K L} - \frac 1 {24} \Omega_{M N} \Omega^{K L} \ , \\
 0 &=& (t_\alpha)_{N L} (t^\alpha)^{M(K} (t_\beta)^{Q)L} - \frac 1 {24} (t_{\beta})_N{}^{(K} \Omega^{Q) M} + \frac 1 {12} (t_\beta)^{M(K} \delta^{Q)}_N \\
 && - \frac 1 2 (t^\alpha)^{M L} (t_\alpha)^{K Q} (t_{\beta})_{N L} + \frac 1 2 (t_\beta)^{M L} (t^\alpha)^{K Q} (t_{\alpha})_{N L} + \frac 1 {24} (t_\beta)^{K Q} \delta^M_N \ .\nn
\end{eqnarray}}

\be
V^M = \Omega^{M N} V_N \ , \ \ \ V_{M} = V^N \Omega_{N M} \ , \ \ \ \Omega^{M K} \Omega_{K N} = - \delta^M_N \ .
\ee
Adjoint indices are instead raised and lowered with the $Cartan-Killing$ form $\kappa_{\alpha \beta} = (t_\alpha)_M{}^N (t_\beta)_N{}^M$, and the $E_{7(7)}$ structure constants can be read from $[t_\alpha , \, t_\beta] = f_{\alpha \beta}{}^\gamma t_\gamma$. The adjoint representation is symmetric for $E_{7(7)}$, i.e. $(t_\alpha)_{[MN]} = 0$.

The internal coordinate dependence of fields and gauge parameters is restricted by the ``section condition'', which states the vanishing of the following duality covariant Laplacians
\be
\Omega^{M N} \, \partial_M \otimes \partial_N \dots = 0 \ , \ \ \ (t_{\alpha})^{M N}  \, \partial_M \otimes \partial_N \dots = 0 \ .
\ee

\subsection{Exceptional generalized diffeomorphisms}

Just so the paper is self-contained, here we discuss very briefly the structure of $E_{7(7)}$ covariant generalized diffeomorphisms \cite{Berman:2012vc},\cite{Hohm:2013uia}. The generalized Lie derivative acting on a contravariant vector is given by
\be
\widehat {\cal L}_\Lambda V^M = \Lambda^K \partial_K V^M - 12 (t^\alpha)^M{}_N (t_\alpha)^K{}_L \, \partial_K \Lambda^L \, V^N + \lambda(V)\, \partial_K \Lambda^K \, V^M \ ,
\ee
with a natural generalization to higher rank tensors. The way this transformation acts on covariant tensors follows from the observation that the symplectic metric $\Omega_{M N}$ is invariant under generalized diffeomorphisms. The same is true for the generators
\be
\widehat {\cal L}_\Lambda \Omega_{M N} = 0 \ , \ \ \ \ \ \widehat {\cal L}_\Lambda (t^\alpha)_{M N} = 0 \ .
\ee

There are two types of trivial parameters
\be
\Lambda_{trivial}^M = (t^\alpha)^{M N} \partial_N T_\alpha \ , \ \ \ \ \ \Lambda_{trivial}^M = \Omega^{M N} T_N \ , \label{trivialEFT}
\ee
with $T_M$ a covariantly constrained vector in the sense of the section condition
\be
(t^\alpha)^{M N}\, T_M\, T_N = (t^\alpha)^{M N}\, T_M\, \partial_N \dots = \Omega^{M N} T_M \otimes T_N = \Omega^{M N} T_M\, \partial_N \dots = 0 \ . \label{covariantlyconstrained}
\ee
The symmetric part of the generalized Lie derivative with weight $\lambda = \frac 1 2$ turns out to be a trivial parameter
\be
\left\{\Lambda_1,\, \Lambda_2\right\} = \widehat {\cal L}^{\lambda = \frac 1 2}_{\Lambda_{(1}} \, \Lambda_{2)} = 12 (t^\alpha)^{M N} \partial_N \gamma_\alpha + \frac 1 2 \Omega^{M N}\, \gamma_N \ , \label{symmprod}
\ee
with
\be
\gamma_\alpha = - \frac 1 2 (t_\alpha)_{K L} \Lambda_1^K \Lambda_2^L  \ , \ \ \ \ \ \gamma_M =  \Omega_{K L}\, \Lambda_{(1}^K \partial_M \Lambda_{2)}^L \ . \label{componentssymmbracket}
\ee
The antisymmetric part is instead the E-bracket with respect to which the generalized Lie derivative algebra closes
\be
\left[\Lambda_1 \, , \Lambda_2 \right]_{(E)}^M  = \widehat {\cal L}^{\lambda = \frac 1 2}_{\Lambda_{[1}} \, \Lambda_{2]} =  2 \Lambda_{[1}^K \partial_K \Lambda_{2]}^M - 12 (t^\alpha)^{M N} (t_\alpha)_{K L} \Lambda_{[2}^K \partial_N \Lambda_{1]}^L - \frac 1 4 \Omega^{M N} \Omega_{K L} \partial_N \left(\Lambda_1^K \Lambda_2^L\right) \ .
\ee
In this sense, $\widehat {\cal L}_{\Lambda_1}^{\lambda = \frac 1 2} \Lambda_2$ plays the role of an exceptional D-bracket.

It was shown in \cite{Hohm:2013uia} that from the definitions above, it follows that a tensor in the adjoint representation $T_\alpha$ transforms as
\be
\delta_\Lambda T_\alpha = \Lambda^K \partial_K T_\alpha + 12 f_{\alpha \beta}{}^\gamma (t^\beta)_L{}^K \partial_K \Lambda^L \, T_\gamma + \lambda'(T) \partial_K \Lambda^K T_\alpha \ . \label{adjointtransf}
\ee
Another useful result is that  a trivial vector
\be
\widetilde T^M= 12 (t^\alpha)^{M N}\partial_N T_\alpha + \frac 1 2 \Omega^{M N} T_N \ , \label{tilde}
\ee
transforms as a vector of weight $\lambda (\widetilde T) = \frac 1 2$, provided $\lambda'(T_\alpha) = 1$ and the covariantly constrained field $T_M$ transforms as follows
\be
\delta_\Lambda T_M = \Lambda^K \partial_K T_M  + 12 (t^\alpha)^N{}_M (t_{\alpha})^K{}_L \, \partial_K \Lambda^L \, T_N + \frac 1 2 \partial_K \Lambda^K \, T_M - 24 (t^\alpha)_L{}^K\, T_\alpha \, \partial_M\partial_K \Lambda^L \ . \label{constrainedtransf}
\ee
Note that the transformation of $T_M$ depends on $T_\alpha$ so both fields must always be considered jointly, forming a covariant pair $(T_M,\, T_\alpha)$.

We emphasize that every vector with a tilde in this paper indicates that it is of the trivial  form (\ref{tilde}), and as such must be considered constrained and with weight $\lambda = \frac 1 2$. Let us also point out that if the vectors $\Lambda_1$ and $\Lambda_2$ in (\ref{componentssymmbracket}) were transformed with respect the generalized Lie derivative with weight $\lambda = \frac 1 2$, then $\gamma_\alpha$ and $\gamma_M$ would transform as (\ref{adjointtransf}) and (\ref{constrainedtransf}) respectively, as expected.

At this point, we would like to make an observation that  will be crucial in the next sections. Given the following adjoint and covariantly constrained tensors
\bea
\eta [\Lambda, T]_\alpha &=& \delta_\Lambda\, T_\alpha + (t_\alpha)_{K L} \, \widetilde T^K \, \Lambda^L \ , \label{EFTredundancy}\\
\eta [\Lambda, T]_M &=& \delta_\Lambda T_M + \Omega_{K L} \left(\partial_M \widetilde T^K \, \Lambda^L - \widetilde T^K \, \partial_M \Lambda^L\right) \ , \nn
\eea
where again $\Lambda^M$ is a vector, $T_\alpha$ is an adjoint tensor of weight $\lambda'(T) = 1$ and $T_M$ is a covariantly constrained tensor, the following trivial parameter vanishes
\be
\widetilde \eta^M = 12 (t^\alpha)^{M N} \partial_N \eta_\alpha + \frac 1 2 \Omega^{M N} \eta_N = 0 \ , \label{EFTredundancy2}
\ee
as can easily be seen from the following rewriting
\be
\widetilde \eta = \widehat {\cal L}_\Lambda \widetilde T - 2 \left\{\widetilde T ,\, \Lambda \right\} = - \widehat {\cal L}_{\widetilde T} \Lambda = 0 \ .
\ee

Let us now state what the Jacobiator looks like in terms of the above defined quantities
\be
\label{jacob-e77-uno}
J(\Lambda_1  , \, \Lambda_2 , \, \Lambda_3) = 3 \left[[\Lambda_{[1}, \, \Lambda_{2}]_{(E)}, \, \Lambda_{3]}\right]_{(E)} = \widetilde N  \ ,
\ee
with $\widetilde N$ given by
\be
\widetilde N =  \left\{\Lambda_{[1}, \, \left[\Lambda_2 , \, \Lambda_{3]}\right]_{(E)} \right\} \ , \label{NijenhuisEFT}
\ee
whose ``Nijenhuis'' components are
\bea
N_{\alpha} &=& - \frac 1 2 (t_\alpha)_{K L} \Lambda_{[1}^K \left[\Lambda_2, \, \Lambda_{3]}\right]^L_{(E)}  \label{NijEFT}\\
N_M &=& \frac 1 2 \Omega_{K L} \left(\Lambda_{[1}^K \partial_M \left[\Lambda_2, \, \Lambda_{3]}\right]^L_{(E)} - \partial_M \Lambda_{[1}^K \, \left[\Lambda_2, \, \Lambda_{3]}\right]^L_{(E)}\right)  \ . \nn
\eea
Of course, these components cannot in general be disentangled due to the ambiguity discussed in (\ref{EFTredundancy}), namely that they can be shifted in such a way to leave $\widetilde N$ invariant. However, in this particular case we will now argue that there is no redundancy. Taking a close look into (\ref{EFTredundancy}) we see that it depends on three objects: a vector $\Lambda$, an adjoint tensor $T_\alpha$ and a covariantly constrained tensor $T_M$. In order to shift the components in (\ref{NijEFT}) such that $\widetilde N$ in (\ref{NijenhuisEFT}) remains invariant, we must find expressions for $\Lambda^M$, $T_{\alpha}$ and $T_M$ depending only on $\Lambda_1$, $\Lambda_2$ and $\Lambda_3$. The only possibility that preserves the way in which the tensors transform is
\be
\Lambda^M = \Lambda_i^M \ , \ \ \ \ \ T_\alpha = -\frac 1 2 (t_\alpha)_{K L} \Lambda_j^K \Lambda_k^L \ , \ \ \ \ T_M = \Omega_{K L} \Lambda_{(j}^K \partial_M \Lambda_{k)}^L \ , \ \ \ \ i,j,k = 1,2,3\ ,
\ee
but the antisymmetry in $[123]$ eliminates it, leaving (\ref{NijEFT}) as the only possible choice.

We end this brief subsection by noting that given a trivial vector $\widetilde T$, one has
\be
\widehat {\cal L}_{\widetilde T} \Lambda = 0 = \left\{\widetilde T,\, \Lambda \right\} + [ \widetilde T,\, \Lambda ]_{(E)}\ , \label{EdeTrivial}
\ee
so the E-bracket between a trivial and a generic vector is itself a trivial vector. This observation will be useful later.

\subsection{The theory and its projected tensor hierarchy}

The fields in $E_{7(7)}$ EFT \cite{Hohm:2013uia} are
\be
e_{\mu}{}^a \ , \ \ \ B_{\mu \nu \alpha} \ , \ \ \ B_{\mu \nu M} \ , \ \ \  A_\mu{}^M \ , \ \ \ {\cal M}_{M N} \ ,
\ee
where the scalar matrix ${\cal M}_{M N}$ is again a duality group-valued constrained field. Note that now the two-form field is non-Abelian, and takes values in the adjoint representation. In addition, this EFT requires extra two-forms $B_{\mu \nu M}$, which are  covariantly constrained as in (\ref{covariantlyconstrained}).

A parent background independent action based on a larger duality group (the analog of $O(10,10)$ for DFT) remains unknown in the case of exceptional duality (though see \cite{west}). Constructions for the internal scalar sector were originally considered in \cite{Pacheco:2008ps}. One must then build the action as envisioned by O. Hohm and H. Samtleben \cite{Hohm:2013uia}, by demanding that ``internal'' generalized diffeomorphisms behave as gauge symmetries in four dimensions and extending the KK-DFT framework to be compatible with exceptional duality symmetries and maximal supergravity (for ExFTs in other dimensions see \cite{EFTs}). In total, the bosonic action enjoys the following symmetries:
\begin{itemize}
\item A global continuous $E_{7(7)}$ symmetry.
\item A local $SO(1,3|\mathbb{R}) \times SU(8|\mathbb{R})$ which is trivial in the bosonic sector.
\item External improved diffeomorphisms, infinitesimally parameterized by $\xi^\mu$.
\item Internal generalized diffeomorphisms, inf. parameterized by $\Lambda^M$.
\item Gauge transformations of the two-forms, parameterized by $\Xi_{\mu \alpha}$ and $\Xi_{\mu M}$.
\end{itemize}

As in DFT, there is a set of parameters that allows to write the gauge transformations in a covariant form, and another set that gives rise to field-independent brackets. As before, we note the former with a hat $\widehat \Lambda^M$, $\widehat \Xi_{\mu \alpha}$, $\widehat \Xi_{\mu M}$, and so on, and the later without hat.
Also as before, and for the same reasons, we will restrict attention to the tensor hierarchy, and so will ignore external improved diffeos, and set the vielbein and scalars to background values, ignoring the perturbations around them. The way the transformations of the fields were obtained by Hohm and Samtleben are the following. The vector fields transform such that the derivative ${\cal D}_\mu = \partial_\mu - \widehat {\cal L}_{A_\mu}$ is covariant with respect to generalized diffeomorphisms. Given that it appears here as the argument of a generalized Lie derivative, this only determines the transformation of $A_\mu{}^M$ up to terms of the form (\ref{trivialEFT}) which we note with $\widehat \Xi_\mu$ bellow
\be
\delta A_{\mu}{}^M = {\cal D}_\mu \Lambda^M + 12 (t^\alpha)^{M N} \partial_N \widehat \Xi_{\mu \alpha} + \frac 1 2 \Omega^{M N} \widehat \Xi_{\mu N} \ .
\ee
These of course will become the gauge parameters of the two-forms. We have written a hat on them to indicate that this is the set for which the gauge transformations are covariant with respect to internal diffeos, as before in DFT. Later one defines a covariant field strength as follows
\be
{\cal F}_{\mu \nu}{}^M = 2 \partial_{[\mu} A_{\nu]}{}^M - \left[A_\mu ,\, A_\nu\right]^M_{(E)} - 12 (t^\alpha)^{M N}\partial_N B_{\mu \nu \alpha} - \frac 1 2 \Omega^{M N} B_{\mu \nu N} \ ,
 \ee
which transforms as
\be
\delta {\cal F}_{\mu \nu}{}^M = 2 {\cal D}_{[\mu} \delta A_{\nu]}{}^M - 12 (t^\alpha)^{M N} \partial_N \Delta B_{\mu \nu \alpha} - \frac 1 2 \Omega^{M N} \Delta B_{\mu \nu N}  \ , \label{DeltaB}
\ee
where $\lambda (\delta A) = \frac 1 2$ and we used the variation symbol $\Delta$ to isolate the covariant part of the transformation of the two-form
\bea
\Delta B_{\mu \nu \alpha} &=& \delta B_{\mu \nu \alpha} + (t_\alpha)_{K L} A_{[\mu}{}^K \delta A_{\nu]}{}^L \\
\Delta B_{\mu \nu M} &=& \delta B_{\mu \nu M} + \Omega_{K L} \left(A_{[\mu}{}^L \partial_M \delta A_{\nu]}{}^K - \partial_M A_{[\mu}{}^L \delta A_{\nu]}{}^K \right) \ .
 \eea
By demanding covariance of the field strength one can obtain the gauge transformations of the two-forms from those of the gauge vector. There is a caveat though. Notice that the transformations of the two-forms enter the variation of the field strength (\ref{DeltaB}) exactly in the same way as in (\ref{EFTredundancy2}). There, we made the observation that the combination $(t^\alpha)^{M N}\partial_M B_{\mu \nu \alpha}$ is not completely independent from $\Omega^{M N} B_{\mu \nu N}$. Then, it is not a priory possible to isolate the transformations $\Delta B_{\mu \nu \alpha}$ from those of $\Delta B_{\mu \nu M}$. At this point one should instead consider a combined two-form
\be
\widetilde B_{\mu \nu M} = 12 (t^\alpha)_M{}^N \partial_N B_{\mu \nu \alpha} + \frac 1 2 B_{\mu \nu M} \ ,
\ee
such that the ambiguity is eliminated, with its corresponding gauge parameter
\be
\widetilde {\widehat \Xi}_{\mu M} = 12 (t^{\alpha})_M{}^N \partial_N \widehat \Xi_{\mu \alpha} + \frac 1 2 \widehat \Xi_{\mu M} \ ,
\ee
and then demanding that $\delta {\cal F}_{\mu \nu}{}^M = \widehat {\cal L}_\Lambda  {\cal F}_{\mu \nu}{}^M$ the unambiguous fields and their gauge transformations are
\bea
\delta A_\mu{}^M \! &=&\! {\cal D}_\mu \Lambda^M + \Omega^{M N} \widetilde {\widehat \Xi}_{\mu N} \label{transftildefields}\\
\Delta \widetilde B_{\mu \nu M} \! &=& \! 2 {\cal D}_{[\mu}\widetilde {\widehat \Xi}_{\nu]M} - 2 \left\{ \Lambda \, , \, {\cal F}_{\mu \nu} \right\}{}_M\nn \\ &=& 2 {\cal D}_{[\mu}\widetilde {\widehat \Xi}_{\nu]M}+ 12 (t^{\alpha})_M{}^N \partial_N \left[(t_{\alpha})_{K L} \Lambda^K {\cal F}_{\mu \nu}{}^L\right] + \frac 1 2 \left[\Omega_{K L} \left(\Lambda^L \partial_M {\cal F}_{\mu \nu}{}^K - \partial_M\Lambda^L {\cal F}_{\mu \nu}{}^K\right)\right]\ , \nn
\eea
where we used the symmetrization $\{  \}$ in (\ref{symmprod}) to show that the transformation is constrained in the same way as the field, and defined the covariant transformation $\Delta \widetilde B_{\mu \nu M} = \delta \widetilde B_{\mu \nu M} - 2 \left\{A_{[\mu},\, \delta A_{\nu]} \right\}{}_M$. The expressions between brackets cannot be directly assigned to the $\Lambda$-transformations of the components of $\widetilde B_{\mu \nu M}$, they must be considered always in this combination. The naive separation would lead to the failure of closure of the gauge transformations, the absence of trivial parameters, etc. We will discuss later how to properly disentangle the components and their transformations.

The Bianchi identities
\bea
3 {\cal D}_{[\mu} {\cal F}_{\nu \rho]}{}^M + \widetilde {\cal H}_{\mu \nu \rho M} &=& 0 \\
{\cal D}_{[\mu} \widetilde {\cal H}_{\nu \rho \sigma] M} - \frac 3 2 \left\{{\cal F}_{[\mu \nu},\, {\cal F}_{\rho \sigma]}\right\}{}_M &=& 0 \ ,
\eea
define the three-form field strength
\be
\widetilde {\cal H}_{\mu \nu \rho M} = 3 \left({\cal D}_{[\mu} \widetilde B_{\nu \rho] M} + 2 \left\{ A_{[\mu}\, , \, \partial_\nu A_{\rho]} - \frac 1 3 [A_\nu,\, A_{\rho]}]_{(E)} \right\}{}_M\right) \ , \label{tilde3form}
\ee
which also is constrained in the same way as the tilded tensors and transforms as a vector under generalized diffeomorphisms, as can be easily computed from
\be
\delta {\widetilde H}_{\mu \nu \rho M} = 3 {\cal D}_{[\mu} \Delta \widetilde B_{\nu \rho] M} + 6 \left\{ \delta A_{[\mu} ,\, {\cal F}_{\nu \rho]}\right\}{}_M \ .  \label{transftilde3form}
\ee

The diffeomorphism covariant transformations (\ref{transftildefields}) admit the following trivial parameters
\bea
\Lambda^M &=& 12 (t^\alpha)^{M N} \partial_N {\chi_\alpha} + \frac 1 2 \Omega^{M N} \chi_N \ = \ \Omega^{M N} \widetilde \chi_N \\
\widetilde {\widehat \Xi}_{\mu M} &=& - {\cal D}_{\mu} \widetilde \chi_M \ ,
\eea
where $\chi_\alpha$ is a generic function in the adjoint of $E_{7(7)}$, $\chi_M$ a covariantly constrained function, such that $\widetilde \chi_M$ carries weight $\lambda = \frac 1 2$. The situation is analogous to that in DFT, where there is a set of parameters (noted with a hat) with respect to which the gauge transformations can be written in a generalized diffeomorphism covariant form, and such that the trivial set of parameters is field dependent. For this set of parameters the brackets with respect to which the algebra closes are field dependent, in full analogy with DFT, so we find it convenient to turn to a redefined set of non-covariant parameters ($\Lambda^M$, $\widetilde \Xi_{\mu M}$)
\be
\widetilde {\widehat \Xi}_{\mu M} = \widetilde \Xi_{\mu M} + 2 \left\{\Lambda, \, A_\mu\right\}{}_M \ ,
\ee
such that for this set the trivial parameters are field independent
\bea
\Lambda^M &=&   \Omega^{M N} \widetilde \chi_N  \ ,\nn\\
\widetilde \Xi_{\mu M} &=& - \partial_{\mu} \widetilde \chi_M \ . \label{trivfieldindependent}
\eea

The gauge transformations are now given by
\bea
\delta A_\mu{}^M &=& \partial_\mu \Lambda^M + \widehat {\cal L}_\Lambda A_\mu{}^M + \Omega^{M N} \widetilde \Xi_{\mu N}\ , \label{transfEFT}\\
\delta \widetilde B_{\mu \nu M} &=& 2 \partial_{[\mu} \widetilde \Xi_{\nu] M} + \widehat {\cal L}_\Lambda \widetilde B_{\mu \nu M} + 2 \left\{ \Sigma_{[\mu} \, , \, A_{\nu]} \right\}{}_M \ , \nn
\eea
where we defined
\be
\Sigma_\mu{}^M = \Omega^{M N} \widetilde\Xi_{\mu N} + \partial_\mu \Lambda^M \ .
\ee

These transformations close
\be
[\delta_1 \, , \, \delta_2] = - \delta_{\Lambda_{12}} - \delta_{\widetilde \Xi_{12}} \ ,
\ee
with respect to  field independent brackets
\bea
\Lambda_{12}^M &=& [\Lambda_1 ,\, \Lambda_2]^M_{(E)} \ ,\label{bracketsEFT}\\
\widetilde \Xi_{12 \mu M} &=& 2 \widehat {\cal L}_{\Lambda_{[1}} \widetilde \Xi_{2]\mu M} - 2 \left\{ \partial_\mu \Lambda_{[1}\, , \, \Lambda_{2]}\right\} \ . \nn
\eea
Merging both parameters into a single one, and defining the brackets
\be
\z = \left(\Lambda^M , \, \widetilde \Xi_{\mu M}\right) \ , \ \ \ \ \ \left[\z_1 ,\, \z_2\right] = \left(\Lambda_{12}^M , \, \widetilde \Xi_{12\mu M}\right) \ ,
\ee
the Jacobiator is given  by
\be
J = 3 \, \left[\left[\z_{[1} ,\, \z_2\right],\, \z_{3]}\right] = \left(\widetilde N^M , \, - \partial_\mu \widetilde N_M \right) \ , \label{JacobiatorEFT}
\ee
with $\widetilde N$ defined in (\ref{NijenhuisEFT}).

\subsection{The $L_\infty$ structure of the $E_{7(7)}$ projected tensor hierarchy} \label{Sec::projectedE7}

We begin here discussing the simplest case of the E-bracket gauge algebra and its formulation in terms of the $L_\infty$ structure. As we will show, we only need to define two graded subspaces, which are
\beq
X_1:&\quad\text{functions:}\quad\chi=\chi_\a+\chi_N\ ,\\
X_0:&\quad\text{gauge parameters:}\quad \L^M\ . \\
\eeq
An arbitrary element $\chi \in X_1$ splits into two different parts: $\chi_\a$ belonging to the adjoint representation {\bf 133} of $E_{7(7)}$, and a covariantly constrained function $\chi_N$.
The non-vanishing products are
\beq
\boxed{
\begin{split}
\ell_2(\L_1,\L_2)=&\left[\L_1,\L_2\right]_{(E)}, &\in X_0\\
\ell_3(\L_1,\L_2,\L_3)=& -N_\a - N_M,&\in X_1\\
\ell_1(\chi)=&\ 12(t^\alpha)^{MN}\partial_N \chi_\alpha+\frac 12 \Omega^{MN}\chi_N,&\in X_0\\
\ell_2(\L,\chi)=&\frac 1 2 \d_\L \chi_\a+\frac 1 2 \d_\L \chi_N &\in X_1\\
\end{split}}
\eeq
where we defined the ``Nijenhuis'' tensors in (\ref{NijEFT}).

The construction goes as follows. We start by defining the graded spaces. Initially, we only define $X_0$ as the space of gauge parameters $\L^M$.
Next we identify
\beq
\ell_2(\L_1,\L_2)=\left[\L_1,\L_2\right]_{(E)}\ ,
\eeq
and we set $\ell_1(\L)=0$, so that the identities $\ell_1^2=0$ and $\ell_1\ell_2-\ell_2\ell_1=0$ acting over gauge parameters are trivially fullfilled.
The first non trivial identity we must verify is
\beq
\label{jacob}
-\ell_1(\ell_3 (\L_1,\L_2, \L_3))=&  \ell_2(\ell_2(\L_1,\L_2),\L_3) + \ell_2(\ell_2(\L_3,\L_1),\L_2)
   +\ell_2(\ell_2(\L_2,\L_3),\L_1) \;\\
  & + \ell_3(\ell_1 (\L_1) ,\L_2, \L_3)
  +    \ell_3( \L_1 ,\ell_1(\L_2), \L_3)
    + \ell_3( \L_1 ,\L_2, \ell_1(\L_3))\ .
\eeq
On the first line of the RHS we recognize the Jacobiator, which for the E-bracket is given by (\ref{NijenhuisEFT})
\beq\label{JacobIntro}
  J^M(\Lambda_1,\Lambda_2,\Lambda_3) \ \equiv \
  3\,\big[\big[\Lambda_{[1},\Lambda_{2}\big], \Lambda_{3]}\big]^M
  \ = \ 12 (t^\alpha)^{MN}\partial_N N_\a +\frac 12\Omega^{MN}N_M\ .
 \eeq
	Comparing the expressions \eqref{JacobIntro} and \eqref{jacob} we get the identifications
\beq
\ell_3(\L_1,\L_2,\L_3)=& - N_\a - N_M,\\
\ell_1(\chi)=&\ 12 (t^\alpha)^{MN}\partial_N \chi_\alpha +\frac 12 \Omega^{MN}\chi_N \equiv \widetilde \chi^M \ .
\label{primeras}
\eeq
Here we continue using the notation that a tilde over a tensor means that it takes the form of a trivial parameter (\ref{tilde}).
In order to accomplish this we have defined a new space $X_1$ because there is a nontrivial $\ell_3(\L_1,\L_2,\L_3)$ which maps into this graded subspace. The elements belonging to this subspace will be denoted as $\chi$. As already pointed out, $\chi \in X_1$ splits into two different parts: $\chi_\a$, which belongs to the adjoint representation of $E_{7(7)}$, and  $\chi_N$, a covariantly constrained field.

Now that we have this new space we have to reconsider the identity $\ell_1\ell_2-\ell_2\ell_1=0$ acting on elements belonging to this new subspace
\beq
\label{idideal}
\ell_1(\ell_2(\L,\chi))=\ell_2(\ell_1(\L),\chi)+\ell_2(\L,\ell_1(\chi))\ .
\eeq
We take the first term in the RHS to zero as it contains $\ell_1(\Lambda) \in X_{-1}$, the space of fields that we are ignoring for the moment. The second term in the RHS takes the form of a trivial parameter by virtue of (\ref{EdeTrivial}). This determines $\ell_2(\L,\chi) \in X_1$, but only up to terms of the form (\ref{EFTredundancy}), which belong to the kernel of $\ell_1$
\bea
\ell_2(\L,\chi)_\alpha&=&\frac 1 2 \d_\L \chi_\a + \dots \nn\\
\ell_2(\L,\chi)_M &=& \frac 1 2\d_\L \chi_M + \dots \ ,
\label{l2lchia}
\eea
The dots represent the ambiguity discussed in (\ref{EFTredundancy}), which is resolved by analyzing the $n = 4$ identity. We will discuss this soon and for the moment take \eqref{l2lchia} without the dots as our definition. We now show that the products defined so far are the only ones needed to guarantee the $L_\infty^{\rm gauge}$ structure of the gauge algebra given by the E-bracket.

We now proceed with the $n = 2$ identity $\ell_1\ell_2 - \ell_2\ell_1 = 0$, but now over $\chi_1$, $\chi_2$ $\in X_1$:
\beq
\ell_1(\ell_2(\chi_1,\chi_2))=\ell_2(\ell_1(\chi_1),\chi_2)+\ell_2(\ell_1(\chi_2),\chi_1)\ .
\label{dacero}
\eeq
The RHS of this identity is zero  because $\ell_1(\chi)$ is trivial, and we already chose $\ell_2(\Lambda, \chi)$ to be identified with a generalized diffeomorphic transformation (\ref{l2lchia}). Then we can safely take
\beq
\ell_2(\chi_{1},\chi_{2})=0\ .
\label{l2chichi}
\eeq

Next, we must check the $n = 4$ identity $\ell_1\ell_4-\ell_2\ell_3+\ell_3\ell_2-\ell_4\ell_1=0$ in the case of four gauge parameters (this identity is trivially fulfilled  when evaluated with arguments in $X_1$). First we note that $\ell_4 \ell_1 = 0$ by crossing (\ref{trivparamQ}) with the fact that the Jacobiator is field independent (\ref{NijenhuisEFT}). We now point out that $\ell_3\ell_2-\ell_2\ell_3$ vanishes per se
\beq
\ell_3 \ell_2 - \ell_2 \ell_3  \ &=
 6\, \ell_3   ([ \L_{[1}, \L_2] , \L_3, \L_{4]}) \ - \
  4 \, \ell_2  (\ell_3 (\L_{[1}, \L_2, \L_3) , \L_{4]} ) = 0 \ ,
	\label{viejaidentidad}
	\eeq
a computation that can be checked with some effort. Then, there is no need to define a $\ell_4(\Lambda_1,\, \L_2 ,\, \L_3 ,\, \L_4) = 0$. Let us however analyze what would have happened if we had chosen a different product in (\ref{l2lchia}), by deforming our previous choice with arbitrary elements $\eta_\alpha$ and $\eta_M$ in the kernel of $\ell_1$ (see \ref{EFTredundancy})
\bea
\ell_2(\L,\chi)_\alpha&=&\frac 1 2 \d_\L \chi_\a + \alpha \,\eta_\a \\
\ell_2(\L,\chi)_M &=& \frac 1 2\d_\L \chi_M + \alpha \, \eta_N \ ,
\eea
with
\bea
\eta[\Lambda,\, \chi]_\alpha &=& \delta_\Lambda\, \chi_\alpha + (t_\alpha)_{K L} \, \widetilde \chi^K \, \Lambda^L \ , \\
\eta[\Lambda, \, \chi]_M &=& \delta_\Lambda \chi_M + \Omega_{K L} \left(\partial_M \widetilde \chi^K \, \Lambda^L - \widetilde \chi^K \, \partial_M \Lambda^L\right) \ , \nn
\eea
 the identity would read
\bea
\ell_1\ell_4&=&
 6\, \ell_3   ([ \L_{[1}, \L_2] , \L_3, \L_{4]}) \ - \
  4 \, \ell_2  (\ell_3 (\L_{[1}, \L_2, \L_3) , \L_{4]} )\nn\\
  &=& 4\alpha \eta[\L_{[1},\, N(\L_2,\L_3,\L_{4]})]_\a + 4\alpha \eta[\L_{[1},\, N(\L_2,\L_3,\L_{4]})]_N\ ,
\eea
where we have used the vanishing of \eqref{viejaidentidad} when the old products \eqref{l2lchia} were considered, and the components of $N$ were defined in (\ref{NijEFT}). We see that in order to satisfy the identity, the $\ell_1\ell_4$ term must absorb the ambiguity.
In consequence, one should then define a new graded subspace $X_2$ and take $\ell_4(\L_1,\L_2,\L_3,\L_4)= 4\alpha \eta_\a + 4\alpha \eta_N$ $\in X_2$. Notice that so defined, $\ell_4(\L_1,\L_2,\L_3,\L_4)$ would live in the kernel of $\widetilde \eta(\eta_a,\eta_N)=12(t^\alpha)^{MN}\partial_N \eta_\a+\frac 12 \Omega^{MN}\eta_N$. One should then take $X_2$ as the space of functions that live in the kernel of $\widetilde \eta$ in order to fulfill the  $n = 1$ $L_\infty$ identity, and also define the action of the $\ell_1$ product over this subspace, which following \cite{Hohm:2017pnh} can be simply taken to be an inclusion map into the subspace $X_1$. In summary, the choice made for $\ell_2(\L,\chi)$ in \eqref{l2lchia} is the simplest one in that requires a lower number of non-vanishing products and graded subspaces.

Including to this analysis the tensors coming from the tensor hierarchy is now an easy task. Following the same steps as before, we defined three graded subspaces
\beq
X_1:&\quad\text{functions:}\quad\chi=\chi_\a+\chi_N\ ,\\
X_0:&\quad\text{gauge parameters:}\quad \z=\L^M+\widetilde\Xi_{\m M}\ . \\
X_{-1}:&\quad\text{fields:}\quad \Psi=A_\m{}^M+\widetilde B_{\m\n}\ . \\
\eeq
We have added a new graded subspace in order to incorporate the fields and their gauge transformations. The only nontrivial products are given by
\beq
\boxed{
\begin{split}
\ell_2(\z_1,\z_2)=&\left[\z_1,\z_2\right], &\in X_0\\
\ell_3(\z_1,\z_2,\z_3)=& -N_\a - N_M,&\in X_1\\
\ell_1(\chi)=&\ 12(t^\alpha)^{MN}\partial_N \chi_\alpha+\frac 12 \Omega^{MN}\chi_N - 12 (t^\alpha)_M{}^{N}\partial_N \del_\m\chi_\alpha-\frac 12\del_\m\chi_M,&\in X_0\\
\ell_2(\z,\chi)=&\frac 1 2 \d_\L \chi_\a+\frac 1 2 \d_\L \chi_N ,&\in X_1\\
\ell_1(\z)=&\del_\m \L^M+\O^{MN}\widetilde\Xi_{\m N}+2\del_{[\m}\widetilde \Xi_{\n]M}, &\in X_{-1}\\
\ell_2(\z,\Psi)=&\widehat\cL_{\L} A_\m^M +\d_{\L}\widetilde B_{\m\n}-12(t^\a)_M^N\del_N\left((t_\a)_{LK}\left(\O^{KP}\widetilde\Xi_{[\m P}+\del_{[\m}\L^K\right)A_{\n]}^L\right)\\
		&+\frac 12 \O_{LK} \left( \del_M A_{[\m}^L   \left(\O^{KP}\widetilde\Xi_{\n] P}+\del_{\n]}\L^K\right) -A_{[\m}^L  \del_M \left(\O^{KP}\widetilde\Xi_{\n] P}+\del_{\n]}\L^K\right)\right) &\in X_{-1}
\end{split}}
\eeq
We have simply added here the $L_\infty$ products coming from the gauge transformations of the fields (\ref{transfEFT}) and their trivial versions (\ref{trivfieldindependent}).

This completes the analysis on how the tensor hierarchy of $E_{7(7)}$ fits into $L_\infty^{\rm gauge + fields}$. We do not consider here the full dynamical theory. Even if restricted to the tensor hierarchy degrees of freedom, we expect similar features as those arising in DFT to appear here. Namely, that the spaces and products defined so far must be extended to include components along the other fields and to depend on their background values, see for example (\ref{background}).

\subsection{The unprojected tensor hierarchy}

So far the two-forms we considered (and the one-form gauge parameters) were a specific combination (or projection) of two different contributions. For a general discussion, let us consider generic projected tensors
\be
\widetilde T^M = 12 (t^{\alpha})^{M N} \partial_N T_\alpha  + \frac 1 2 \Omega^{M N} T_N \ . \label{intertwining operator}
\ee
While $T_\alpha$ is a generic adjoint tensor, $T_M$ is covariantly constrained (i.e. as if it were a derivative) and ${\widetilde T}^M$ is also constrained to be of the form above. We saw that when the two-forms $\widetilde B_M$ and their gauge parameters $\widetilde \Xi_\mu$ were grouped into this form, the hierarchy closes exactly. This projection is well known in the context of gauged supergravities, in which case the responsible is usually called the {\it intertwining tensor}. Here, the projector is given in terms of a differential operator, and so we will refer to it as the {\it intertwining operator}.

The particular projection discussed so far is the one entering the action of $E_{7(7)}$ EFT, a duality covariantization of maximal supergravity. Un-projecting points towards new M-theoretical degrees of freedom not present in the supergravity limit. A full hierarchy contains in principle a tower of $p$-forms in different representations $A \in R_1$, $B \in R_2$, $C \in R_3$, etc. Then, $R_1$ here is the {\bf 56} of $E_{7(7)}$. The intertwining operator noted with a tilde ``$\ \widetilde \ \ $'' maps between these spaces $\ \widetilde{}\ : R_{p+1} \to R_p$, such that $\ \widetilde \widetilde = 0$. So far we considered $\widetilde B \in R_1$, and here we are interested in $B\in R_2$ and $\widetilde C \in R_2$. One could in principle continue the hierarchy until the $p$-forms saturate the space-time dimensionality. Modulo covariantly constrained fields, the representations we are expecting here are $R_2 = {\bf 133}$, $R_3 = {\bf 912}$ and $R_4 = {\bf 8645 \oplus 133}$.

The intertwining operator has a non-vanishing kernel, and so it is not invertible. At this stage it might seem a little confusing, but we will call $(\widetilde T_\alpha, \, \widetilde T_M)$ the elements of the kernell of $\widetilde T^M$, namely
\be
12 (t^{\alpha})^{M N} \partial_N \widetilde T_\alpha  + \frac 1 2 \Omega^{M N} \widetilde T_N = 0  \ . \label{projection}
\ee
 The reason of this notation will become clear soon. In the meantime it is important to keep in mind that $\widetilde T_M$ is not $\widetilde T^M$ with its index lowered, but simply a constrained tensor that belongs to the kernel of $\widetilde T^M$.  In the previous sections, we identified two tensors that belong to the kernel of the intertwining operator \label{projection}
\bea
\widetilde T [\Lambda, V]_\alpha &=& \delta_\Lambda\, V_\alpha + (t_\alpha)_{K L} \, \widetilde V^K \, \Lambda^L \ , \label{Kernell1}\\
\widetilde T [\Lambda, V]_M &=& \delta_\Lambda V_M + \Omega_{K L} \left(\partial_M \widetilde V^K \, \Lambda^L - \widetilde V^K \, \partial_M \Lambda^L\right) \ , \nn
\eea
namely, for this specific choice one has $\widetilde T^M = 0$.

It turns out however that there is a more general way to parameterize the kernel of the intertwining operator. It is in terms of new tensors in different representations than the ones considered so far. On the one hand, a tensor in the {\bf 912} $T_\alpha{}^M$ (more on representations and projectors can be found in \cite{deWit:2002vt},\cite{Riccioni:2007au})
\be
T_\alpha{}^M = P_\alpha^{(912)\, M\, \beta}{}_N \, T_\beta{}^N \ , \ \ \ \ P_\alpha^{(912)\, M \, \beta}{}_N = \frac 1 7 \left(\delta_\alpha^\beta \delta_N^M - 12\, (t_\alpha t^\beta)_N{}^M + 4 \,(t^{\beta} t_\alpha)_N{}^M \right) \ ,
\ee
and a covariantly constrained tensor  $T_M{}^N$ (with unconstrained upper index)
\be
\Omega^{N M} \partial_N \otimes T_M{}^K = \Omega^{M N} \, T_M{}^K\, T_N{}^L =  (t^\alpha)^{N M} \partial_N \otimes T_M{}^K = (t^\alpha)^{M N} \, T_M{}^K\, T_N{}^L = 0 \ .
\ee
Taking into account that due to the section condition
\be
(t^{\alpha})_M{}^K \, P^{(912) L\, \beta}_\alpha{}_N \, \partial_{(K} \otimes \partial_{L)}= 0 \ ,
\ee
more generally the kernel of $\widetilde T^M$ is given by \cite{Hohm:2013uia}
\bea
\widetilde T_\alpha &=& 2\, \partial_M T_\alpha{}^M + (t_\alpha)_N{}^M \, T_M{}^N \nn \\
\widetilde T_M &=& - 2\, \partial_N T_M{}^N - \partial_M T_N{}^N \ ,\label{Kernell2}
\eea
namely, for {\it any} $T_{\alpha}{}^M$ and $T_M{}^N$ one has $\widetilde T^M = 0$.

The way to recover the specific case (\ref{Kernell1}) is by making the choice
\bea
T_\alpha{}^M &=& \frac 7 2 P_\alpha^{(912)\, M \, \beta}{}_N \, \Lambda^N \, V_\beta \\
T_M{}^N &=& (t^\alpha)_K{}^N \left(8\, \partial_M \Lambda^K\, V_\beta - 4\, \Lambda^K\, \partial_M V_\beta\right) - \frac 1 2 \Lambda^N\, V_M \ . \nn
\eea

It should be clear now why we chose the tilde notation again. The equation (\ref{Kernell2}) is the analog to (\ref{intertwining operator}) for the next level in the hierarchy.
An interesting point is that from the definition (\ref{Kernell2}), it can be checked that $(\widetilde T_\alpha , \, \widetilde T_M$) transform as in (\ref{adjointtransf}) with weight $\lambda' = 1$ and (\ref{constrainedtransf}), provided
\bea
\delta T_\alpha{}^M \!\!\!&=&\!\!\! \Lambda^K \partial_K T_\alpha{}^M - 12\, (t^\beta)^M{}_N (t_\beta)^K{}_L\, \partial_K \Lambda^L \, T_\alpha{}^N  + 12\, f_{\alpha \gamma}{}^\beta (t^\gamma)^K{}_L \, \partial_K \Lambda^L \, T_\beta{}^M + \frac 3 2 \partial_K \Lambda^K \,T_\alpha{}^M \nn \\
\delta T_M{}^N \!\!\!&=&\!\!\! \Lambda^K \partial_K T_M{}^N  - 12\, (t^\alpha)^N{}_P (t_\alpha)^K{}_L \, \partial_K \Lambda^L \, T_M{}^P  + 12 \, (t^\alpha)^P{}_M (t_\alpha)^K{}_L \, \partial_K \Lambda^L \, T_P{}^N \nn\\ && + \,\partial_K \Lambda^K \, T_M{}^N - 24 \, (t^\alpha)^N{}_L \, \partial_M \partial_K\Lambda^L \, T_\alpha{}^K \ .
\eea
Of course, these transformations preserve the constrained nature of both tensors.

Given that the intertwining operator is not invertible, it is not possible to read $T_\alpha$ and $T_M$ from a given $\widetilde T^M$, except {\it up to} terms of the form (\ref{Kernell2}). Let us apply this lesson to disentangle the components of the projected two-form of the previous sections. From the transformation of $\widetilde B$ in (\ref{transftildefields}) we can write the covariant form of the transformations of its components
\bea
\Delta B_{\mu \nu \alpha} &=& 2 {\cal D}_{[\mu} \widehat \Xi_{\nu] \alpha} + (t_{\alpha})_{K L} \Lambda^K {\cal F}_{\mu \nu}{}^L + \widetilde {\widehat \Sigma}_{\mu \nu \alpha} \label{CovTransfB}\\
\Delta B_{\mu \nu M} &=& 2 {\cal D}_{[\mu} \widehat \Xi_{\nu] M} - \Omega_{K L} \left(\Lambda^K \partial_M {\cal F}_{\mu \nu}{}^L + {\cal F}_{\mu \nu}{}^K \partial_M \Lambda^L\right) + \widetilde {\widehat \Sigma}_{\mu \nu M} \ , \nn
\eea
which are related to the genuine transformation through the relation $\Delta \widetilde B_{\mu \nu M} = \delta \widetilde B_{\mu \nu} - 2 \left\{A_{[\mu},\, \delta A_{\nu]}\right\}$
\bea
\Delta B_{\mu \nu \alpha} &=& \delta B_{\mu \nu \alpha} + (t_\alpha)_{K L} A_{[\mu}{}^K \delta A_{\nu]}{}^L \\
\Delta B_{\mu \nu M} &=& \delta B_{\mu \nu M} - \Omega_{KL} \left(A_{[\mu}{}^K \partial_M \delta A_{\nu]}{}^L + \delta A_{[\nu}{}^K \partial_M A_{\mu]}{}^L\right) \ . \nn
\eea
The last terms in both lines of  (\ref{CovTransfB}) parameterize the ambiguity in removing the projection, but will obviously correspond to the transformation of the two-form with respect to the parameter of the next field in the hierarchy (a three-form). The hat on it is to continue with the notation that parameters with a hat are those that allow to write the gauge transformations covariantly with respect to generalized diffeomorphisms. The tilde indicates that the parameters are of the form (\ref{Kernell2}), namely they belong to the kernel of $\Delta \widetilde B_{\mu \nu M}$.

We now remove the projection from the three-form field strength  (\ref{tilde3form})
\bea
{\cal H}_{\mu \nu \rho \alpha} &=& 3 {\cal D}_{[\mu} B_{\nu \rho] \alpha} - 3 (t_{\alpha})_{K L} A_{[\mu}{}^K \Gamma_{\nu \rho]}{}^L - \widetilde C_{\mu \nu \rho \alpha} \\
{\cal H}_{\mu \nu \rho M} &=& 3 {\cal D}_{[\mu} B_{\nu \rho] M} + 3 \Omega_{K L} \left(A_{[\mu}{}^K \partial_M \Gamma_{\nu \rho]}{}^L + \Gamma_{[\nu\rho}{}^K \partial_M A_{\mu]}{}^L\right) - \widetilde C_{\mu \nu \rho M} \ ,\nn
\eea
where we defined the following combination relevant for the Chern-Simons contributions
\be
\Gamma_{\mu \nu}{}^M = \partial_{[\mu} A_{\nu]}{}^M - \frac 1 3 \left[A_\mu , \, A_\nu\right]{}_{(E)}^M \ .
\ee
The transformation of the three-form components (which can also be derived from (\ref{transftilde3form})) are
\bea
\delta {\cal H}_{\mu \nu \rho \alpha} &=& 3 {\cal D}_{[\mu} \Delta B_{\nu \rho] \alpha} - 3 (t_{\alpha}) \delta A_{[\mu}{}^K {\cal F}_{\nu \rho]}{}^L - \Delta \widetilde C_{\mu \nu \rho \alpha} \\
\delta {\cal H}_{\mu \nu \rho M} &=& 3 {\cal D}_{[\mu} \Delta B_{\nu \rho] M} + 3 \Omega_{K L} \left(\delta A_{[\mu}{}^K \partial_M {\cal F}_{\nu \rho]}{}^L + {\cal F}_{[\nu\rho}{}^K \partial_M A_{\mu]}{}^L\right) - \Delta \widetilde C_{\mu \nu \rho M} \ ,\nn
\eea
where
\bea
\Delta \widetilde C_{\mu \nu \rho \alpha} &=& \delta \widetilde C_{\mu \nu \rho \alpha} + 3 \widetilde T_{\alpha}[\delta A_{[\mu},\, B_{\nu \rho]}] + 2 \widetilde T_\alpha [A_{\mu}\, ,\, \{A_\nu,\delta A_{\rho]}\}] \nn\\
\Delta \widetilde C_{\mu \nu \rho M} &=& \delta \widetilde C_{\mu \nu \rho M} + 3 \widetilde T_{M}[\delta A_{[\mu},\, B_{\nu \rho]}] + 2 \widetilde T_M [A_{\mu}\, ,\, \{A_\nu,\delta A_{\rho]}\}] \ ,
 \eea
and the requirement of these transforming covariantly determines the covariant transformations of the three-form fields
\bea
\Delta \widetilde C_{\mu\nu\rho \alpha} &=& 3 {\cal D}_{[\mu} \widetilde {\widehat \Sigma}_{\nu \rho] \alpha} - \widetilde T_{\alpha} [\Lambda ,\, {\cal H}_{\mu \nu \rho}] - 3 \widetilde T_{\alpha} [{\cal F}_{[\mu \nu}, \, \widehat \Xi_{\rho]}] \\
\Delta \widetilde C_{\mu\nu\rho M} &=& 3 {\cal D}_{[\mu} \widetilde {\widehat \Sigma}_{\nu \rho] M} - \widetilde T_M [\Lambda ,\, {\cal H}_{\mu \nu \rho}] - 3 \widetilde T_M [{\cal F}_{[\mu \nu}, \, \widehat \Xi_{\rho]}] \ ,
\eea
where $\widetilde T$ was defined in (\ref{Kernell1}).

The new three-form field and its parameter are now constrained to take the form (\ref{Kernell2})
\bea
\widetilde C_{\mu \nu \rho \alpha} &=& 2 \partial_M C_{\mu \nu \rho \alpha}{}^M + (t_\alpha)_N{}^M\, C_{\mu \nu \rho M}{}^N\\
\widetilde C_{\mu \nu \rho M} &=& -2 \partial_N C_{\mu \nu \rho M}{}^ N - \partial_M C_{\mu \nu \rho N}{}^N \ , \nn
\eea
and
\bea
\widetilde {\widehat \Sigma}_{\mu \nu  \alpha} &=& 2 \partial_M {\widehat \Sigma}_{\mu \nu  \alpha}{}^M + (t_\alpha)_N{}^M\, {\widehat \Sigma}_{\mu \nu  M}{}^N\\
\widetilde {\widehat \Sigma}_{\mu \nu  M} &=& -2 \partial_N {\widehat \Sigma}_{\mu \nu M}{}^ N - \partial_M {\widehat \Sigma}_{\mu \nu  N}{}^N \ . \nn
\eea
This is the projection required to truncate the hierarchy to this level. This new intertwining operator has its own non-vanishing kernel\footnote{ \label{Footnote}
It is easy to anticipate some representations of the next and final level of the hierarchy. Consider for instance a tensor $\Theta_\alpha{}^{M N} \in {\bf 8645} \oplus {\bf 133}$, namely satisfying
\be\Theta_\alpha{}^{M N} = (t^\beta)^{M N} \Theta_{[\alpha \beta]}  = P^{(912) M\, \beta}_\alpha{}_P \, \Theta_\beta{}^{P N} \ .
\ee
Due to the second identity we can choose a candidate in the $\bf 912$ that because of the first one belongs to the kernel of $\widetilde T_\alpha$ and $\widetilde T_M$
\be
\widetilde \Sigma_\alpha{}^M = \partial_N \Theta_\alpha{}^{M N} \ , \ \ \ \widetilde  \Sigma_M{}^N = 0 \ ,
\ee
namely $\widetilde T_\alpha = 2 \partial_M \widetilde \Sigma_\alpha{}^M = 0$, thanks to the section condition.
}, and so the projection is non-removable unless higher tensors enter the game.

We have been able to move a step upwards in the hierarchy, so let us resume the main results. The closed hierarchy now contains fields $A_\mu{}^M$, $B_{\mu \nu \alpha}$, $B_{\mu \nu M}$, $\widetilde C_{\mu \nu \rho \alpha}$ and $\widetilde C_{\mu \nu \rho M}$, the last three being constrained fields. Their gauge parameters are $\Lambda^M$, $\widehat \Xi_{\mu \alpha}$, $\widehat \Xi_{\mu M}$, $\widetilde {\widehat \Sigma}_{\mu \nu \alpha}$ and $\widetilde {\widehat \Sigma}_{\mu \nu M}$, and again the last three are also constrained. The gauge transformations are
\bea
\delta A_\mu{}^M &=& {\cal D}_\mu \Lambda^M + \Omega^{M N} \widetilde {\widehat \Xi}_{\mu N} \nn \\
\Delta B_{\mu \nu \alpha} &=& 2 {\cal D}_{[\mu} \widehat \Xi_{\nu] \alpha} + (t_{\alpha})_{K L} \Lambda^K {\cal F}_{\mu \nu}{}^L + \widetilde {\widehat \Sigma}_{\mu \nu \alpha} \nn\\
\Delta B_{\mu \nu M} &=& 2 {\cal D}_{[\mu} \widehat \Xi_{\nu] M} - \Omega_{K L} \left(\Lambda^K \partial_M {\cal F}_{\mu \nu}{}^L + {\cal F}_{\mu \nu}{}^K \partial_M \Lambda^L\right) + \widetilde {\widehat \Sigma}_{\mu \nu M}   \\
\Delta \widetilde C_{\mu\nu\rho \alpha} &=& 3 {\cal D}_{[\mu} \widetilde {\widehat \Sigma}_{\nu \rho] \alpha} - \widetilde T_{\alpha} [\Lambda ,\, {\cal H}_{\mu \nu \rho}] - 3 \widetilde T_{\alpha} [{\cal F}_{[\mu \nu}, \, \widehat \Xi_{\rho]}] \nn\\
\Delta \widetilde C_{\mu\nu\rho M} &=& 3 {\cal D}_{[\mu} \widetilde {\widehat \Sigma}_{\nu \rho] M} - \widetilde T_M [\Lambda ,\, {\cal H}_{\mu \nu \rho}] - 3 \widetilde T_M [{\cal F}_{[\mu \nu}, \, \widehat \Xi_{\rho]}] \ .\nn
\eea
They admit the following field-dependent trivial parameters
\bea
\Lambda^M_{trivial} &=& \widetilde \chi^M = 12 (t^\alpha)^{M N} \partial_N \chi_\alpha + \frac 1 2 \Omega^{M N} \chi_N \nn \\
\widehat \Xi_{\mu \alpha}^{trivial} &=& - {\cal D}_\mu \chi_\alpha + \widetilde {\widehat \chi}_{\mu \alpha} \nn \\
\widehat \Xi_{\mu M}^{trivial} &=& - {\cal D}_\mu \chi_M + \widetilde {\widehat \chi}_{\mu M} \\
\widetilde {\widehat \Sigma}_{\mu \nu \alpha}^{trivial} &=& - 2 {\cal D}_{[\mu} \widetilde {\widehat \chi}_{\nu]\alpha} - \widetilde T_\alpha[{\cal F}_{\mu \nu} , \, \chi] \nn \\
\widetilde {\widehat \Sigma}_{\mu \nu M}^{trivial} &=& - 2 {\cal D}_{[\mu} \widetilde {\widehat \chi}_{\nu]M} - \widetilde T_M[{\cal F}_{\mu \nu} , \, \chi]  \ ,\nn
\eea
where
\be
12 (t^\alpha)^{M N} \partial_N \widetilde {\widehat \chi}_{\mu \alpha} + \frac 1 2 \Omega^{M N} \widetilde {\widehat \chi}_{\mu N} = 0 \ . \label{constChi}
\ee
The covariant transformations close
\be
[\delta_1 ,\, \delta_2] = - \delta_{\Lambda_{12}} - \delta_{{\widehat \Xi}_{12}} - \delta_{{\widetilde{\widehat \Sigma}}_{12}} \ ,
\ee
with respect to field-dependent brackets
\bea
\Lambda_{12}^M &=& [\Lambda_1,\, \Lambda_2]_{(E)}^M \nn \\
\widehat \Xi_{12\mu \alpha} &=& (t_\alpha)_{K L} \Lambda^K_{[1} {\cal D}_\mu \Lambda^L_{2]} \nn\\
\widehat \Xi_{12 \mu M} &=& \Omega_{K L} ({\cal D}_\mu \Lambda_{[1}^K \partial_M \Lambda_{2]}^L + \Lambda_{[2}^K \partial_M {\cal D}_\mu \Lambda_{1]}^L )\\
\widetilde {\widehat \Sigma}_{12 \mu \nu \alpha} &=& 4 \widetilde T_\alpha [{\cal D}_{[\mu} \Lambda_{[1},\, \widehat \Xi_{2]\nu]}] + 2 \widetilde T_\alpha [\widetilde {\widehat \Xi}_{[1[\mu},\, \widehat \Xi_{2]\nu]}] + 2 \widetilde T_\alpha [\Lambda_{[1},\, \{\Lambda_{2]},\, {\cal F}_{\mu \nu}\}] \nn \\
\widetilde {\widehat \Sigma}_{12 \mu \nu M} &=& 4 \widetilde T_M [{\cal D}_{[\mu} \Lambda_{[1},\, \widehat \Xi_{2]\nu]}] + 2 \widetilde T_M [\widetilde {\widehat \Xi}_{[1[\mu},\, \widehat \Xi_{2]\nu]}] + 2 \widetilde T_M [\Lambda_{[1},\, \{\Lambda_{2]},\, {\cal F}_{\mu \nu}\}] \ .\nn
 \eea
If we wished to eliminate the field dependence of the brackets, we have to perform the following field-dependent redefinitions of the gauge parameters
\bea
\Xi_{\mu \alpha} &=& \widehat \Xi_{\mu \alpha} + (t_{\alpha})_{K L} A_\mu{}^K \Lambda^L \nn\\
\Xi_{\mu M} &=& \widehat \Xi_{\mu M} + \Omega_{K L} \left(\partial_M \Lambda^K A_\mu{}^L - \Lambda^K \partial_M A_\mu{}^L\right) \\
\widetilde \Sigma_{\mu \nu \alpha} &=& \widetilde {\widehat \Sigma}_{\mu \nu \alpha} - 2 \widetilde T_\alpha [A_{[\mu},\, \Xi_{\nu]}] - \widetilde T_\alpha [\Lambda,\, B_{\mu \nu}] - 2 \widetilde T_\alpha [A_{[\mu},\, \{A_{\nu]},\, \Lambda\}] \nn \\
\widetilde \Sigma_{\mu \nu M} &=& \widetilde {\widehat \Sigma}_{\mu \nu M} - 2 \widetilde T_M [A_{[\mu},\, \Xi_{\nu]}] - \widetilde T_M [\Lambda,\, B_{\mu \nu}] - 2 \widetilde T_M [A_{[\mu},\, \{A_{\nu]},\, \Lambda\}] \ .\nn
\eea
The redefinition is non-covariant with respect to generalized diffeomorphisms, and the trivial parameters are now given by field-independent quantities
\bea
\Lambda^M_{trivial} &=& 12 (t^\alpha)^{M N} \partial_N \chi_\alpha + \frac 1 2 \Omega^{M N} \chi_N \nn \\
\Xi_{\mu \alpha}^{trivial} &=& - \partial_\mu \chi_\alpha + \widetilde \chi_{\mu \alpha} \nn \\
\Xi_{\mu M}^{trivial} &=& - \partial_\mu \chi_M + \widetilde \chi_{\mu M} \label{trivialfieldindp}\\
\widetilde \Sigma_{\mu \nu \alpha}^{trivial} &=& - 2 \partial_{[\mu} \widetilde \chi_{\nu] \alpha} \nn \\
\widetilde \Sigma_{\mu \nu M}^{trivial} &=& - 2 \partial_{[\mu} \widetilde \chi_{\nu] M} \ , \nn
\eea
where we have made the redefinitions
\bea
\widetilde \chi_{\mu \alpha} &=& \widetilde {\widehat \chi}_{\mu \alpha} + \widetilde T_\alpha[A_\mu,\, \chi] \nn \\
\widetilde \chi_{\mu M} &=& \widetilde {\widehat \chi}_{\mu M} + \widetilde T_M[A_\mu,\, \chi] \ ,
 \eea
that preserve the constraint (\ref{constChi}).

The brackets for the redefined parameters are now independent of the fields
\bea
\Lambda_{12}^M &=& [\Lambda_1,\, \Lambda_2]_{(E)}^M \nn \\
\Xi_{12 \mu \alpha} &=& 2 \delta_{\Lambda_{[1}} \Xi_{2]\mu \alpha}  + (t_\alpha)_{KL} \partial_\mu \Lambda^K_{[1} \Lambda_{2]}^L \nn \\
\Xi_{12 \mu M} &=& 2 \delta_{\Lambda_{[1}} \Xi_{2]\mu M}  + \Omega_{KL} \left(\partial_\mu \Lambda_{[2}^K \partial_M \Lambda_{1]}^L - \partial_M \partial_\mu \Lambda_{[2}^K
\Lambda_{1]}^L\right) \label{bracketfieldindp}\\
\widetilde \Sigma_{12 \mu \nu \alpha} &=& 2 \widetilde T_\alpha [\Lambda_{[1},\, \widetilde \Sigma_{2]\mu\nu}] - 4 \widetilde T_\alpha [\partial_{[\mu}\Lambda_{[1},\, \Xi_{2]\nu]}] - 2 \widetilde T_\alpha [\widetilde \Xi_{[1[\mu},\, \Xi_{2]\nu]}] \nn \\
\widetilde \Sigma_{12 \mu \nu M} &=& 2 \widetilde T_M [\Lambda_{[1},\, \widetilde \Sigma_{2]\mu\nu}] - 4 \widetilde T_M [\partial_{[\mu}\Lambda_{[1},\, \Xi_{2]\nu]}] - 2 \widetilde T_M [\widetilde \Xi_{[1[\mu},\, \Xi_{2]\nu]}] \ .\nn
\eea
One can then compute the Jacobiator components as we did repeatedly before
\bea
J^M &=& \widetilde N^M = 12 (t^\alpha)^{M P} \partial_P N_\alpha + \frac 1 2 \Omega^{M P} N_P  \nn \\
J_{\mu \alpha} &=& - \partial_\mu N_\alpha + \widetilde N_{\mu \alpha} \nn \\
J_{\mu M} &=& - \partial_\mu N_M + \widetilde N_{\mu M} \label{Jactriv}\\
\widetilde J_{\mu \nu \alpha} &=& - 2 \partial_{[\mu} \widetilde N_{\nu] \alpha} \nn \\
\widetilde J_{\mu \nu M} &=& - 2 \partial_{[\mu} \widetilde N_{\nu] M} \ ,\nn
\eea
with the Nijenhuis tensors $N_\alpha$ and $N_M$ defined in (\ref{NijEFT}), and
\bea
\widetilde N_{\mu \alpha} &=& -2 \widetilde T_\alpha [\Lambda_{[1}, \, \{\Lambda_2,\, \partial_\mu \Lambda_{3]}\}] \nn \\
\widetilde N_{\mu M} &=& -2 \widetilde T_M [\Lambda_{[1}, \, \{\Lambda_2,\, \partial_\mu \Lambda_{3]}\}]  \ .
\eea

We can now compute the bracket between  a trivial parameter $\z_t$ of the form (\ref{trivialfieldindp}) and a generic parameter, finding
\bea
\left[\z_t ,\, \z\right]^M &=& \widetilde \chi'{}^M \nn \\
\left[\z_t ,\, \z \right]_{\mu \alpha} &=& - \partial_\mu \chi'_\alpha + \widetilde \chi'_{\mu \alpha} \nn \\
\left[\z_t ,\, \z\right]_{\mu M} &=& - \partial_\mu \chi'_M + \widetilde \chi'_{\mu M}  \label{TheoremEFT}\\
\left[\z_t ,\, \z\right]_{\mu \nu \alpha} &=& - \partial_{[\mu} \widetilde \chi'_{\nu] \alpha} \nn \\
\left[\z_t ,\, \z\right]_{\mu \nu M} &=& - \partial_{[\mu} \widetilde \chi'_{\nu] M} \ ,\nn
\eea
where
\bea
\chi'_{\alpha} &=& \frac 1 2 (t_\alpha)_{K L} \Lambda^L \widetilde \chi^L \nn \\
\chi'_{M} &=&  - \frac 1 2 \Omega_{KL } \left(\Lambda^K \partial_M \widetilde \chi^L - \partial_M \Lambda^K \widetilde \chi^L\right) \\
\widetilde \chi'_{\mu \alpha} &=& - \widetilde T_\alpha[\Lambda, \Xi_{t \, \mu}] \nn \\
\widetilde \chi'_{\mu M} &=& - \widetilde T_M[\Lambda, \Xi_{t \, \mu}] \ .\nn
\eea

We have then shown that the next level in the hierarchy admits a set of parameters for which the criteria required by the theorem discussed in Sections \ref{sec::theorem} and \ref{sec::theorem2} is met. Namely, that the theory admits trivial field independent parameters (\ref{trivfieldindependent}), that for this set of parameters the bracket is field independent (\ref{bracketfieldindp}), that the bracket between a trivial parameter and a generic one is trivial (\ref{TheoremEFT}), and that the Jacobiator is a trivial parameter (\ref{Jactriv}). The theorem then states that the gauge algebra again fits into an $L_\infty^{\rm gauge + fields}$ structure. This procedure could be extended all the way up to space-time saturation, starting by the next level discussed in footnote \ref{Footnote}.

\section{$L_\infty$ and gauged supergravity} \label{sec::gs}

As discussed in the previous sections, the gauge algebra of DFT and EFT does not satisfy the Jacobi identity, requiring a deformation therein produced by a non-vanishing triple product (so-called the Nijenhuis tensor). This is a manifestation of the fact that duality covariant theories contain trivial parameters, or said in a more fancy way, symmetries for symmetries.

On the other hand, generalized compactifications of these theories (originally introduced in \cite{Aldazabal:2011nj} in the context of DFT, see also \cite{Grana:2012rr},\cite{SSDFT}, and later considered in an EFT context \cite{GSSEFT}) give rise to gauged supergravities. For reviews see \cite{reviews}. It is then natural to ask the question of whether the $L_\infty$ algebraic structure of the parent double and exceptional field theories leaves some imprint after compactification. Of course, this question can be asked and answered independently, without any mention to parent theories and compactifications, so we will address these questions from both ends.

Here we show that the most salient aspects of the algebraic structure of DFT and EFT are indeed reproduced by gauged supergravities. The failure of the Jacobi identity to be satisfied is governed by the so-called intertwining tensor, a particular projection of the embedding tensor. We explicitly compute the Jacobiator, and find that gauged supergravities admit novel symmetries for symmetries in which the trivial parameters are deformed by gaugings. Our goal is to show that gauged supergravities admit (via field redefinitions) a set of parameters that satisfy sufficient criteria (see Subsections \ref{sec::theorem} and \ref{sec::theorem2}) for the algebra to fit into an $L_\infty^{\rm gauge + fields}$ structure.

The presentation proceeds as follows. First we perform a generalized Scherk-Schwarz compactification of the KK formulation of DFT, ending with a gauged supergravity corresponding to the electric sector of a theory with half-maximal supersymmetry. This is the simplest possible scenario to explore these issues, serving as a prototype to confirm our expectations. We then move to the general case of generic gauged supergravities, without any mention to compactifications, for which the structure of the gauge transformations of the tensor hierarchy is well know. In this context we perform convenient redefinitions in order to identify field-independent brackets, their Jacobiator, the Nijenhuis tensors, and the general form of trivial gauge parameters.

\subsection{Generalized Scherk-Schwarz compactification of KK-DFT}

Let us briefly discuss the generalized Scherk-Schwarz (SS) compactification of the KK-DFT action so that our presentation is self-contained. The generalized metric and frame formulations were compactified in \cite{Aldazabal:2011nj}, and here we use the same technique for this formulation. The KK-formulation is particularly useful to perform KK tower expansions in tori compactifictations \cite{Aldazabal:2016yih}. We start by recalling the KK-DFT action (\ref{KKDFT})
\be
S  =\int d^{n}xd^{2d}X\sqrt{-g}e^{-2\phi}\mathcal{L}\ ,
\ee
\bea \mathcal{L}  &=&\widehat{\mathcal{R}}-4\, g^{\mu\nu}\mathcal{D}_{\mu}\phi\mathcal{D}_{\nu}\phi+4\, \nabla_{\mu}\left(g^{\mu\nu}\mathcal{D}_{\nu}\phi\right)-\frac{1}{12}\mathcal{H}_{\mu\nu\rho}\mathcal{H}^{\mu\nu\rho}\label{DFT-KK lagrangian}\\
 && +\frac{1}{8}g^{\mu\nu}\mathcal{D}_{\mu}\mathcal{M}_{MN}\mathcal{D}_{\nu}{\cal M}^{MN}-\frac{1}{4}\mathcal{M}_{MN}\mathcal{F}_{\mu\nu}{}^M\mathcal{F}^{\mu\nu N}-V \ .\nonumber
\eea
We will only discuss the case of $n=4$ space-time dimensions: some results will suffer modifications in other dimensions. We can distinguish three types of global transformations (to be complemented with  $\partial_M \to  U_M{}^P \partial_P$ and $\partial_\mu \to e^{\frac \gamma 2} \partial_\mu$)
\bea
g_{\mu\nu} &\to& e^{\gamma}\, {g}_{\mu\nu}\nonumber \\
{\cal M}^{MN} &\to& U^{M}{}_P\, U^{N}{}_Q\, {\mathcal{M}}^{PQ} \nonumber \\
\phi &\to& \phi + c + \gamma \label{GlobalSymm}\\
A_{\mu}{}^M  &\to& e^{\frac{\gamma}{2}}\, U^{M}{}_P \, {A}_{\mu}{}^P\nonumber \\
B_{\mu\nu} &\to& e^{\gamma} {B}_{\mu\nu} \ ,\nonumber
\eea
where $c, \gamma \in \mathbb{R}$ and $U \in O(d,d|\mathbb{R})$. When a field is shifted with a warp factor $e^{\omega \gamma}$ we call $\omega$ the ``external weight'', such  that
\be
 \omega(g) = \omega(B) = 1 \ , \ \ \ \omega(A) =\frac 1 2\ , \ \ \ \omega(e^{-2\phi}) = -2\ , \ \ \  \omega({\cal M}) = 0 \ . \label{externalweights}
 \ee
 While $U$-rotations (standard duality symmetry) and $\gamma$ and $c$-sifts leave the Lagrangian invariant, the $c$-transformations re-scale the measure
\be
\sqrt{-g} e^{-2\phi} \to e^{-2c} \, \sqrt{-g} e^{-2\phi} \ ,
\ee
and so are not a symmetry of the action, but of the equations of motion only. The Scherk-Schwarz compactification procedure consists in proposing a compactification ansatz based on the global transformations
\begin{align}
g_{\mu\nu}(x,X) & =e^{\gamma(X)}\, {g}_{\mu\nu}(x)\nonumber \\
{\cal M}^{MN}(x,X) & =U^{M}{}_A(X)U^{N}{}_B(X)\, {\mathcal{M}}^{AB}(x)\nonumber \\
\phi(x,X) & = {\phi}(x)+c(X)+\gamma(X)\label{GSansatz}\\
A_{\mu}{}^M(x,X) & =e^{\frac{\gamma(X)}{2}}U^{M}{}_A(X)\, {A}_{\mu}{}^A(x)\nonumber \\
B_{\mu\nu}(x,X) & =e^{\gamma(X)} \, {B}_{\mu\nu}(x)\ ,\nonumber
\end{align}
where in this context $U^{M}{}_A$, $\gamma$ and $c$ are know as ``twists''.  We distinguish between indices $M,N,P,\dots$ transforming under $O(d,d)$ transformations in the parent DFT action, and those $A,B,C,\dots$ transforming under the $O(d,d)$ in the effective gauge supergravity. Equivalently, we propose an ansatz for the gauge parameters
\begin{align}
\Lambda^{M}(x,X) & =e^{\frac{\gamma(X)}{2}}U^{M}{}_A(X) \, {\Lambda}^{A}(x)\nonumber \\
\Xi_{\mu}(x,X) & =e^{\gamma(X)}\, {\Xi}_{\mu}(x) \ ,\label{parametersansatz}
\end{align}
by assigning them external weights $\omega(\Lambda) = \frac 1 2$ and $\omega(\Xi) = 1$.

In general, the SS ansatz for a tensor $T^{M}(x,X)$ with weight $\lambda$ and external weight $\omega$ is
\be
T^M(x,X) = e^{\omega\, \gamma(X) - 2 \lambda \, c(X)} \, U^M{}_A(X) \, T^A(x) \ .
\ee
The generalized diffeomorphisms determine the form of the effective gauge transformation of $T^A(x)$ through the relation
\be
\widehat {\cal L}_\Lambda T^M (x,X) = e^{\omega\, \gamma(X) - 2 \lambda \, c(X)} \, U^M{}_A(X) \,{\cal L}_\Lambda T^A (x) \ ,
\ee
yielding
\be
{\cal L}_\Lambda T^A = - X_{B C}{}^A \, \Lambda^B \, T^C + \left(\lambda + \omega - \frac 1 2\right)\, f_B \Lambda^B\, T^A \ , \label{gaugetransf}
\ee
where
\be
X_{B C}{}^A = - f_{B C}{}^A + \, f_{[C}\, \delta^A_{B]} - \frac 1 2 \, f^A\, \eta_{B C} \ ,
\ee
and we finally arrive to the gaugings or fluxes $f^{A}$ and $f_{AB}{}^C$, defined as
\begin{align*}
f^{A} & \equiv e^{\frac{\gamma}{2}}\, U_{M}{}^A\partial^{M}\gamma=-2\, e^{\frac{\gamma}{2}}\left(2U_{M}{}^A\partial^{M}c-\partial^{M}U_{M}{}^A\right)\\
f_{ABC} & \equiv 3 \, e^{\frac{\gamma}{2}}\, U_{M[A}\partial^{M} U^P{}_{B}U^{Q}{}_{C]} \eta_{P Q} \ .
\end{align*}
 Despite of being built from $X$-dependent twists, when
they appear in this particular combination we demand that they are constant. Demanding that the fluxes are constant, plus strong constrained twists, is known to imply the quadratic constraints
\begin{align}
f^{A}f_{A} & = 0 \nonumber \\
f^{A}f_{AB}{}^C & = 0\label{cuadraticconstraint} \\
f_{[AB}{}^E f_{C]DE} & = \frac{2}{3}f_{[ABC}f_{D]} \ . \nonumber
\end{align}
The reverse implication is not true: the quadratic constraints are weaker than the strong constraint, which in the context of generalized Scherk-Schwarz compactifications is know to be unnecessary. We will not discuss this issue here, see \cite{Aldazabal:2011nj},\cite{Grana:2012rr} for discussions on this point.

The gauge transformations of the fields in the parent action (\ref{from})-(\ref{to}) can also be {\it twisted} in this way to  extract their analogs in the effective theory. From now on, all the quantities we display correspond to those in the effective gauged supergravity unless the opposite is explicitly mentioned
\begin{align}
\delta g_{\mu\nu} & = {\cal L}_\Lambda g_{\mu \nu} \nonumber \\
\delta {\cal M}^{AB} & = {\cal L}_\Lambda {\cal M}^{A B}\nonumber \\
\delta e^{-2\phi} & = {\cal L}_\Lambda e^{-2 \phi} \label{transformation}\\
\delta A_{\mu}{}^A & = \partial_{\mu} {\Lambda}^{A} +  {\cal L}_\Lambda A_{\mu}{}^A + f^{A} \,{\Xi}_{\mu}\nonumber \\
\delta B_{\mu\nu} & = 2\partial_{[\mu} {\Xi}_{\nu]}+ {\cal L}_\Lambda B_{\mu \nu} + \partial_{[\mu}{\Lambda}^{A}{A}_{\nu]A} - f_{A}\, A_{[\mu}{}^ A{\Xi}_{\nu]} \ , \nn
\end{align}
where all fields have vanishing weight except for $\lambda (e^{-2 \phi}) = 1$, and the external weights are listed in (\ref{externalweights}).

Let us point out that similarly to DFT (\ref{TransfHierarchyDFT}), the gauge transformations can be taken to a covariant form
\begin{align}\label{eq:KKDFTtransformations}
{\delta} {A}_{\mu}{}^A & = {\mathcal{D}}_{\mu} {\Lambda}^{A} + f^{A} \widehat \Xi_\mu \\
{\delta} {B}_{\mu\nu} & = 2  {\mathcal{D}}_{[\mu}\widehat \Xi_{\nu]} - {\mathcal{F}}_{\mu\nu}{}^A \Lambda_A + {A}_{[\mu}{}^A {\delta}{A}_{\nu]A} \ ,
\end{align}
 by redefining the one-form parameter as in (\ref{redefDFT})
\be
\widehat \Xi_\mu = \Xi_\mu + A_\mu{}^A {\Lambda}_{A} \ .
\ee
The brackets for the parameters with a hat are expected to be field dependent, while those without a hat depend on parameters only. The non-covariant ones are then more convenient to deal with when analyzing the gauge algebra, which will be discussed later.

The covariant derivatives (\ref{CovDer}) can also be twisted, leading to their analogs in gauged supergravity, namely
\be
{\cal D}_\mu  = \partial_\mu - {\cal L}_{A_\mu} \ ,
\ee
such that for a covariant tensor transforming as in  (\ref{gaugetransf}) one has
\be
{\cal L}_\Lambda {\cal D}_\mu T^A = - X_{B C}{}^A \, \Lambda^B \, {\cal D}_\mu T^C + \left(\lambda + \omega - \frac 1 2 \right)\, f_B \Lambda^B\, {\cal D}_\mu T^A \ .
\ee

For the field strengths (\ref{FieldstrengthsDFT}) we have the same story
\bea
{\cal F}_{\mu \nu}{}^A &=& 2 \partial_{[\mu} A_{\nu]}{}^A - f_{B C}{}^A\, A_\mu{}^B A_\nu{}^C - f_B\, A_{[\mu}{}^B A_{\nu]}{}^A - f^A\, B_{\mu \nu}\\
{\cal H}_{\mu \nu \rho} &=& 3 \left({\cal D}_{[\mu} B_{\nu \rho]} + A_{[\mu}{}^A \partial_\nu A_{\rho]A} - \frac 1 3 f_{ABC} A_\mu{}^A A_\nu{}^B A_\rho{}^C \right)\ .
\eea
and they satisfy the following Bianchi identities
\bea
3 {\cal D}_{[\mu} {\cal F}_{\nu \rho]}{}^A  + f^A\, {\cal H}_{\mu\nu\rho} &=& 0 \\
{\cal D}_{[\mu} {\cal H}_{\nu \rho \sigma]} - \frac 3 4 {\cal F}_{[\mu \nu}{}^A {\cal F}_{\rho \sigma]A} &=& 0 \ .
\eea

Then, we automatically see that the compactified effective action takes exactly the same form as the parent one (\ref{DFT-KK lagrangian}), where the covariant derivatives and the field strengths must be replaced by their lower dimensional gauged versions. The only subtlety is a local overall shift $\int dX \, e^{-2c - \gamma}$ which integrates in the internal space to modify the effective action's Planck constant.  There is also the non-trivial task of compactifying  the scalar potential, which gives
\be
V  = \frac{1}{12} {\mathcal{M}}^{AB} {\mathcal{M}}^{CD} {\mathcal{M}}_{EF}f_{AC}{}^E f_{BD}{}^ F + \frac{1}{4} {\mathcal{M}}^{AB} \left(f_{AC}{}^Df_{BD}{}^C + 3f_{A}f_{B}\right) + \frac{1}{6} f_{ABC}f^{ABC} \ .\label{potential}
\ee
The last term can be seen to vanish due to the strong constraint \cite{Aldazabal:2011nj},\cite{Grana:2012rr}, so we put it in by hand. When the term is non-vanishing the theory cannot be uplifted to maximal supergravity \cite{Max}.

The gauge invariance of the effective action is achieved due to the fact that the measure transforms like
\begin{align}
\delta\left(\sqrt{- {g}}e^{-2 {\phi}}\right) & = {\Lambda}^{A} f_{A} \sqrt{- {g}}e^{-2 {\phi}} \ ,\label{deltamessure}
\end{align}
which is compensated by the variation of the Lagrangian
\begin{align}
\delta {\cal L} = - \Lambda^A f_A {\cal L} \ .
\end{align}

The brackets with respect to which the gauge transformations close can be either obtained by direct computation, or by twisting those in (\ref{Brackets})
\bea
\Lambda^M_{12}(x,X) &=& e^{\frac {\gamma(X)} 2} U^M{}_A(X) \, \Lambda_{12}^A(x)\\
\Xi_{\mu 12}(x,X) &=& e^{\gamma(X)}\, \Xi_{\mu 12}(x)\ ,
\eea
giving rise to
\bea
\Lambda_{12}^A &=& f_{B C}{}^A \Lambda_1^B \Lambda_2^C + f_B\, \Lambda^B_{[1} \Lambda^A_{2]}\\
\Xi_{\mu 12} &=& 2 f_A\, \Lambda^A_{[1} \Xi_{2]\mu} + \Lambda_{[1}^A \partial_\mu \Lambda_{2]A} \ ,
\eea
which  satisfy
\be
\left[\delta_1,\, \delta_2\right]  = - \delta_{12} \ ,
\ee
for the compactified gauge transformations (\ref{transformation}).

As before, merging the two parameters into a single one and defining the bracket notation
\be
\z = (\Lambda^A,\, \Xi_\mu) \ , \ \ \ [\z_1,\, \z_2]^A = \Lambda^A_{12}\ , \ \ \ [\z_1,\, \z_2]_\mu = \Xi_{\mu 12}\ ,
\ee
one can readily compute the Jacobiator
\be
J = 3 \left[\left[\z_{[1},\, \z_{2}\right],\, \z_{3]}\right] = (f^A\, N , \, - \partial_\mu N)\ , \label{effectiveJ}
\ee
with the Nijenhuis scalar defined by
\be
N(\Lambda_1, \Lambda_2, \Lambda_3) = \frac 1 2\, f_{A B C}\, \Lambda_1^A \Lambda_2^B \Lambda_3^C \ . \label{effectiveN}
\ee
The same result is obtained by compactifying the parent quantities (\ref{Jacobiators})-(\ref{Nijenhuis}) with external weights $\omega(J^A) = \frac 1 2$, $\omega(J_\mu) = 1$, $\omega(N) = 1$.

We now have a clear indication that the electric sector of half-maximal supergravity admits trivial parameters
\be
\Lambda^A_{trivial} = f^A \, \chi \ , \ \ \ \ \ \Xi_{\mu, trivial} = - \partial_\mu \chi\ ,
\ee
which indeed can be verified by direct inspections using the quadratic constraints (\ref{cuadraticconstraint}). Moreover, it is easy to see that the commutator of a trivial parameter with a generic one, yields a new trivial parameters characterized by the function
\be
\chi' = - \frac 1 2 f_A \Lambda^A \, \chi \ .
\ee

We then see that the electric sector of half-maximal gauged supergravity inherits from its parent theory the sufficient conditions for its gauge algebra to fit into $L_{\infty}^{\rm gauge + fields}$.

\subsection{The tensor hierarchy in gauged supergravity} \label{sec:3-form TH}

The gauge structure of the tensor hierarchy in gauged supergravities has the nice advantage of being writable in a universal form, regardless of the dimension and the amount of supersymmetry. Following \cite{Weidner:2006rp}, and motivated by the fact that our discussion so far reached the three-form in this paper only, we will project the hierarchy to this level here, and discuss the full hierarchy relevant for $n = 4$ space-time dimensions in the Appendix. The conclusion will be that gauged supergravities admit symmetries for symmetries whose trivial parameters are deformed by the embedding tensor. We will afterwards show in the next subsection that the gauge algebra of the tensor hierarchy meets the criteria of Subsections \ref{sec::theorem} and \ref{sec::theorem2}  implying that it has an $L_\infty^{\rm gauge + fields}$ structure.

Starting from a generic ungauged supergravity with global symmetry
group $G_{0}$, we can turn some subgroup $G \in G_0$  into local using
the embedding tensor formalism, which maintains at least formally the $G_{0}$ covariance
by treating the embedding tensor as a spurionic object. In this procedure, the generators
of the algebra of the local subgroup $X_{M}$ are parameterized as
a projection of the $\mathfrak{g}_{0}$ generators, $t_{\alpha}$
\begin{align}
X_{M} & \equiv\Theta_{M}{}^{\alpha}t_{\alpha}\ , \label{eq:Xm}
\end{align}
 with the embedding tensor $\Theta_{M}{}^{\alpha}$ a mixed index
tensor with $M=1,...,dim(\mathfrak{g})$ in some $\overline{V}$ representation,
and $\alpha$ in the adjoint of $G_{0}$. This embedding tensor has
to satisfy linear and quadratic constraints. The former is a supersymmetric requirement which can be
written as a projection that selects some particular irreducible
representations of the product $\overline{V}\otimes\mathfrak{g}_{0}$
to which the embedding tensor belongs. The latter is needed
to ensure the closure of the gauge algebra and can be obtained by demanding
gauge invariance of $\left(X_{M}\right)_{N}{}^O=\Theta_{M}{}^{\alpha}\left(t_{\alpha}\right)_{N}{}^O$,
which leads to
\begin{align}
\left[X_{M},X_{N}\right] & =-X_{MN}{}^OX_{O}\ .\label{eq:cuadratic_constraint}
\end{align}

The gauge transformations act on vectors as follows
\begin{align}
\delta_{ \Lambda }T^{M}= & - X_{NP}{}^M\, \Lambda^{N}  T^{P} \label{vectors}\\
\delta_{ \Lambda }T_{M}= & X_{NM}{}^P\, \Lambda^{N}  T_{P}\ , \nn
\end{align}
with respect to local gauge parameters $\Lambda^{M}(x)$. In
order to preserve gauge invariance it is necessary to replace derivatives
$\partial_{\mu}$ by covariant derivatives $\mathcal{D}_{\mu}$ through the minimal
coupling procedure
\begin{align}
\mathcal{D}_{\mu} & =\partial_{\mu}-A_{\mu} {}^MX_{M}\ ,\label{eq:D_general}
\end{align}
 where $A_{\mu} {}^M $ are the gauge vectors. The derivative is covariant provided the vectors transform as follows
\be
\delta A_\mu{}^M = {\cal D}_\mu \Lambda^M + \dots\ ,
\ee
where the dots represent terms that vanish when contracted with the embedding tensor. The story continues by introducing a field strength
\be
\left[{\cal D}_\mu,\, {\cal D}_\nu\right] = - {\cal F}_{\mu \nu}{}^M X_M = - \left(2 \partial_{[\mu} A_{\nu]}{}^M + X_{N P}{}^M A_{[\mu}{}^N A_{\nu]}{}^P + \dots\right) X_M\ ,
\ee
which again is defined up to a projection with the embedding tensor. It turns out that the dots are fixed by demanding covariance of the embedding tensor, namely that it transforms as a vector (\ref{vectors}), for which one must introduce new degrees of freedom $B_{\mu \nu I}$
\be
{\cal F}_{\mu \nu}{}^M = 2 \partial_{[\mu} A_{\nu]}{}^M + X_{N P}{}^M A_{[\mu}{}^N A_{\nu]}{}^P + Z^{M I} B_{\mu \nu I} \ , \label{fieldstrengthGS}
\ee
where $Z^{M I}$, the so-called  intertwining tensor, is constrained to vanish when projected with $X_M$
\be
Z^{M I}\, X_M = 0 \ .
\ee
This object can then be used to fill in the dots in the variation of the vector fields
\be
\delta A_\mu{}^M = {\cal D}_\mu \Lambda^M - Z^{M I} \widehat \Xi_{\mu I}\ . \label{ProjectedA}
\ee
The gauge transformation of the two form is then forced to compensate the failure of the first two terms in (\ref{fieldstrengthGS}) to transform covariantly, but then of course  is defined only up to terms that vanish when projected in this case with the intertwining tensor. So, covariance of ${\cal F}_{\mu \nu}{}^M$ only teaches how to transform a projection of the two-form with respect to a projection of its one-form parameter
\be
\widetilde B_{\mu \nu}{}^M = Z^{M I} B_{\mu \nu I}\ , \ \ \ \widetilde {\widehat \Xi}_{\mu}{}^M = Z^{M I} \widehat \Xi_{\mu I}\ ,
\ee
giving
\bea
\delta A_\mu{}^M &=& {\cal D}_\mu \Lambda^M - \widetilde {\widehat \Xi}_{\mu}{}^M \label{projA}\\
\delta \widetilde B_{\mu \nu}{}^M &=& 2 {\cal D}_{[\mu} \widetilde {\widehat \Xi}_{\nu]}{}^M - 2 X_{(N P)}{}^M \left(\Lambda^N \, {\cal F}_{\mu \nu}{}^P - A_{[\mu}{}^N \delta A_{\nu]}{}^P\right)\ . \label{projB}
\eea
Here we have used the same notation as in the rest of the paper, namely that tensors with a tilde are projected by intertwiners, and tensors with a hat allow to write the gauge transformations in gauge covariant form and involve brackets that are field dependent. In fact a quick computation shows that this {\it projected} tensor hierarchy is self consistent, as it closes with respect to the brackets
\bea
\Lambda_{12}{}^M &=& - X_{N P}{}^M \Lambda_{[1}^N \Lambda_{2]}^P \\
\widetilde {\widehat \Xi}_{12\mu}{}^M &=& 2 X_{(N P)}{}^M {\cal D}_{\mu} \Lambda_{[1}^N \Lambda_{2]}^P \ .
\eea
By redefining the parameters
\be
\widetilde {\widehat \Xi}_{\mu}{}^M = \widetilde \Xi_\mu{}^M + 2 X_{(N P)}{}^M A_\mu{}^N \Lambda^P \ ,
\ee
the one-form bracket becomes field independent
\be
\widetilde \Xi_{12\mu}{}^M = 2 X_{(N P)}{}^M \left(\Lambda_{[1}^N\partial_\mu \Lambda_{2]}^P - 2 Z^{N I} \Lambda_{[1}^P\, \Xi_{2]\mu I}\right) \ .
\ee

We conclude this brief discussion by showing the projected three-form field strength
\be
\widetilde {\cal H}_{\mu \nu \rho}{}^M = 3 {\cal D}_{[\mu} \widetilde B_{\nu \rho]}{}^M + 6 X_{(N P)}{}^M A_{[\mu}{}^N \left(\partial_\nu A_{\rho]}{}^P + \frac 1 3 X_{O Q}{}^P A_\nu{}^O A_{\rho]}{}^Q\right) \ ,
\ee
and the projected Bianchi identities
\be
3 {\cal D}_{[\mu} {\cal F}_{\nu \rho]}{}^M - \widetilde {\cal H}_{\mu \nu \rho}{}^M  = 0\ , \ \ \ {\cal D}_{[\mu} \widetilde {\cal H}_{\nu \rho \sigma]}{}^M - \frac 3 2 X_{(N P)}{}^M {\cal F}_{[\mu \nu}{}^N {\cal F}_{\rho \sigma]}{}^P = 0\ .
\ee

Let us now move one step upwards in the hierarchy, by un-projecting the two-form. This requires including a three-form which will now be projected. One could keep going until the space-time dimension puts an upper limit to the hierarchy, but we will stop here and discuss the general case in the Appendix. Besides $A_{\mu}{}^M$ and $B_{\mu\nu I}$ we now introduce a three-form $C_{\mu\nu\rho}{}^{A}$, which in this case is required by demanding covariance of the un-projected three-form field strength ${\cal H}_{\mu \nu \rho I}$. For the sake of briefly we will not pursue the whole discussion, but simply state the results and refer to the details in \cite{Weidner:2006rp}. The gauge transformations are given by
\begin{align}
\delta A_{\mu}{}^M & =\mathcal{D}_{\mu}{\Lambda}^{M}-Z^{MI}\,\widehat{\Xi}_{\mu I}\label{eq:A_transformation}\\
\delta B_{\mu\nu I} & =2\mathcal{D}_{[\mu}\widehat{\Xi}_{\nu]I}-2d_{IMN}{\Lambda}^{M}\mathcal{F}_{\mu\nu}{}^N+2d_{IMN}A_{[\mu}{}^ M\delta A_{\nu]}{}^N -Y_{IA}\,\widehat{\Sigma}_{\mu\nu}{}^A\label{eq:B_transformation}\\
\delta \left(Y_{I A}\, C_{\mu\nu\rho}{}^A\right) & =Y_{I A} \left( 3\mathcal{D}_{[\mu}\widehat{\Sigma}_{\nu\rho]}{}^A+3g_{M} {}^{AI}\mathcal{F}_{[\mu\nu} {}^{M}\widehat{\Xi}_{\rho]I}+g_{M} {}^{AI} {\Lambda}^{M}\mathcal{H}_{\mu\nu\rho I} \right. \nn\\
&\left.\ \ \ \ \ \ \ \ \ \ \ +3g_{M} {}^{AI}B_{I[\mu\nu}\delta A_{\rho]} {}^{M}+2g_{M} {}^{AI}d_{INP}A_{[\mu} {}^{M}A_{\nu} {}^{N}\delta A_{\rho]} {}^{P} \right) \ ,\label{eq:C_transformation}
\end{align}
where we have introduced the field strength tensors
\begin{align}
\mathcal{F}_{\mu\nu} {}^{M} & =2\partial_{[\mu}A_{\nu]} {}^{M}+X_{NO}{}^MA_{[\mu} {}^{N}A_{\nu]} {}^{O}+Z^{MI}B_{\mu\nu I}\label{eq:1form_Stensor}\\
\mathcal{H}_{\mu\nu\rho I} & =3\mathcal{D}_{[\mu}B_{\nu\rho]I}+6d_{IMN}A_{[\mu} {}^{M}\left(\partial_{\nu}A_{\rho]} {}^{N}+\frac{1}{3}X_{OP}{}^NA_{\nu} {}^{O}A_{\rho]} {}^{P}\right)+Y_{IA}C_{\mu\nu\rho} {}^{A}\ , \label{eq:2form_Stensor}
\end{align}
with the covariant derivative
\begin{align}
\mathcal{D}_{\mu}B_{\nu\rho I} & \equiv\partial_{\mu}B_{\nu\rho I}-A_{\mu} {}^MX_{MI}{}^JB_{\nu\rho J} \ .
\end{align}
The field strengths satisfy the generalized Bianchi identities
\begin{align}
3 \mathcal{D}_{\left[ \mu \right.}\mathcal{F}_{\left. \nu \rho \right]} {}^{M} - Z^{MI} \mathcal{H}_{\mu\nu\rho I} & = 0\\
4 \mathcal{D}_{\left[ \mu \right.}\mathcal{H}_{\left. \nu \rho \sigma \right] I} - 6 d_{IMN} \mathcal{F}_{\left[ \mu \nu \right.} {}^{M} \mathcal{F}_{\left. \rho \sigma \right]} {}^N - Y_{IA} \mathcal{G}_{\mu \nu \rho \sigma} {}^{A} &= 0\ ,
\end{align}
with $ \mathcal{G}_{\mu \nu \rho \lambda} {}^{A} $ the 3-form strength tensor, defined by this equation up to a projection with $Y_{I A}$.

Let us explain the notation.  The tensors $d_{I M N}$ and $g_M{}^{A I}$ are gauging independent and represent the $G_0$-invariants of the theory. Just to put an example, in $n=4$  maximal supergravity, where the gauge group is $E_{7(7)}$, these would the related to $\Omega_{M N}$ and $(t_\alpha)^{M N}$. The tensors $Z^{M I}$ and $Y_{I A}$ on the other hand do depend on the gaugings, and so together with $X_M$ are responsible for its deformation. The intertwining tensor $Z^{M I}$ for example is related to the symmetric part of the embedding tensor $X_{(M N)}{}^P$. The fields are $A_{\mu}{}^M$, $B_{\mu \nu I}$ and $C_{\mu \nu \rho}{}^A$, their associated gauge parameters are $\Lambda^M$, $\widehat \Xi_{\mu I}$ and $\widehat \Sigma_{\mu \nu}{}^A$, and their field strengths are ${\cal F}_{\mu \nu}{}^M$, ${\cal H}_{\mu \nu \rho I}$ and ${\cal G}_{\mu \nu \rho \sigma}{}^A$.  The indices $M$, $I$, $A$,  denote the representations to which the tensors belong. Then, $X_{M I}{}^J$ corresponds to the embedding tensor in the representation of the two-form field, $X_{M A}{}^B$ of the three-form field, etc.

Gauge covariance is achieved provided the following constraints are imposed
\begin{align}
d_{I[MN]} & =0\nonumber \\
d_{I(MN}g_{O)} {}^{AI} & =0\nonumber \\
Z^{MI}d_{INO} & =X_{(NO)}{}^M\nonumber \\
X_{MI}{}^J+2Z^{NJ}d_{IMN} & =Y_{IA}g_{M} {}^{AJ}\nonumber \\
Z^{MI}Y_{IA} & =0\nonumber \\
Z^{MI}X_{M} & =0\nonumber \\
Y_{IA}g_{M} {}^{AJ}Z^{MK} & =2Z^{MK}Z^{NJ}d_{IMN}\ , \label{eq:constraints}
\end{align}
where the last two follow from the others and the quadratic constraints.

The vector fields $A_\mu{}^M$ in (\ref{eq:A_transformation}) transform exactly as in (\ref{projA}). Instead, the transformation of the two-form in (\ref{projB}) corresponds to a projection of (\ref{eq:B_transformation}) with the intertwining tensor $Z^{M I}$. When removing the projection in (\ref{eq:B_transformation}) one has to include terms that vanish when projected, which in this case are represented by $Y_{I A} \widehat \Sigma_{\mu \nu}{}^A$. The bracket of this parameter is responsible of absorbing the failure of closure of the unprojected gauge transformation, and its associated field $Y_{I A} C_{\mu \nu \rho}{}^A$ is responsible for guaranteing the covariance of ${\cal H}_{\mu \nu \rho I}$, namely to enforce
\be
\delta {\cal H}_{\mu\nu\rho I} = X_{M I}{}^J \Lambda^M  {\cal H}_{\mu\nu\rho J} \ .
\ee
In fact, if the three-form fields $C_{\mu \nu \rho}{}^A$ were absent in (\ref{eq:2form_Stensor}), then ${\cal H}_{\mu \nu \rho I}$ would fail to transform covariantly and the failure would be proportional to $X_{M I}{}^J  + 2 Z^{N J} d_{I M N}$, which in turn is proportional to $Y_{I A}$, as can be seen in (\ref{eq:constraints}). To follow the notation we have been using, we could put a tilde on the three-form and its gauge parameter to denote that both are projected
\be
\widetilde C_{\mu \nu \rho I} = Y_{I A} C_{\mu \nu \rho}{}^A \ , \ \ \ \ \widetilde {\widehat \Sigma}_{\mu \nu  I} = Y_{I A} \widehat \Sigma_{\mu \nu}{}^A \ ,
\ee
and we have to remember that now the tensors with tilde satisfy
\be Z^{M I} \widetilde C_{\mu \nu \rho I} = 0 \ , \ \ \ \ Z^{M I} \widetilde {\widehat \Sigma}_{\mu \nu  I} = 0 \ .
\ee

If we decided to move a step forward and removed the projection with $Y_{I A}$, then the  strength tensor $ \mathcal{G}_{\mu \nu \rho \lambda} {}^{A} $ would  not be covariant, and its failure would be proportional to $X_{MB}{}^A+g_{M} {}^{AJ}Y_{JB}$, forcing the inclusion of a four-form and
 so on. The same mechanism is repeated over and over. Then, if we want to analyze
the tensor hierarchy up to the three-form only, a truncation
scheme must be considered, which is possible given that
\begin{align}
Y_{IA}\left(X_{MB}{}^A+g_{M} {}^{AJ}Y_{JB}\right) & =0\ ,
\end{align}
vanishes thanks to the $G$-invariance of $Y_{IA}$ and \eqref{eq:constraints}.
We then cut the p-form chain at $p=3$ and only consider $\left\{ A_{\mu} {}^{M},B_{\mu\nu I}, \widetilde C_{\mu\nu\rho I}\right\} $, their parameters $\left\{\Lambda^M, \widehat {\Xi}_{\mu I} , \widetilde {\widehat \Sigma}_{\mu \nu I} \right\}$ and their field strengths $\left\{{\cal F}_{\mu \nu}{}^M, {\cal H}_{\mu \nu\rho I}, \widetilde{\cal G}_{\mu \nu\rho \sigma I} \right\}$.

To summarize, the gauge transformations of the projected tensor hierarchy up to the three-form are
\begin{align}
\delta A_{\mu}{}^M & =\mathcal{D}_{\mu}{\Lambda}^{M}-Z^{MI}\,\widehat{\Xi}_{\mu I}\\
\delta B_{\mu\nu I} & =2\mathcal{D}_{[\mu}\widehat{\Xi}_{\nu]I}-2d_{IMN}{\Lambda}^{M}\mathcal{F}_{\mu\nu}{}^N+2d_{IMN}A_{[\mu}{}^ M\delta A_{\nu]}{}^N -\widetilde {\widehat{\Sigma}}_{\mu\nu I} \\
\delta \widetilde  C_{\mu\nu\rho I} & = 3\,\mathcal{D}_{[\mu}\widetilde{\widehat \Sigma}_{\nu\rho] I} + Y_{I A}\, g_M{}^{A J} \left(3 \,\mathcal{F}_{[\mu\nu} {}^{M}\widehat{\Xi}_{\rho]J} + {\Lambda}^{M}\mathcal{H}_{\mu\nu\rho J} \right. \\
& \left. \ \ \ \ \ \ \ \ \ \ \ \ \ \ \ \ \ \ \ \ \ \ \ \ \ \ \ \ \ \ \ \ \ \ + \,3  \,B_{[\mu\nu J}\delta A_{\rho]} {}^{M}+2\, d_{JNP}A_{[\mu} {}^{M}A_{\nu} {}^{N}\delta A_{\rho]} {}^{P} \right) \ , \nn
\end{align}
the field strengths are defined as
\begin{align}
\mathcal{F}_{\mu\nu} {}^{M} & =2\partial_{[\mu}A_{\nu]} {}^{M}+X_{NO}{}^MA_{[\mu} {}^{N}A_{\nu]} {}^{O}+Z^{MI}B_{\mu\nu I}\nn \\
\mathcal{H}_{\mu\nu\rho I} & =3\mathcal{D}_{[\mu}B_{\nu\rho]I}+6d_{IMN}A_{[\mu} {}^{M}\left(\partial_{\nu}A_{\rho]} {}^{N}+\frac{1}{3}X_{OP}{}^NA_{\nu} {}^{O}A_{\rho]} {}^{P}\right)+ \widetilde C_{\mu\nu\rho I} \\
\widetilde {\cal G}_{\mu \nu \rho\sigma I} & = 4\mathcal{D}_{[\mu}\widetilde{C}_{\nu\rho\sigma]I}-Y_{IA}\, g_{M}{}^{AJ}\left(6B_{[\mu\nu J}\mathcal{F}_{\rho\sigma]}{}^M-3Z^{MK}B_{[\mu\nu J}B_{\rho\sigma]K}\right. \nn \\& \left.\ \ \ \ \ \ \ \ \ \ \ \ \ \ \ \ \ \ \ +8d_{JNP}A_{[\mu}{}^MA_{\nu}{}^N\partial_{\rho}A_{\sigma]}{}^P+2d_{JPN}X_{QR}{}^PA_{[\mu}{}^MA_{\nu}{}^NA_{\rho}{}^QA_{\sigma]}{}^R\right) \ , \nn
\end{align}
and satisfy the following Bianchi identities
\begin{align}
3 \,\mathcal{D}_{[ \mu}\mathcal{F}_{\nu \rho]}{}^{M} -  Z^{MI} \mathcal{H}_{\mu\nu\rho I} &= 0\nn\\
4 \,\mathcal{D}_{[ \mu}\mathcal{H}_{\nu \rho \sigma] I} - 6 d_{IMN} \mathcal{F}_{[\mu\nu}{}^M \mathcal{F}_{\rho \sigma]}{}^N
- \widetilde {\cal G}_{\mu \nu \rho \sigma I} & = 0 \\
5\, \mathcal{D}_{[\mu}\widetilde{\mathcal{G}}_{\nu\rho\sigma\lambda]I}+10Y_{IA}g_{M}{}^{AJ}\mathcal{F}_{[\mu\nu}{}^M\mathcal{H}_{\rho\sigma\lambda]J} & =0 \ .\nn
\end{align}

Moving to the closure of the gauge algebra, we can now find the unprojected one-form bracket $\widehat \Xi_{12 \mu I}$, and the projected two-form bracket $\widetilde {\widehat \Sigma}_{12 \mu \nu I}$
\begin{align}
\Lambda_{12}^M & = -X_{NP}{}^M \Lambda_{[1}^{N} \Lambda_{2]}^{P}\nonumber \\
\widehat{\Xi}_{12\mu I} & = 2\,d_{IMN}\, \mathcal{D}_{\mu} \Lambda_{[1}^{M} \Lambda_{2]}^{N}\nonumber \\
\widetilde {\widehat{\Sigma}}_{12 \mu \nu I} & = - 2\, Y_{I A}\,g_{M} {}^{AJ}\left(d_{JNP}\, \Lambda_{[1}^{M} \Lambda_{2]}^{N}\,\mathcal{F}_{\mu\nu} {}^{P}+2\, \widehat{\Xi}_{[1 [\mu J}\, \mathcal{D}_{\nu]}\Lambda_{2]}^{M}-Z^{MK}\, \widehat{\Xi}_{[1[\mu J}\, \widehat{\Xi}_{2]\nu]K}\right)\ .\label{eq:field_dependent_brackets}
\end{align}

As expected, these brackets are those of the set of parameters for which the gauge transformations are written covariantly, and so are field dependent. We show in the following subsection that via redefinitions one can reach the hypothesis of the theorem presented in Subsections \ref{sec::theorem} and \ref{sec::theorem2}, implying that the gauge algebra of
 the tensor hierarchy in gauged supergravities fits into an $L_\infty^{\rm gauge + fields}$ algebra.

 We have projected the hierarchy to the three-form field to match these results with those of the EFT section, but it can of course be continued more generally to higher forms. In the Appendix we show how to extend the hierarchy exhaustively for the case we are interested in here: $n = 4$ space-time dimensions.

\subsection{$L_\infty$ algebra and gauged supergravity}

In order to obtain field
independent brackets we propose the following gauge parameter redefinitions
\begin{align}
\widehat{\Xi}_{\mu I} & = \Xi_{\mu I}+2d_{IMN}A_{\mu} {}^{M}\Lambda^{N}\nonumber \\
\widetilde {\widehat{\Sigma}}_{\mu\nu I} & = \widetilde \Sigma_{\mu\nu I} - \, Y_{I A}\, g_{M} {}^{AJ}\left(2A_{[\mu} {}^{M}\Xi_{\nu]J}+B_{\mu\nu J}\Lambda^{M}+2A_{[\mu} {}^{M} A_{\nu]} {}^{N} d_{JNP}\Lambda^{P}\right)\ ,\label{eq:parameter_redefinition}
\end{align}
now yielding the brackets
\begin{align}
\Lambda_{12}^M & = -X_{NP}{}^M\Lambda_{[1}^{N}\Lambda_{2]}^{P}\nonumber \\
\Xi_{12\mu I} & = 2\,d_{IMN}\left(\Lambda_{[1}^{M}\partial_{\mu}\Lambda_{2]}^{N}+\,2\,Z^{MJ}\,\Xi_{[1\mu J}\Lambda_{2]}^{N}\right)\nonumber \\
\widetilde \Sigma_{12 \mu\nu I} & = -2\, Y_{I A}\, g_{M} {}^{AJ}\left(2\Lambda_{[1}^{M}\, \partial_{[\mu}\Xi_{2]\nu]J}+Z^{MK}\,\Xi_{[1[\mu J}\,\Xi_{2]\nu]K}+ \widetilde \Sigma_{[1 \mu\nu J}\Lambda_{2]}^{M}\right) \ .\label{eq:field_independent_brackets}
\end{align}

Defining a composed parameter with its corresponding bracket
\be\label{eq:composed_parameter}
\z  =\left( \Lambda^{M},\,  \Xi_{\mu I},\, \widetilde \Sigma_{\mu\nu I} \right) \ , \ \ \ \left[\z_1,\, \z_2\right]  =\left( \Lambda_{12}^{M},\,  \Xi_{12\mu I},\, \widetilde \Sigma_{12\mu\nu I} \right) \ ,
\ee
we construct the Jacobiator
\be
J(\zeta_{1},\zeta_{2},\zeta_{3}) = 3 \left[\left[\zeta_{[1},\zeta_{2}\right],\zeta_{3}\right]\ ,
\ee
with components
\begin{align}
J^{M} & =Z^{MI} N_{I}\nonumber \\
J_{\mu I} & =\partial_{\mu} N_{I}+ \widetilde N_{\mu I} \label{eq:Jacobiators} \\
\widetilde J_{\mu\nu I} & = 2\,\partial_{[\mu} \widetilde N_{\nu] I} \ , \nn
\end{align}
where we defined the Nijenhuis tensors
\begin{align}
N_{I} & = d_{IMN}X_{OP}{}^M\Lambda_{[1} {}^{O}\Lambda_{2} {}^{P}\Lambda_{3]} {}^{N}\\
\widetilde N_{\mu I} & = 2\, Y_{I A} \, g_{M}{}^{AJ}\, d_{JNP}\, \Lambda_{[1}^{M}\left(2\Lambda_{2}^{N}\partial_{\mu}\Lambda_{3]}^{P}+3Z^{NK}\Xi_{2\mu K}\Lambda_{3]}^P\right) \ .
\end{align}

We then have a strong indication that the following are trivial parameters
\begin{align}
\Lambda^{M}_{trivial} & =Z^{MI} \chi_{I}\nonumber \\
\Xi_{\mu I\, trivial} & =\partial_{\mu} \chi_{I}+ \widetilde \chi_{\mu I} \label{TrivGS}\\
\widetilde \Sigma_{\mu\nu I\, trivial} & = 2\,\partial_{[\mu} \widetilde \chi_{\nu] I} \ , \nn
\end{align}
where now $\chi_I$ is arbitrary and $\widetilde \chi_{\mu I}$ is constrained to satisfy $Z^{M I} \widetilde \chi_{\mu I} = 0$. It can be checked that this is indeed the case.  For the other set of parameters one has instead
\begin{align}
\Lambda^{M}_{trivial} & =Z^{MI} \chi_{I}\nonumber \\
\widehat \Xi_{\mu I\, trivial} & ={\cal D}_{\mu} \chi_{I}+ \widetilde {\widehat \chi}_{\mu I} \\
\widetilde {\widehat\Sigma}_{\mu\nu I\, trivial} & = 2\,{\cal D}_{[\mu} \widetilde {\widehat \chi}_{\nu] I} - Y_{I A}\, g_M{}^{A J}\, {\cal F}_{\mu \nu}{}^M \, \chi_J \ , \nn
\end{align}
for a redefined but still constrained $\widetilde {\widehat \chi}_{\mu I}$.

We conclude this section by pointing out that gauged supergravities share with DFT and EFT the interesting feature that the bracket between a generic parameter and a trivial one, gives another trivial one
\begin{align*}
\left[\zeta_{trivial},\zeta\right] & =\zeta'_{trivial}\ ,
\end{align*}
where $\z_{trivial}$ is defined in (\ref{TrivGS}) and
\be
\z'_{trivial} = \left(Z^{MI} \chi'_{I},\, \partial_{\mu} \chi'_{I}+ {\widetilde \chi}'_{\mu I}, \,  2\,\partial_{[\mu} \widetilde \chi'_{\nu] I}\right) \ ,
\ee
with
\begin{align}
\chi'_{I} & =Z^{MJ}d_{IMN}\Lambda^{N}\chi_{J}\\
{\widetilde \chi}'_{\mu I} & = - Y_{IA}\, g_{M}{}^{AJ}\, Z^{MK}\, \Xi_{\mu J}\, \chi_{K} \ .
\end{align}
Once again we see that there is a set of parameters that meet the hypothesis of the theorem in Subsections \ref{sec::theorem} and \ref{sec::theorem2}, implying that the tensor hierarchy in gauged supergravities lie within an $L_\infty^{\rm gauge + fields}$ structure.

The concrete graded spaces and products are obtained following the procedure described in previous sections. For completion we give here some products in the $L_\infty^{\rm gauge}$ sector, which are trivial to read from the results above. The only graded subspaces in this case are
\bea
X_2 \ \ &:& \ \ \eta = (\eta_I , \ \widetilde \eta_{\mu I}) = \left(Y_{I A}  C^A \, , \ - Y_{I A} \partial_\mu C^A\right) \ \ \ \forall C^A \ , \\
X_1 \ \ &:& \ \ \chi=(\chi_I,\ \widetilde \chi_{\mu I})\ , \\
X_0 \ \ &:& \ \ \zeta=(\Lambda^M, \ \Xi_{\mu I} ,\ \widetilde \Sigma_{\mu \nu I})\ .
\eea
The product $\ell_1$ maps $X_2 \to X_1$ through inclusion. Some explicit products are
\bea
\ell_1(\eta) &=& \eta_I + \widetilde \eta_{\mu I} \ , \\
\ell_1(\chi) &=& \Lambda^M_{trivial} + \Xi_{\mu I\, trivial} + \widetilde \Sigma_{\mu \nu I\, trivial}\ ,\\
\ell_2(\zeta_1, \, \zeta_2) &=& \Lambda_{12}^M + \Xi_{12\mu I} + \widetilde \Sigma_{12 \mu \nu I} \ ,\\
\ell_2 (\zeta,\, \chi) &=& - \chi'_I - \widetilde \chi'_{\mu I} + {\cal K}_I + \widetilde {\cal K}_{\mu I}\ , \\
\ell_3 (\zeta_1,\, \zeta_2,\, \zeta_3) &=& - N_I - \widetilde N_{\mu I} \ .
\eea
As done in a previous section, the product  $\ell_2 (\zeta,\, \chi)$ is defined here up to terms ${\cal K}_I$, $\widetilde {\cal K}_{\mu I}$ taking values in the kernel of $\ell_1$, namely
\be
0 = Z^{M I} {\cal K}_I + \partial_\mu {\cal K}_I + \widetilde {\cal K}_{\mu I} + 2 \partial_{[\mu} \widetilde {\cal K}_{\nu]I} \ .
\ee
These can be chosen in such a way that\footnote{Note that the last terms proportional to the $Y$-tensor would vanish if the hierarchy were only projected to the two-form field, as expected from the results in Section \ref{Sec::projectedE7}.}
\be
\ell_2 (\zeta,\, \chi) = \frac 1 2 \delta_\Lambda \chi_I + Y_{I A} g_M{}^{A J} \left( Z^{M K} \Xi_{\mu J} \chi_K - \frac 1 2 \partial_\mu (\Lambda^M \chi_J)\right) \ ,
\ee
which in turn implies
\bea
\ell_2 (\chi_1 , \chi_2 ) &=& 0 \ , \\
\ell_3 (\zeta_1 , \zeta_2 , \chi) &=& Y_{I A}  C'{}^A - Y_{I A} \partial_\mu C'{}^A \ ,
\eea
with
\be
C'{}^A = \frac 1 3 g_{M}{}^{A J}\, \left( \Lambda_{12}^M \chi_J - \frac 1 2 Y_{J B} g_N{}^{B K} \Lambda_{[1}^M \Lambda_{2]}^N \chi_K \right)\ ,
 \ee
among higher products defined from higher identities. As stated by the theorem discussed in Section \ref{sec::theorem} no further graded subspaces are required beyond $X_2$. Instead, the effect of unprojecting the tensor hierarchy is to populate $X_2,\, X_1$ and$ X_0$ with the new representations.

\subsection{Maximal and half-maximal gauged supergravity}
Here we discuss how the general framework introduced in Section  \ref{sec:3-form TH} reproduces the tensor hierarchy of maximal and half-maximal gauged supergravities in four space-time dimensions. In Table \ref{Reps} we write the representations for each theory.
\begin{table}[ht]
\centering
    \begin{tabular}{|l|l|l|}
    \hline \ \ \ \ \ \ \   \ \ \ \ \ \ \ \ \ \ \  & \ \ \ \ {\bf Maximal} \ \ \ \   & \ \ \ \ {\bf Half-maximal} $n = 6$\ \ \ \ \ \   \\ \hline
$G_{0}$ & $E_{7(7)}$ & $SL(2) \times O(6,6)$ \\ \hline
Embedding tensor &  {\bf 912} & ({\bf 2}, {\bf 12}) + ({\bf 2}, {\bf 220}) \\ \hline
1-forms & {\bf 56} & ({\bf 2}, {\bf 12})  \\ \hline
2-forms & {\bf 133} & ({\bf 3}, {\bf 1}) + ({\bf 1}, {\bf 66} )  \\ \hline
3-forms & {\bf 912} & {\rm Projected out in \cite{Schon:2006kz}} \\ \hline
4-forms & {\bf 133} + {\bf 8645} & {\rm Projected out in \cite{Schon:2006kz}}\\ \hline
    \end{tabular}
    \caption{Duality groups and representations of the tensor hierarchy in maximal and half-maximal supergravities in four space-time dimensions. }
\label{Reps}
\end{table}

The general discussion in Section \ref{sec:3-form TH} perfectly matches the standard formulation of the maximal theory \cite{deWit:2002vt}, and so there is not much to say beyond the general discussion. The formulation of the half-maximal theory \cite{Schon:2006kz} requires instead a closer inspection. We can read from Table \ref{Reps} that the global symmetry of the ungauged maximal supergravity is a continuous $SL(2)\times O(6,n)$. We will take the fundamental representation to have split indices $(\alpha,M)$, where $\alpha,\beta,\gamma$ are $SL(2)$ indices and $M,N,P$ are fundamental $O(6,n)$ indices. This slightly modifies the notation in Section \ref{sec:3-form TH}, there the whole fundamental index was noted by $M,N,P$. Hopefully this will not cause confusion. The invariants of the global group are the antisymmetric Levi-Civita tensor $\epsilon_{\alpha\beta}$ with the conventions $\epsilon_{+-}=\epsilon^{+-}=1$ and $\epsilon_{\alpha \gamma}\epsilon^{\beta\gamma}=\delta^{\beta}_{\alpha}$. The $O(6,n)$ invariant tensor is $\eta_{M N}$. Both tensors are used like metrics to raise and lower indices.  In terms of these, the generators are\begin{align}
\left(t_{MN}\right)_{P}{}^Q & =\delta_{[M}^Q\eta_{N]P}\nonumber \\
\left(t_{\alpha\beta}\right)_{\gamma}{}^{\delta} & =\delta^{\delta}_{(\alpha}\epsilon_{\beta)\gamma} \ .\label{Generators}
\end{align}

 In order to gauge the supergravity we promote a subgroup $G\subset G_{0}$
to local with the help of the embedding tensor as we did in Section
\ref{sec:3-form TH}. The precise identifications are listed in Table \ref{table:1}.

\begin{table}[ht]
    \begin{tabular}{|l|l|}
	\hline
	{\bf General} &  {\bf Half-Maximal}   \\	\hline
	\centering $M$ & $\alpha M$   \\    \hline
	\centering  $I$ & $(\alpha \beta)$ \ \ \ $[M N]$  \\
	\hline
	\centering
	$\Lambda{}^M \ , \
	\widehat{\Xi}_{\mu I}$ & $\Lambda{}^{\alpha M} \ , \ \widehat{\Xi}_{\mu, \alpha\beta }\ , \
	\widehat{\Xi}_{\mu, MN }$  \\ 	\hline
	$ A_{\mu}{}^M\ , \ B_{\mu\nu I} $ & $ A_{\mu}{}^{\alpha M}\ , \ 	B_{\mu\nu, \alpha\beta } \ , \ 	B_{\mu\nu, MN }$  \\
	\hline
	\centering
	$d_{IMN}$ & $d_{\gamma\delta,\alpha M\beta N}=\eta_{MN}\epsilon_{\alpha(\gamma}\epsilon_{\delta)\beta}\ , \ \ \
	d_{OP,\alpha M\beta N}=\epsilon_{\alpha\beta}\eta_{M[O}\eta_{P]N}$ \\
	\hline
	\centering
	$\Theta_{M}{}^I$ & $\Theta_{\alpha M}{}^{\beta\gamma}=f_{\delta M}\epsilon^{\delta(\beta}\delta^{\gamma)}_{\alpha} \ , \ \ \
	\Theta_{\alpha M}{}^{NP} =f_{\alpha M}{}^{NP}+\delta_{M}^{[N}f_{\alpha}{}^{P]}$ \\
	\hline
	\centering
	$X_{MN}{}^O$ & $\begin{array}{c l}
	&X_{\alpha M\beta N}{}^{\gamma O}=\Theta_{\alpha M}{}^{\delta\lambda}\left(t_{\delta\lambda}\right)_{\beta}{}^{\gamma}\delta_{N}^O  +\Theta_{\alpha M}{}^{PQ}\left(t_{PQ}\right)_{N}{}^O\delta_{\beta}^{\gamma}\\
	&=-\delta_{\beta}^{\gamma}f_{\alpha MN}{}^O+\frac{1}{2}\left(\delta_{M}^O\delta_{\beta}^{\gamma}f_{\alpha N}
	 -\delta_{N}^O\delta_{\alpha}^{\gamma}f_{\beta M}-\eta_{MN}\delta_{\beta}^{\gamma}f_{\alpha}{}^O +\epsilon_{\alpha\beta}\epsilon^{\delta\gamma}\delta_{N}^O f_{\delta M}\right)
	\end{array}$ \\
	\hline
	$Z^{MI}$ & $\begin{array}{c l}
	Z^{\alpha M,\beta\gamma}&=\frac{1}{2}\eta^{MN}\epsilon^{\alpha\delta}\Theta_{\delta N}{}^{\beta\gamma}=-\frac{1}{2}\epsilon^{\alpha(\beta}\epsilon^{\gamma)\delta}f_{\delta}{}^M\\
	Z^{\alpha M,NO}&=\frac{1}{2}\eta^{MP}\epsilon^{\alpha\beta}\Theta_{\beta P}{}^{NO}=\frac{1}{2}\epsilon^{\alpha\beta}\left(f_{\beta}{}^{MNO}+\eta^{M[N}f_{\beta}{}^{O]}\right)
	\end{array}$  \\
	\hline
\end{tabular}
\caption{Identification between tensors in the general structure of the tensor hierarchy (Section \ref{sec:3-form TH}) and the $n=4$ half-maximal gauged supergravity.}
\label{table:1}
\end{table}

The gaugings now take the form $f_{\alpha M}$ and $f_{\alpha MNP}=f_{\alpha[MNP]}$,  in terms of which the quadratic constraints (\ref{eq:cuadratic_constraint}) take the form
\begin{align}
f_{\alpha}{}^Mf_{\beta M} & =0\nonumber \\
f_{(\alpha}{}^Pf_{\beta)PMN} & =0\nonumber \\
3f_{\alpha[MN}{}^Pf_{\beta R]OP}-2f_{(\alpha[MNR}f_{\beta)O]} & =0\nonumber \\
\epsilon^{\alpha\beta}\left(f_{\alpha}{}^Pf_{\beta PMN}+f_{\alpha M}f_{\beta N}\right) & =0\nonumber \\
\epsilon^{\alpha\beta}\left(f_{\alpha MNR}f_{\beta PQ}{}^R-f_{\alpha}{}^Rf_{\beta R[M[P}\eta_{Q]N]}-f_{\alpha[M}f_{N][PQ]\beta}+f_{\alpha[P}f_{Q][MN]\beta}\right) & =0\ .\label{eq:gaugings_constraint}
\end{align}

From the Table \ref{table:1} and the general expressions in Section (\ref{sec:3-form TH}) we can compute the gauge transformations of the vector fields and two-forms
\bea
\delta A_{\mu}{}^{\alpha M} & =&\mathcal{D}_{\mu}\Lambda^{\alpha M}-\frac{1}{2}\epsilon^{\alpha\beta}\left[f_{\gamma}{}^M\epsilon^{\gamma\delta}\widehat{\Xi}_{\mu, \delta\beta}+\left(f_{\beta}{}^{MNP}+\eta^{MN}f_{\beta}{}^P\right)\widehat{\Xi}_{\mu, NP}\right]
\nn \\
	\delta B_{\mu\nu,\alpha\beta}&=& 2\mathcal{D}_{[\mu}\widehat{\Xi}_{\nu],\alpha\beta} -2\eta_{MN}\epsilon_{\gamma(\alpha}\epsilon_{\beta)\delta}(\Lambda^{\gamma M}\mathcal{F}_{\mu\nu}{}^{\delta N} - A_{[\mu}{}^{\gamma M}\delta A_{\nu]}{}^{\delta N})\\
		\delta B_{\mu\nu,MN}&=& 2\mathcal{D}_{[\mu}\widehat{\Xi}_{\nu],MN}-2\epsilon_{\alpha\beta}\eta_{O[M}\eta_{N]P}(\Lambda^{\alpha O}\mathcal{F}_{\mu\nu}{}^{\beta P} -A_{[\mu}{}^{\alpha O}\delta A_{\nu]}{}^{\beta P}) \ , \nn
\eea
their field strengths
\bea
	\mathcal{F}_{\mu\nu}{}^{\alpha M} &=& 2\mathcal{\partial}_{[\mu}A_{\nu]}{}^{\alpha M}+X_{\beta N\gamma P}{}^{\alpha M}A_{[\mu}{}^{\beta N}A_{\nu]}{}^{\gamma P} \nn \\ && +\frac{1}{2}\epsilon^{\alpha\beta}\left[f_{\gamma}{}^M\epsilon^{\gamma\delta}B_{\mu\nu,\delta\beta}+\left(f_{\beta}{}^{MNP}+\eta^{MN}f_{\beta}{}^P\right)B_{\mu\nu,NP}\right] \nn \\
	\mathcal{H}_{\mu\nu\rho,\alpha\beta}&=&3\mathcal{D}_{[\mu}B_{\nu\rho],\alpha\beta}+6\eta_{MN}\epsilon_{\gamma(\alpha}\epsilon_{\beta)\delta}A_{[\mu}{}^{\gamma M}\left(\partial_{\nu}A_{\rho]}{}^{\delta N} +\frac{1}{3}X_{\lambda O\epsilon P}{}^{\delta N}A_{\nu}{}^{\lambda O}A_{\rho]}{}^{\epsilon P} \right)\\
	\mathcal{H}_{\mu\nu\rho,MN}&=&3\mathcal{D}_{[\mu}B_{\nu\rho],MN}+6\epsilon_{\alpha\beta}\eta_{O[M}\eta_{N]P}A_{[\mu}{}^{\alpha O}\left(\partial_{\nu}A_{\rho]}{}^{\beta P} +\frac{1}{3}X_{\gamma Q\delta R}{}^{\beta P}A_{\nu}{}^{\gamma Q}A_{\rho]}{}^{\delta R}\right) \ , \nn
\eea
and the covariant derivatives
\be
\mathcal{D}_{\mu}=\partial_{\mu}-A_{\mu}{}^{\alpha M}\Theta_{\alpha M}{}^{NO}t_{NO}-A_{\mu}{}^{\alpha M}\Theta_{\alpha M}{}^{\beta\gamma}t_{\beta\gamma} \ .
\ee

With these results we can also compute the $Y_{I A}$ tensor using its definition in (\ref{eq:constraints})
\be
Y_{I A} g_{M}{}^{A J} = X_{M I}{}^J + 2 Z^{N J} d_{I M N} \ ,
\ee
with components
\begin{align}
Y_{\alpha\beta,A}\, g_{\gamma M}{}^{A,\delta\lambda} & =-f_{(\alpha M}\delta_{\beta)}^{(\delta}\delta_{\gamma}^{\lambda)}\label{intertwininghalfmax}\\
Y_{\alpha\beta,A}\, g_{\gamma M}{}^{A,NP} & =\left(f_{(\alpha M}{}^{NP}+\delta_{M}^{[N}f_{(\alpha}{}^{P]}\right)\epsilon_{\beta)\gamma}\nn\\
Y_{NP,A}\, g_{\gamma M}{}^{A,\delta\lambda} & =-\frac{1}{2}\delta_{\gamma}^{(\delta}\epsilon^{\lambda)\beta}f_{\beta[N}\eta_{P]M}\nn\\
Y_{NP,A}\, g_{\gamma M}{}^{A,OQ} & =-2f_{\gamma M[N}{}^{[O}\delta_{P]}^{Q]}+f_{\gamma [N}{}^{OQ}\eta_{P]M}-f_{\gamma[N}\delta_{P]}^{[O}\delta_{M}^{Q]} \ . \nn
\end{align}
The fact that these are non-vanishing implies that in half-maximal gauged supergravity the algebra closes because the two-forms are projected with the intertwiner $Z^{M I}$. In principle the full algebra can be obtained un-projecting and adding higher tensors through the standard procedure.

We can now as a final check see how to relate these expressions to those in the electric sector computed from DFT. The identifications for the gaugings are
\begin{align}
f_{-M} & =0 \nn\\
f_{-MNP} & =0 \label{electricgaugings}\\
f_{+M} & = f_{M}\nn\\
f_{+MNO} & = f_{MNO} \ .\nn
\end{align}
Then,  the
last two equations of (\ref{eq:gaugings_constraint}) vanish trivially
and the remaining constraints reduce to
\begin{align}\label{eq:electric_cuadratic_constraint}
f{}^M\,f_{M} & =0 \nn\\
f{}^P\, f_{PMN} & =0\\
3f_{[MN}{}^P\,f_{R]OP}-2f_{[MNR}\,f_{O]} & =0 \ ,\nn
\end{align}
and match (\ref{cuadraticconstraint}) exactly.

To obtain full identification the parameters must be related as
\bea
\Lambda^{+ M} = \Lambda^M \ , \ \ \ \widehat \Xi_{\mu \, --} = - 2 \widehat \Xi_{\mu} \ ,
\eea
and the  fields  as follows
\be
A_\mu{}^{+ M} = A_\mu{}^M \ , \ \ \ B_{\mu \nu\, --} = - 2 B_{\mu \nu} \ .
\ee
 We call this selection the ``electric'' section of the theory. It turns out that by selecting the electric gaugings (\ref{electricgaugings}), the transformations of the electric components above do not depend on the magnetic ones (namely, the parameters $\Lambda^{- M}$, $\widehat \Xi_{\mu\, - +}$, $\widehat \Xi_{\mu\, + +}$ and $\widehat \Xi_{\mu\, M N}$ and the fields $A_{\mu}{}^{- M}$, $B_{\mu \nu \, - +}$, $B_{\mu \nu \, ++}$ and $B_{\mu \nu\, M N}$ disappear from the transformations of $A_\mu{}^{+ M}$ and $B_{\mu \nu\, --}$). Then, they decouple forming a closed subalgebra. With these identifications the gauge transformations of the effective action of DFT are recovered (\ref{eq:KKDFTtransformations}).

 We are now in the position of understanding why the small electric sector of gauged supergravity, consisting in vector fields and single two-form, closes exactly without projection with an intertwining tensor nor the inclusion of higher forms. The reason is that the electric section yields the following vanishing components of the intertwining tensor
 \be
  Y_{--,A}\, g_{\gamma M}{}^{A, I}  =0 \ ,
 \ee
 and this particular combination is responsible for the closure of the gauge algebra of the two-form to fail.

We can also re-derive the
 Jacobiator starting with (\ref{eq:Jacobiators}) for half-maximal
\begin{align}
J^{\alpha M} & =Z^{\alpha M\beta\gamma}N_{\beta\gamma}+Z^{\alpha MNO}N_{NO} \nn\\
J_{\mu,\alpha\beta} & =\partial_{\mu}N_{\alpha\beta}\\
J_{\mu, MN} & =\partial_{\mu}N_{MN} \ ,\nn
\end{align}
with the Nijenhuis tensors given by
\begin{align}
N_{\alpha\beta} & =\eta_{MN}\epsilon_{\gamma(\alpha}\epsilon_{\beta)\delta}X_{\lambda O\epsilon P}{}^{\gamma M}\Lambda_{[1}{}^{\lambda O}\Lambda_{2}{}^{\epsilon P}\Lambda_{3]}{}^{\delta N} \nn\\
N_{MN} & =\epsilon_{\alpha\beta}\eta_{O[M}\eta_{N]P}X_{\gamma Q\delta R}{}^{\alpha O}\Lambda_{[1}{}^{\gamma Q}\Lambda_{2}{}^{\delta R}\Lambda_{3]}{}^{\beta P} \ .
\end{align}
Then, imposing the electric section, it is easy to see that
\begin{align}
N_{--} & =f_{MNO} \Lambda_{[1}{}^M \Lambda_{2}{}^N \Lambda_{3]}{}^O = 2 N\ ,
\end{align}
where in the last equality we established the relation with (\ref{effectiveN}). Finally, using that the only non-vanishing components of the intertwining tensor are
\begin{align}
Z^{-M,OP} & =-\frac{1}{2}\left(f^{MOP}+\eta^{M[O}f^{P]}\right) \nn \\
Z^{-M,+-} & =-\frac{1}{4}f^{M}\\
Z^{+M,--} & =\frac{1}{2}f^{M} \ , \nn
\end{align}
the following relevant components of the Jacobiator survive
\begin{align}
J^{+M} & = \frac{1}{2}f^{M} \, N_{--} = J^M\\
J_{\mu,--} & =\partial_{\mu} N_{--} = - 2 J_\mu \ ,
\end{align}
which are trivial parameters, as expected, and in the last equality we established the relation with (\ref{effectiveJ}). This completes the embedding of the tensor hierarchy for the electric and full half-maximal supergravity in the general structure discussed before.

\section*{Acknowledgments} We thank S. Iguri and C. Nu\~nez for comments. Our work is supported by CONICET.

\section*{Appendix}

\begin{appendix}

\section{Relating the generalized Metric and the KK formulations of DFT}

In this Appendix we relate the degrees of freedom and parameters of the generalized metric formulation of DFT \cite{Hohm:2010pp} (${\cal H}_{{\cal M} {\cal N}}$ , $d$ , $\xi^{\cal M}$) with those in the KK formulation \cite{Hohm:2013nja} ($g_{\mu \nu}$, $B_{\mu \nu}$, $A_\mu{}^M$, ${\cal M}_{M N}$, $\phi$, $\xi^\mu$, $\Lambda^M$, $\Xi_\mu$). The $O(D,D)$ indices ${\cal M}, {\cal N} = 1,\dots,2D$ with $D = n + d$  are here distinguished from the $O(d,d)$ indices $M, N = 1 , \dots,2d$.  The relation is given by
\be
\eta_{\cal M N} = \left(\begin{matrix} 0 & \delta^\mu_\nu & 0 \\ \delta_\mu^\nu & 0 & 0 \\ 0 & 0 & \eta_{M N}
\end{matrix}\right) \ , \ \ \ \ \ \partial_{\cal M} = \left(0 , \partial_\mu , \partial_M \right)\ ,
\ee
such that the strong constraint in $O(D,D)$ $\partial_{\cal M} \partial^{\cal M} = 0$ becomes that in $O(d,d)$ $\partial_M \partial^M = 0$. The parameters are related by \be
 \xi^{\cal M} = \left(\begin{matrix} - \Xi_\mu \\ \xi^\mu \\ \Lambda^M \end{matrix}\right)\ ,
\ee
and the dynamical degrees of freedom by
\be
{\cal H}_{\cal M N} = \left(\begin{matrix} g^{\mu \nu} &  - g^{\mu \rho} C_{\rho \nu} & - g^{\mu \rho} A_{\rho N} \\ - g^{\nu \rho} C_{\rho \mu} & g_{\mu \nu} + C_{\rho \mu} C_{\sigma \nu} g^{\rho \sigma} + A_\mu{}^P {\cal M}_{P Q} A_\nu{}^Q & C_{\rho \mu} g^{\rho \sigma} A_{\sigma N} + A_{\mu}{}^P {\cal M}_{P N} \\
-g^{\nu \rho} A_{\rho M} & C_{\rho \nu} g^{\rho \sigma} A_{\sigma M} + A_\nu{}^P {\cal M}_{P M} & {\cal M}_{M N} + g^{\rho \sigma} A_{\rho M} A_{\sigma N}
\end{matrix}\right)\ ,
\ee
where $C_{\mu \nu} = - B_{\mu \nu} + \frac 1 2 A_\mu{}^P A_{\nu P}$. From here it is straightforward to reproduce the KK-formulation from the $O(D,D)$ covariant expressions of the generalized Lie derivative and DFT action.

\section{Checking some $L_\infty$ identities}
\label{app-identidades}

Here we check some non-trivial identities for Section \ref{sct-Linf}. We first identify the terms that are non vanishing and construct the possible lists of arguments from them.

\paragraph{All identities that include at least one $E\in X_{-2}$ hold true.} The argument is very similar to the one presented in [1701.08824], but there are some differences. Each term of an $L_\infty$ relation is of the form $\ell_i(...\ell_j(...))$, where $...$ represents a list of arguments. The only non vanishing products involving an $E \in X_{-2}$ are $\ell_2(E,\z)$ and $\ell_3(E,\z,\Psi)$, see (\ref{lf}). If we consider that there is an $E$ in the second list $\ell_i(...\ell_j(E,...))$ then the possibilities are:
	\bea
	&\ell_i(...\underbrace{\ell_2(E,\z_1)}_{\equiv \tilde E\in X_{-2}})\Rightarrow \ell_2(\z_2,\tilde E) \text{  or  } \ell_3(\z_2,\Psi,\tilde E),\label{primercaso}\\
	&\ell_i(...\underbrace{\ell_3(E,\z_1,\Psi_1)}_{\equiv \hat  E\in X_{-2}})\Rightarrow \ell_2(\z_2,\hat E) \text{  or  } \ell_3(\z_2,\Psi_2,\hat E).\label{segundocaso}
	\eea
	From this we learn that the possible lists are $(\z_1,E,\z_2)$, $(\z_1,E,\z_2,\Psi)$, or $(E,\z_1,\Psi_1,\z_2,\Psi_2)$. If instead we consider that there is an $E$ in the first list $	\ell_i(E,...\ell_j(...))$, then the possibilities are:
	\beq
	&\ell_2(E,\underbrace{\ell_j(...)}_{\in X_0})\Rightarrow \ell_2(E,\ell_1(\chi)) \text{  or  } \ell_2(E,\ell_2(\z_1,\z_2)),\\
	&\ell_3(E,\z_1,\underbrace{\ell_j(...)}_{\in X_{-1}})\Rightarrow \ell_3(E,\z_1,\ell_1(\z_2)) \text{  or  } \ell_3(E,\z,\ell_2(\z_2,\Psi)),\\
	&\ell_3(E,\Psi,\underbrace{\ell_j(...)}_{\in X_0})\Rightarrow \ell_3(E,\Psi,\ell_1(\chi)) \text{  or  } \ell_3(E,\Psi,\ell_2(\z_1,\z_2))\ .
	\eeq
	There are lists of arguments that are the same to those of the previous case and we have the new ones $(\chi,E)$ and  $(\chi,\Psi,E)$.
	
	Now we consider the list $(E,\chi)$, for which the identity reads:
	\beq
	\ell_1(\ell_2(E,\chi))&=\ell_2(\ell_1(E),\chi)+\ell_2(\ell_1(\chi),E) \ ,
	\eeq
	which holds because $\ell_2(E,\chi)=0$, $\ell_1(E) = 0$ and
\beq
\ell_2(\cD\chi,E)=&
\begin{dcases}
   \widehat \cL_{\del^M\chi} \Delta B^{\m\n}=0\ ,\\
   \Delta B^{\mu \nu} \left(\del_\m\del^M\chi-\del^M\del_\m\chi\right) + \widehat \cL_{\del^M\chi} \Delta_{c} A^\mu{}_M =0\ .\\
\end{dcases}
\eeq
	Now we consider the list $(\chi,\Psi,E)$ with its corresponding identity:
\beq
0\stackrel{?}{=}&\ell_1\left(\ell_3(\chi ,\Psi ,E)\right)-\ell_2\left(\ell_2(\chi ,E),\Psi \right)+\ell_2\left(\ell_2(\chi ,\Psi ),E\right)-\ell_2\left(\ell_2(\Psi ,E),\chi \right)\\
&+\ell_3\left(\ell_1(E),\chi ,\Psi \right)-\ell_3\left(\ell_1(\chi ),E,\Psi \right)-\ell_3\left(\ell_1(\Psi ),E,\chi \right),\\
0\stackrel{?}{=}&\ell_3\left(\ell_1(\chi ),E,\Psi \right)=\ell_3\left(\cD \chi ,E,\Psi \right)=\widehat \cL_{\del^N\chi} \Delta B^{\mu \nu}  A_{\nu M} + \Delta B^{\mu \nu} \widehat \cL_{\del^N\chi} A^\mu{}_M =0\ .
\eeq

 The remaining lists, $(\z_1,E,\z_2)$, $(\z_1,E,\z_2,\Psi)$, or $(E,\z_1,\Psi_1,\z_2,\Psi_2)$, also satisfy the identities as can be seen from the closure of gauge transformations over field equations
\beq
\left[\d_{\z_1},\d_{\z_2}\right] \cF = -\d_{[\z_1,\z_2]} \cF\ .
\label{closuref}
\eeq
Expanding in powers of $\Psi$ we arrive at the $L_{\infty}$ relations over the list of the form $(\z_1,\z_2,E,\Psi^n)$, with $n\geq 0$.
To lowest order in $\Psi$ we get		\beq
	\ell_2(\z_2,\ell_2(\z_1,\cF))+\ell_3(\z_2, \cF,\ell_1(\z_1))-\ell_2(\z_1,\ell_2(\z_2,\cF))-\ell_3(\z_1, \cF,\ell_1(\z_2))=-\ell_2(\ell_2(\z_1,\z_2),\cF)\ .
	\eeq
This is the $L_{\infty}$ relation needed for the list $(\z_1,E,\z_2)$, once one takes into account that the missing term $\ell_1(\ell_3(\z_1,\z_2,\cF))$ expected in such identity does not appears as it was set equal to zero. The linear order yields \beq
	-\ell_3(\ell_2(\z_1,\z_2),\cF,\Psi)&=\ell_2(\z_2,\ell_3(\z_1,\cF,\Psi))+\ell_3(\z_2,\ell_2(\z_1,\cF),\Psi)+\ell_3(\z_2, \cF,\ell_2(\z_1,\Psi))\\
		&\quad -\ell_4(\z_2,\cF,\ell_1(\z_1),\Psi)-\ell_2(\z_1,\ell_3(\z_2,\cF,\Psi))-\ell_3(\z_1,\ell_2(\z_2,\cF),\Psi)\\
		&\quad-\ell_3(\z_1, \cF,\ell_2(\z_2,\Psi))+\ell_4(\z_1,\cF,\ell_1(\z_2),\Psi)\ ,
	\eeq
	which again is an $L_\infty$ relation once one considers the terms that are equal to zero.
	Finally the quadratic $\Psi^2$ order has an extra difficulty because from \eqref{closuref} one gets the identity evaluated on diagonal arguments $(\z_1,\z_2,E,\Psi,\Psi)$ instead of $(\z_1,\z_2,E,\Psi_1,\Psi_2)$. However, this is not a problem because an $L_\infty$ relation which holds true on diagonal arguments also holds true on non-diagonal arguments, which can be proved using the polarization identities.

\paragraph{All identities that include at least one $\chi \in X_1$ hold true.} We recall first that there are only two non trivial products involving $\chi$: $\ell_1(\chi)$ and $\ell_2(\chi,\z)$. This means that we can have at most two $\chi$'s.  If we had more than two $\chi$'s we would have had two or more $\chi$'s in the same product which is zero. Take for example the case of three $\chi$'s:
\beq
\underbrace{\ell_i(\chi_a,\chi_b,\ell_j(\chi_c...))}_{=0}\qquad\text{or}\qquad \ell_i(\chi_a,\underbrace{\ell_j(\chi_b,\chi_c...)}_{=0})\ .
\eeq
Now take the list $(\chi_1,\chi_2,...)$. A term in the identities would look like $	\ell_i(\chi_1...\ell_j(\chi_2,...))$.
	The possibilities are:
	\bea
	&&\ell_i(\chi_1...\underbrace{\ell_2(\chi_2,\z)}_{\equiv \tilde \chi \in X_{1}})\Rightarrow \ell_i(\chi_1,...\tilde \chi)=0, \\
	&&\ell_i(\chi_1...\underbrace{\ell_1(\chi_2)}_{\equiv \tilde  \z\in X_{0}})\Rightarrow \ell_2(\chi_1,\tilde \z)\ .
	\eea
	The only list of arguments is then $(\chi_1,\chi_2)$. The identity reads
	\beq
	\ell_1(\underbrace{\ell_2(\chi_1,\chi_2)}_{=0})&=\ell_2(\ell_1(\chi_1),\chi_2)+\ell_2(\ell_1(\chi_2),\chi_1)=\ell_2(\cD\chi_1,\chi_2)+\ell_2(\cD\chi_2,\chi_1)=0\ .
	\eeq
	Now take the list with only one $\chi$, $(\chi,...)$. If $\chi$ is in the first sublist, we have
	\beq
	\ell_i(\chi...\ell_j(...))\Rightarrow\ell_2(\chi,\underbrace{\ell_j(...)}_{\in X_0})\Rightarrow\ell_2(\chi,\ell_2(\z_1,\z_2))\ .
	\eeq
	One can then check that the identity for the list $(\chi,\z_1,\z_2)$ holds.
	
Returning to the confection of lists,	if $\chi$ is in the second sublist, we have more possibilities. In the first place we could have
	\beq
	\ell_i(...,\underbrace{\ell_2(\chi,\z)}_{\tilde\chi\in X_1})\Rightarrow
	\begin{dcases}
   \ell_2(\tilde \z,\ell_2(\chi,\z)),\quad\text{same list as before},\\
   \ell_1(\ell_2(\chi,\z_2))\ .
\end{dcases}
	\eeq
	The last case was also checked when we discussed the pure gauge structure. In the second place
		\beq
	\ell_i(...,\underbrace{\ell_1(\chi)}_{\tilde\z\in X_0})\Rightarrow
	\begin{dcases}
   \ell_2(\z,\ell_1(\chi)),\quad\text{same list as before},\\
   \ell_1(\ell_1(\chi)),\quad\text{checked when we discussed pure gauge structure},\\
	 \ell_2(\Psi,\ell_1(\chi)),\quad\text{with list } (\Psi,\chi).\\
\end{dcases}
	\eeq
This last case gives the following identity which is true
\beq
0=&\ell_1\left(\ell_2(\chi ,\Psi )\right)-\ell_2\left(\ell_1(\chi ),\Psi \right)-\ell_2\left(\ell_1(\Psi ),\chi \right)\ .
\eeq

\paragraph{Finally, all identities which include at least 3 $\z$'s and any number of $\Psi$ hold true.}

The list with only three gauge parameters was already checked when we analized the pure gauge structure (it was our starting point). Next we could consider four gauge parameters.
The identity is $\ell_1\ell_4-\ell_2\ell_3+\ell_3\ell_2-\ell_4\ell_1=0$ and without any $\ell_4$, the following expression must vanish
\beq
\ell_3 \ell_2 - \ell_2 \ell_3  \ = \  \
 6\, \ell_3   ([ \z_{[1}, \z_2] , \z_3, \z_{4]}) \ - \
  4 \, \ell_2  (\ell_3 (\z_{[1}, \z_2, \z_3) , \z_{4]} )  \ .
\eeq
Taking into account that $\ell_3$ over three gauge parameters of the form $\z=\L^M+\Xi_\m$ discards the $\Xi_\m$ part, we get for the first term on the RHS the following expression
\beq
\label{cancelate}
6\, \ell_3   ([ \L_{[1}, \L_2]_{(C)} , \L_3, \L_{4]}) &=
					2 \left[ \,[\L_{[1}, \L_2]_{(C)},\L_3\,\right]_{(C)}^M \L_{4]\,M}
					+[ \L_{[3}, \L_4]_{(C)}^M [ \L_{1}, \L_{2]}]_{(C)\,M} \ .
\eeq
On the other hand
\beq
- 4 \, \ell_2  (\ell_3 (\z_{[1}, \z_2, \z_3) , \z_{4]} )&= \widehat \cL_{\L_{[4}}\left([ \L_{1}, \L_2]_{(C)}^M \L_{3]\,M}\right) \ ,
\eeq
so using the Leibnitz rule for the Generalized Lie derivative and the fact that the D-bracket can be replaced by the C-bracket because of the antisymmetrization the identity can be proved.

Now lets take into account gauge parameters and fields: $(\z_1,\z_2,\z_3, \Psi^n)$. We first notice that we cannot use the products $\ell_1 (\z), \ell_2 (\z ,\Psi),\ell_3 (\z,\Psi,\Psi)$, which give an element $\tilde \Psi\in X_ {-1}$. If they were in the second product $\ell_j$ we would have
\beq
&\ell_i(...\ell_j(...))=\ell_3(\z_1,\z_2,\tilde\Psi)=0 \ .
\eeq
Trying to use them in the first product does not work either. For $\ell_1 (\z)$ we would need the second product to give an element of $ X_0$:
\beq
\ell_1 (\underbrace{\ell_{n+3}(\z_1,\z_2,\z_3, \Psi^n)}_{\notin X_0})\ .
\eeq
For $\ell_2 (\z ,\Psi)$ we would need the second product to give an element of $ X_0$ or $ X_{-1}$ :
\beq
\ell_2 (\underbrace{\ell_{n+2}(\z_1,\z_2,\z_3,\Psi^{n-1}}_{\notin X_0}) ,\Psi)\ ,
\qquad
\ell_2 (\z_1,\underbrace{ \ell_{n+2}(\z_2,\z_3,\Psi^{n}}_{\notin X_{-1}}) )\ .
\eeq
For $\ell_3 (\z,\Psi,\Psi)$ we would need the second product to give an element of $ X_0$ or $ X_{-1}$ :
\beq
\ell_3 (\underbrace{\ell_{n+1}(\z_1,\z_2,\z_3,\Psi^{n-2}}_{\notin X_0}) ,\Psi,\Psi),
\qquad
\ell_3 (\z_1,\underbrace{ \ell_{n+1}(\z_2,\z_3,\Psi^{n-1}}_{\notin X_{-1}}),\Psi )\ .
\eeq
Thus, one of the products must be $\ell_3(\z_1,\z_2,\z_3)\equiv\chi\in X_1$. If it was in the second product $\ell_j$ we would have
\beq
&\ell_i(...\ell_j(...))=\ell_{n+1}(\Psi^n,\ell_3(\z_1,\z_2,\z_3))=\ell_{n+1}(\Psi^n,\chi)=0\ .
\eeq
 If it was in the first product, we would need the second one to give a $X_0$ element to get a nontrivial term, but this is not the case
\beq
\ell_3(\z_1,\z_2,\underbrace{\ell_{n+1}(\z_3,\Psi^n)}_{\notin X_0})\ .
\eeq
Considering more than three gauge parameters and repeating the arguments as before one also finds that the $L_\infty$ identities are fulfilled trivially.

\section{Full hierarchy in four space-time dimensions}
Having discussed in detail the three-form tensor hierarchy in generic gauged supergravities we can
move a step upwards and introduce the four-form field which corresponds
to the last field in the p-form chain for $n=4$ space-time dimensions (some of these results can also be found in \cite{Bergshoeff:2009ph}). To do this
we continue by un-projecting the three-form fields
\begin{align}
\widetilde{C}_{\mu\nu\rho I} & \equiv Y_{IA}C_{\mu\nu\rho}{}^ A\ ;\ \widetilde{\widehat{\Sigma}}_{\mu\nu I}\equiv Y_{IA}\widehat{\Sigma}_{\mu\nu}{}^ A\ ;\ \mathcal{\widetilde{G}}_{\mu \nu \rho\sigma I}\equiv Y_{IA}\mathcal{G}_{\mu \nu\rho\sigma}{}^ A \ ,
\end{align}
at the expense of introducing a new projected four-form field $\widetilde{D}_{\mu\nu\rho\sigma}{}^A$
and its projected shift parameter $\widetilde{\widehat{\Theta}}_{\mu\nu\rho}{}^ A$, modifying
the previous results
\begin{align}
\delta C_{\mu\nu\rho}{}^ A & =3D_{[\mu}\widehat{\Sigma}_{\nu\rho]}{}^ A+g_{M}{}^{AI}\left(3\mathcal{F}_{[\mu\nu}{}^ M\widehat{\Xi}_{\rho]I}+\Lambda^{M}\mathcal{H}_{\mu\nu\rho I}+3B_{[\mu\nu I}\delta A_{\rho]}{}^ M \right. \nn \\ & \left. \ \ \ \ \ \ \ \ \ \ \ \ \ \ \ \ \ \ \ \ \ \ \ \ \ \ \ \ \ \ +2d_{IMN}A_{[\mu}{}^ MA_{\nu}{}^ N\delta A_{\rho]}{}^ P\right)-\widetilde{\widehat{\Theta}}_{\mu\nu\rho}{}^ A\\
\mathcal{\mathcal{G}}_{\mu\nu\rho\sigma}{}^ A & =4\mathcal{D}_{[\mu}C_{\nu\rho\sigma]}{}^ A-g_{M}{}^{AI}\left(6B_{[\mu\nu I}\mathcal{F}_{\rho\sigma]}{}^ M-3Z^{MJ}B_{[\mu\nu I}B_{\rho\sigma]J}+8d_{INP}A_{[\mu}{}^ MA_{\nu}{}^ N\partial_{\rho}A_{\sigma]}{}^ P\right. \nn \\ & \left. \ \ \ \ \ \ \ \ \ \ \ \ \ \ \ \ \ \ \ \ \ \ \ \ \ \ \ \ \ \ +2d_{IPN}X_{QR}{}^ PA_{[\mu}{}^ MA_{\nu}{}^ NA_{\rho}{}^ QA_{\sigma]}{}^ R\right) +\widetilde{D}_{\mu\nu\rho\sigma}{}^ A \ ,
\end{align}
and satisfying
\begin{align}
Y_{IA}\widetilde{D}_{\mu\nu\rho\sigma}{}^ A & =Y_{IA}\widetilde{\widehat{\Theta}}_{\mu\nu\rho}{}^ A=0 \ .
\end{align}

Then, imposing gauge covariance of the unprojected four-form field strength
tensor
\begin{align}
\delta\mathcal{G}_{\mu\nu\rho\sigma}{}^ A & =-X_{MB}{}^ A\Lambda^{M}\mathcal{G}_{\mu\nu\rho\sigma}{}^ B\ ,
\end{align}
we find the transformation of the new field
\begin{align*}
\delta\widetilde{D}{}_{\mu\nu\rho\sigma}{}^ A= & 4\mathcal{D}_{[\mu}\widetilde{\widehat{\Theta}}_{\nu\rho\sigma]}{}^ A-W^{AX}h_{XBM}\left(6\mathcal{F}_{[\mu\nu}{}^ M\widehat{\Sigma}_{\rho\sigma]}{}^ B+\Lambda^{M}\mathcal{G}_{\mu\nu\rho\sigma}{}^ B-4C_{[\mu\nu\rho}{}^ B\delta A_{\sigma]}{}^ M\right. \nn\\& \left.+2g_{N}{}^{BI}d_{IPQ}A_{[\mu}{}^ MA_{\nu}{}^ NA_{\rho}{}^ P\delta A_{\sigma]}{}^ Q\right) +4V^{AIJ}\left(\widehat{\Xi}_{[\mu I}\mathcal{H}_{\nu\rho\sigma]J}+3d_{INP}B_{[\mu\nu J}A_{\rho}{}^ N\delta A_{\sigma]}{}^ P\right)\ ,
\end{align*}
where we introduced the new intertwining tensor $W^{AX}$, which has
to satisfy
\begin{align*}
W^{AX}h_{XBM} & = X_{MB}{}^ A+g_{M}{}^{AI}Y_{IB}\\
W^{AX}Y_{IA} & =0 \ ,
\end{align*}
 and the $G_{0}$-invariant tensor, $h_{XBM}$. In addition we defined
\begin{align*}
V^{AIJ} & \equiv2g_{M}{}^{A[I}Z^{J]M} \ ,
\end{align*}
which also vanishes when contracted with $Y_{KA}$ thanks to the
constraints. Due to space-time saturation in four dimensions the field strength of the four-form field and its Bianchi identity vanish.

It can be shown that the gauge algebra of the new system $\left\{ A_{\mu}{}^ M,B_{\mu\nu I},C_{\mu\nu\rho}{}^ A,\widetilde{D}_{\mu\nu\rho\sigma}{}^ A\right\} $
closes up to terms proportional to $h_{XB(M}g_{N)}{}^{BI}$ and $V^{AIJ}$, which must then be set to zero as new constraints. In fact these are related by the equation
\begin{align}
W^{AX} & \left(h_{XB(M}g_{N)}{}^{BI}\right)=V^{AIJ}d_{JMN}\ , \label{eq:Vcero}
\end{align}
and the new constraints are
\bea
h_{XB(M}g_{N)}{}^{BI} &=& 0 \nn \\
V^{AIJ} = 2g_{M}{}^{A[I}Z^{J]M} &=& 0 \ .
\eea
The brackets that arise from closure of the algebra are given by
\begin{align}
\Lambda_{12}^ M & =-X_{NP}{}^ M\Lambda_{[1}^ N\Lambda_{2]}^ P\\
\widehat{\Xi}_{12\mu I} & =2d_{IMN}\mathcal{D}_{\mu}\Lambda_{[1}^ M\Lambda_{2]}^ N\\
\widehat{\Sigma}_{12\mu\nu}{}^ A & =-2g_{M}{}^{AI}\left(d_{INP}\Lambda_{[1}^ M\Lambda_{2]}^ N\mathcal{F}_{\mu\nu}{}^ P+2\widehat{\Xi}_{I[\mu[1}\mathcal{D}_{\nu]}\Lambda_{2]}^ M-Z^{MJ}\widehat{\Xi}_{I[\mu[1}\widehat{\Xi}_{\nu]2]J}\right)\\
\widetilde{\widehat{\Theta}}_{12\mu\nu\rho}{}^ A & =-W^{AX}h_{XBM}\left(g_{N}{}^{BI}\Lambda_{[1}^ M\Lambda_{2]}^ N\mathcal{H}_{\mu\nu\rho I}+6\widehat{\Sigma}_{[\mu\nu}{}^ B_{[1}\mathcal{D}_{\rho]}\Lambda_{2]}^ M\right) \ .
\end{align}

To finish this chain we can unproject the 4-form
\begin{align}
\widetilde{D}_{\mu\nu\rho\sigma}{}^A & =W^{AX}D_{\mu\nu\rho\sigma X}\ ;\ \widetilde{\widehat{\Theta}}_{\mu\nu\rho}{}^ A=W^{AX}\widehat{\Theta}_{\mu\nu\rho X}\ ,
\end{align}
and introduce a new gauge parameter $\widehat{\Phi}_{\mu\nu\rho\sigma X}$
into its transformation
\begin{align*}
\delta D_{\mu\nu\rho\sigma X}= & 4\mathcal{D}_{[\mu}\widehat{\Theta}_{\nu\rho\sigma]X}-h_{XBM}\left(6\mathcal{F}_{[\mu\nu}{}^ M\widehat{\Sigma}_{\rho\sigma]}{}^ B+\Lambda^{M}\mathcal{G}_{\mu\nu\rho\sigma}{}^ B-4C_{[\mu\nu\rho}{}^ B\delta A_{\sigma]}{}^ M\right. \nn \\ & \left.+2g_{N}{}^{BI}d_{IPQ}A_{[\mu}{}^ MA_{\nu}{}^ NA_{\rho}{}^ P\delta A_{\sigma]}{}^ Q\right)-\widehat{\Phi}_{\mu\nu\rho\sigma X}\ .
\end{align*}
In higher dimensions the new parameter would be identified with the gauge parameter of the five-form field. Consistency requires that it vanishes under the $W^{A X}$ projection
\begin{align*}
W^{AX}\widehat{\Phi}_{\mu\nu\rho\sigma X} & =0\ .
\end{align*}

We can finally summarize all the results for the full tensor hierarchy in generic gauged supergravities in four space-time dimensions:
\begin{itemize}

\item Parameters, fields and curvatures in four-dimensions

\begin{table}[ht]
\begin{center}
    \begin{tabular}{|l|l|l|l|}
    \hline  {Representation} &  {\bf Parameter } \ \ \ \ \ \ \  \ \ \ \   & {\bf Field} \ \ \ \  \ \ \ \ \ \ \  \ \ \ \   &  {\bf Curvature} \ \ \ \ \ \   \\ \hline
$R_1$ & $\Lambda^M$ & $A_\mu{}^M$ & ${\cal F}_{\mu \nu}{}^M$ \\ \hline
$R_2$ & $\widehat \Xi_{\mu I}$ & $B_{\mu \nu I}$ & ${\cal H}_{\mu \nu \rho I}$ \\ \hline
$R_3$ & $\widehat \Sigma_{\mu \nu}{}^A$ & $C_{\mu \nu \rho}{}^A$ & ${\cal G}_{\mu \nu \rho \sigma}{}^A$ \\ \hline
$R_4$ & $\widehat \Theta_{\mu \nu\rho X}$ & $D_{\mu\nu\rho\sigma X}$ & 0 \\ \hline
$R_4$ & $\widehat \Phi_{\mu \nu\rho\sigma X}$ & 0 & 0 \\ \hline
    \end{tabular}
    \end{center}
\end{table}

\item Gauge transformations
\begin{align*}
\delta A_{\mu}{}^ M & =\mathcal{D}_{\mu}\Lambda^{M}-Z^{MI}\widehat{\Xi}_{\mu I}\\
\delta B_{\mu\nu I} & =2\mathcal{D}_{[\mu}\widehat{\Xi}_{\nu]I}-2d_{IMN}\left(\Lambda^{M}\mathcal{F}_{\mu\nu}{}^ N-A_{[\mu}{}^ M\delta A_{\nu]}{}^ N\right)-Y_{IA}\widehat{\Sigma}_{\mu\nu}{}^ A\\
\delta C_{\mu\nu\rho}{}^ A & =3D_{[\mu}\widehat{\Sigma}_{\nu\rho]}{}^ A+g_{M}{}^{AI}\left(3\mathcal{F}_{[\mu\nu}{}^ M\widehat{\Xi}_{\rho]I}+\Lambda^{M}\mathcal{H}{}_{\mu\nu\rho}{}_ I+3B_{[\mu\nu I}\delta A_{\rho]}{}^ M \right. \\ & \ \ \ \left.+2d_{IMN}A_{[\mu}{}^ MA_{\nu}{}^ N\delta A_{\rho]}{}^ P\right)-W^{AX}\widehat{\Theta}_{\mu\nu\rho X}\\
\delta D_{\mu\nu\rho\sigma X} & =4\mathcal{D}_{[\mu}\widehat{\Theta}_{\nu\rho\sigma]X}-h_{XBM}\left(6\mathcal{F}_{[\mu\nu}{}^ M\widehat{\Sigma}_{\rho\sigma]}{}^ B+\Lambda^{M}\mathcal{G}_{\mu\nu\rho\sigma}{}^ B-4C_{[\mu\nu\rho}{}^ B\delta A_{\sigma]}{}^ M\right. \\ & \ \ \  \left.+2g_{N}{}^{BI}d_{IPQ}A_{[\mu}{}^ MA_{\nu}{}^ NA_{\rho}{}^ P\delta A_{\sigma]}{}^ Q\right)-\widehat{\Phi}_{\mu\nu\rho\sigma X}\ .
\end{align*}

\item Field Strengths
\begin{align}
\mathcal{F}_{\mu\nu} {}^{M} & =2\partial_{[\mu}A_{\nu]} {}^{M}+X_{NO}{}^MA_{[\mu} {}^{N}A_{\nu]} {}^{O}+Z^{MI}B_{\mu\nu I}\nn \\
\mathcal{H}_{\mu\nu\rho I} & =3\mathcal{D}_{[\mu}B_{\nu\rho]I}+6d_{IMN}A_{[\mu} {}^{M}\left(\partial_{\nu}A_{\rho]} {}^{N}+\frac{1}{3}X_{OP}{}^NA_{\nu} {}^{O}A_{\rho]} {}^{P}\right)+ Y_{I A} C_{\mu\nu\rho}{}^A \\
\mathcal{\mathcal{G}}_{\mu\nu\rho\sigma}{}^ A & =4\mathcal{D}_{[\mu}C_{\nu\rho\sigma]}{}^ A-g_{M}{}^{AI}\left(6B_{[\mu\nu I}\mathcal{F}_{\rho\sigma]}{}^ M-3Z^{MJ}B_{[\mu\nu I}B_{\rho\sigma]J}+8d_{INP}A_{[\mu}{}^ MA_{\nu}{}^ N\partial_{\rho}A_{\sigma]}{}^ P\right. \nn \\ & \ \ \ \left.+2d_{IPN}X_{QR}{}^ PA_{[\mu}{}^ MA_{\nu}{}^ NA_{\rho}{}^ QA_{\sigma]}{}^ R\right) +W^{AX}D_{\mu\nu\rho\sigma X}\ . \nn
\end{align}

\item Bianchi identities
\begin{align}
2\mathcal{D}_{[\mu}\mathcal{D}_{\nu]} + {\cal F}_{\mu \nu}{}^M X_M & = 0\\
3 \mathcal{D}_{\left[ \mu \right.}\mathcal{F}_{\left. \nu \rho \right]} {}^{M} - Z^{MI} \mathcal{H}_{\mu\nu\rho I} & = 0\\
4 \mathcal{D}_{\left[ \mu \right.}\mathcal{H}_{\left. \nu \rho \sigma \right] I} - 6 d_{IMN} \mathcal{F}_{\left[ \mu \nu \right.} {}^{M} \mathcal{F}_{\left. \rho \sigma \right]} {}^N - Y_{IA} \mathcal{G}_{\mu \nu \rho \sigma} {}^{A} &= 0\ .
\end{align}

\item Constraints
\begin{align*}
Z^{MI}d_{INP} & =X_{\left(NP\right)}{}^ M\\
Z^{MI}X_{M} & =0\\
d_{I[MN]} & =0\\
Y_{IA}g_{M}{}^{AJ} & =X_{MI}{}^ J+2Z^{NJ}d_{IMN}\\
Y_{IA}Z^{MI} & =0\\
d_{I(MN}g_{P)}{}^{AI} & =0\\
W^{AX}h_{XBM} & =X_{MB}{}^ A+g_{M}{}^{AI}Y_{IB}\\
W^{AX}Y_{IA} & =0\\
h_{XB(M}g_{N)}{}^{BI} & =0\\
W^{AX}\widehat{\Phi}_{\mu\nu\rho\sigma X} & =0
\end{align*}

\item Brackets
\begin{align*}
\Lambda_{12}^ M & =-X_{NP}{}^ M\Lambda_{[1}^ N\Lambda_{2]}^ P\\
\widehat{\Xi}_{12\mu I} & =2d_{IMN}\mathcal{D}_{\mu}\Lambda_{[1}^ M\Lambda_{2]}^ N\\
\widehat{\Sigma}_{12\mu\nu}{}^ A & =-2g_{M}{}^{AI}\left(d_{INP}\Lambda_{[1}^ M\Lambda_{2]}^ N\mathcal{F}_{\mu\nu}{}^ P+2\widehat{\Xi}_{I[\mu[1}\mathcal{D}_{\nu]}\Lambda_{2]}^ M-Z^{MJ}\widehat{\Xi}_{I[\mu[1}\widehat{\Xi}_{\nu]2]J}\right)\\
\widehat{\Theta}_{12\mu\nu\rho X} & =-h_{XBM}\left(g_{N}{}^{BI}\Lambda_{[1}^ M\Lambda_{2]}^ N\mathcal{H}_{\mu\nu\rho I}+6\widehat{\Sigma}_{[\mu\nu}{}^ B_{[1}\mathcal{D}_{\rho]}\Lambda_{2]}^ M\right)\\
\widehat{\Phi}_{12\mu\nu\rho\sigma X} & =-\left(X_{MX}{}^ Y+W^{AY}h_{XAM}\right)\left(h_{YBN}\Lambda_{[1}^ M\Lambda_{2]}^ N\mathcal{G}_{\mu\nu\rho\sigma}{}^ B+8\widehat{\Theta}_{Y[\mu\nu\rho[1}\mathcal{D}_{\sigma]}\Lambda_{2]}^ M\right)\\
 & +4h_{XBM}Z^{MI}\left[2W^{BY}\widehat{\Theta}_{Y[\mu\nu\rho[1}\widehat{\Xi}_{2]\sigma]I}-6\widehat{\Xi}_{J[\mu[1}\mathcal{D}_{\nu}\widehat{\Sigma}_{\rho\sigma]2]}{}^ B+2g_{N}{}^{BJ}\mathcal{H}_{J[\mu\nu\rho}\widehat{\Xi}_{\sigma]I[1}\Lambda_{2]}^ N \right. \\ & \ \ \ \ \ \ \ \ \ \ \ \ \ \ \ \ \ \ \ \ \ \left. +3\mathcal{F}_{[\mu\nu}{}^ N\left(2d_{INP}\Lambda_{[1}^ P\widehat{\Sigma}_{2]\rho\sigma]}{}^ B-g_{N}{}^{BJ}\widehat{\Xi}_{J \rho[1}\widehat{\Xi}_{2]\sigma]I}\right)\right]
\end{align*}

\item Trivial parameters
\begin{align*}
\Lambda_{trivial}^{M} & =Z^{MJ}\chi_{J}\\
\widehat{\Xi}_{\mu I\ trivial} & =\mathcal{D}_{\mu}\chi_{I}+Y_{IA}\chi_{\mu}{}^ A\\
\widehat{\Sigma}_{\mu\nu}{}^ A{}_{\ trivial} & =2\mathcal{D}_{[\mu}\chi_{\nu]}{}^ A-g_{M}{}^{AJ}\mathcal{F}_{\mu\nu}{}^ M\chi_{J}+W^{AX}\chi_{\mu\nu X}\\
\widehat{\Theta}_{\mu\nu\rho X}{}_{\ trivial} & =3\mathcal{D}_{[\mu}\chi_{\nu\rho]X}+3h_{XMB}\mathcal{F}_{[\mu\nu}{}^ M\chi_{\rho]}{}^ B+\chi_{\mu\nu\rho X}\\
\widehat{\Phi}_{\mu\nu\rho\sigma X\ trivial} & =4\mathcal{D}_{[\mu}\chi_{\nu\rho\sigma]X}-6\left(X_{MX}{}^ Y+W^{AY}h_{XAM}\right)\mathcal{F}_{[\mu\nu}{}^ M\chi_{\rho\sigma] Y}\\
 & \ \ \ +h_{XMB}Z^{MI}\left(4\mathcal{H}_{[\mu\nu\rho I}\chi_{\sigma]}{}^ B-\mathcal{G}_{\mu\nu\rho\sigma}{}^ B\chi_{I}\right) \ .
\end{align*}

\end{itemize}

We then see that the symmetry for symmetries effect in gauged supergravities is a generic feature even for dimensionally saturated tensor hierarchies.
\end{appendix}

\end{document}